\documentclass[prd,tightenlines,showpacs,groupedaddress,superscriptaddress,11pt,nofootinbib,floatfix,preprintnumbers,letterpaper]{revtex4}
\usepackage{hyperref,amssymb,amsmath,graphicx,xcolor,cancel}
\usepackage{bm}

\newcommand{\gsim}{\raisebox{-0.7ex}{$\stackrel{\textstyle >}{\sim}$ }}

\def\si{^1 \hskip -0.03in S _0}
\def\siii{^3 \hskip -0.025in S _1}
\def\diii{^3 \hskip -0.03in D _1}

\def\pionmass{450~{\rm MeV}}
\def\pionmassFULL{449.9(0.3)(0.3)(4.6) ~{\rm MeV} }
\def\pionmassFULLa{449.9(0.3)(0.3)(4.6)}

\def\kaonmassFULL{595.9(0.2)(0.2)(6.1) ~{\rm MeV} }
\def\kaonmassFULLa{595.9(0.2)(0.2)(6.1)  }

\def\rhomassFULLa{887.3(2.4)(2.5)(9.1)  }

\def\kstarmassFULLa{962.4(1.5)(0.9)(9.9)  }

\def\MNFULLa{1226(01)(01)(12)}

\def\lammassFULLa{1312(01)(01)(13) }
\def\sigmassFULLa{1346(01)(01)(14) }
\def\ximassFULLa{1415(01)(01)(15) }

\def\deltamassFULLa{1486(02)(03)(15) }
\def\sigstarmassFULLa{1557(03)(03)(16) }
\def\xistarmassFULLa{1629(02)(03)(17) }
\def\omegamassFULLa{1700(01)(02)(17) }

\def\revcolor{black}
\def\revrevcolor{blue}
\def\revrevcolor{black}
\def\revrevrevcolor{red}
\def\revrevrevcolor{black}

\def\bFULL{0.1167(12)}
\def\bFULL{0.1167(16)}

\def\La{2.801(29)}
\def\Lb{3.734(38)}
\def\Lc{5.602(58)}

\def\Ta{7.469(77)}
\def\Tb{11.20(12)}
\def\Tc{11.20(12)}
\def\NcfgA{4407} 
\def\NscrA{\textcolor{\revcolor}{ $1.16\times 10^6$ }}
\def\NcfgB{4142} 
\def\NscrB{\textcolor{\revcolor}{ $3.95\times 10^5$}}
\def\NcfgC{\textcolor{\revcolor}{1047}} 
\def\NscrC{\textcolor{\revcolor}{$6.8\times 10^4$}}

\def\cfga{$24^3\times 64$\ }
\def\cfgb{$32^3\times 96$\ }
\def\cfgc{$48^3\times 96$\ }

\def\Bd{14.4^{+(1.6)(2.7)(0.2)(0.2)}_{-(1.8)(1.8)(0.2)(0.2)} }
\def\BdSUMMARY{14.4^{+3.2}_{-2.6} }

\def\Bnn{15.2^{+(1.7)(4.4)(0.2)(0.2)}_{-(1.7)(3.7)(0.2)(0.2)}}
\def\BnnSUMMARY{15.2^{+4.7}_{-4.1}}
\def\Bnn{12.5^{+(1.7)(2.5)(0.2)(0.2)}_{-(1.9)(4.5)(0.2)(0.2)}}
\def\BnnSUMMARY{12.5^{+3.0}_{-5.0}}

\def\Bdo{14.4^{+(1.6)(2.7)(0.2)(0.2)}_{-(1.8)(1.8)(0.2)(0.2)} }
\def\Bdn{14.3^{+(1.8)(3.7)(0.2)(0.2)}_{-(2.0)(3.7)(0.2)(0.2)} }

\def\BdSUMMARYn{14.3^{+(4.1)(0.3)}_{-(4.2)(0.3)} }

\begin{document}

\title{Two Nucleon Systems at $m_\pi\sim 450~{\rm MeV}$ from Lattice QCD}

\author{Kostas~Orginos} \affiliation{Department of Physics, College of
  William and Mary, Williamsburg, VA 23187-8795, USA}
\affiliation{Jefferson Laboratory, 12000 Jefferson Avenue, Newport
  News, VA 23606, USA}

\author{Assumpta~Parre\~no} \affiliation{Dept. d'Estructura i
  Constituents de la Mat\`eria.  Institut de Ci\`encies del Cosmos
  (ICC), Universitat de Barcelona, Mart\'{\i} Franqu\`es 1,
  E08028-Spain}

\author{Martin J. Savage}
\affiliation{Institute for Nuclear Theory, University of Washington, Seattle, WA 98195-1550, USA}

\author{Silas~R.~Beane} \affiliation{Department of Physics, University
  of Washington, Box 351560, Seattle, WA 98195, USA}
  
\author{Emmanuel~Chang} 
\affiliation{Institute for Nuclear Theory, University of Washington, Seattle, WA 98195-1550, USA}

\author{William Detmold} \affiliation{Center for Theoretical Physics,
  Massachusetts Institute of Technology, Cambridge, MA 02139, USA}

\collaboration{NPLQCD Collaboration}

\date{\today}

\preprint{INT-PUB-15-042}  
\preprint{NSF-KITP-15-120}  
\preprint{NT@UW-15-09}
\preprint{MIT-CTP-4709}

\pacs{11.15.Ha, % Lattice gauge theory
  12.38.Gc, % Lattice QCD calculations
}

%%%%%%%%%%%%%%%%%%%%%%%%%%%%%%%%%%%%%%%%%%%%%%
\begin{abstract}
Nucleon-nucleon systems are studied 
with lattice quantum chromodynamics
at a pion mass of $m_\pi\sim\pionmass$
in three spatial volumes using  $n_f=2+1$ flavors of light quarks.
At the quark masses employed in this work, 
the deuteron binding energy is calculated to be  $B_d = \BdSUMMARY~{\rm MeV}$,
while the dineutron is bound by  $B_{nn} = \BnnSUMMARY~{\rm MeV}$.
Over 
the range of energies that are studied, 
the S-wave scattering phase shifts calculated in 
the $\si$ and $\siii$-$\diii$
 channels are found to be 
similar to those in nature,
and indicate repulsive short-range  components of the interactions, consistent with
phenomenological nucleon-nucleon interactions.
In both channels, 
the phase shifts are determined at three energies that
lie within the radius of convergence of the effective range expansion, 
allowing for constraints to be placed on the inverse scattering lengths and effective ranges.
The extracted phase shifts  allow for 
matching to nuclear effective field theories, 
from which low energy counterterms are extracted and issues of convergence are investigated.
As part of the analysis, a detailed investigation of the single hadron sector is performed, enabling a 
precise determination of the violation of the Gell-Mann--Okubo mass relation.

\vskip 0.1in
\noindent
[An Erratum to the published version is included as an appendix. It details the impact of an error discovered in 2020 and corrects typographical errors.]
\end{abstract}
\maketitle

%%%%%%%%%%%%%%%%%%%%%%%%%%%%%%%%%%%%%%%%%%%%%%%
\section{Introduction}
\label{sec:intro}
Calculating the interactions between nucleons and the properties of multi-nucleon systems directly from quantum chromodynamics (QCD) 
will be an important milestone in the development of nuclear physics.
While Lattice QCD (LQCD) calculations of simple hadronic systems 
are now being performed at the physical light-quark masses and the effects of quantum electrodynamics (QED) 
are beginning to be included 
(see, e.g. Ref.~\cite{Borsanyi:2014jba}),
such calculations 
have not yet been presented 
for more complex systems such as nuclei.
However, remarkable progress has been made in the  ongoing efforts to calculate the lowest-lying energy levels of the simplest nuclei and hypernuclei (with $A\le 4$) and the nucleon-nucleon scattering S-matrix elements~\cite{Beane:2006mx,Ishii:2006ec,Beane:2009py,Beane:2009gs,Yamazaki:2009ua,Beane:2010hg,Inoue:2010hs,Beane:2010em,Yamazaki:2011nd,Beane:2011iw,Murano:2011nz,Doi:2011gq,Beane:2012vq,Yamazaki:2012hi,Aoki:2012tk,Beane:2013br,Ishii:2013ira,Yamazaki:2013rna,Murano:2013xxa,Yamazaki:2015asa,Berkowitz:2015eaa}.
The magnetic moments  and polarizabilities
of the light nuclei have recently been calculated~\cite{Beane:2014ora,Chang:2015qxa}, and
by determining the short-range interaction between nucleons and the electromagnetic field,
the first LQCD calculation of the radiative capture process $np\rightarrow d\gamma$~\cite{Beane:2015yha}
was recently  performed and  the experimentally measured cross section was recovered 
within the uncertainties of the calculation after extrapolation to the physical quark masses.
These calculations   represent crucial steps toward verifying LQCD as a 
useful technique with which to calculate the properties of nuclear systems.
However, it will take significant computational resources to reduce   
the associated 
uncertainties  below those of experiment.
Near term advances in the field will come from calculations of quantities that are challenging or impossible to access experimentally, such as multi-nucleon forces,   hyperon-nucleon interactions, 
rare weak matrix elements and exotic nuclei, such as hypernuclei and charmed nuclei, that are of modest computational complexity.
Further, performing calculations  specifically to match LQCD results to low-energy effective field theories (EFTs) will provide a means to  make first 
predictions at the physical quark masses and to make predictions of quantities beyond those calculated with LQCD.
Such calculations are now underway, using the results of our previous works and those of Yamazaki {\it et al}., 
with the first efforts described, for example,  in Ref.~\cite{Beane:2012ey} for hyperon-nucleon interactions
and Ref.~\cite{Barnea:2013uqa,Baru:2015ira} for nucleon-nucleon interactions and light nuclei.

In this work, we present the results of LQCD calculations of two-nucleon systems performed 
at a pion mass of $m_\pi\sim\pionmass$ in three lattice volumes of spatial extent 
$L= 2.8~{\rm fm}, 3.7 ~{\rm fm}$ and $ 5.6~{\rm fm}$ at a lattice spacing of $b\sim 0.12~{\rm fm}$.
\textcolor{\revrevrevcolor}{
As only one lattice spacing has been employed, extrapolations of the results to the continuum limit have not been performed,
although the uncertainties that we finally present encompass the expected effects of these extrapolations.
}
In Section~\ref{sec:methods}, we introduce the LQCD methods that are used to determine correlation functions and 
Section~\ref{sec:SingleHadrons} reports the results of precision studies of the single hadron systems. 
Section~\ref{sec:3s1} explores the $\siii$-$\diii$ coupled-channel systems in detail, 
while the $\si$ channel is discussed in 
Section~\ref{sec:1s0}. 
In Section~\ref{sec:NNEFT},  these channels are further investigated in the context of 
nucleon-nucleon effective field theory (NNEFT) 
before the conclusions of the study are presented in  
Section~\ref{sec:concs}.

%%%%%%%%%%%%%%%%%%%%%%%%%%%%%%%%%%%%%%%%%%%%%%%
\section{Methodology}
\label{sec:methods}

\subsection{Calculational Details}
\label{subsec:calcs}

LQCD calculations were performed on three ensembles of $n_f=2+1$ isotropic  
gauge-field configurations with $L=24,32$ and $48$ lattice sites in each spatial direction, 
$T=64,96,96$ sites in the temporal direction, respectively,  and with a lattice spacing of $b=0.1167(16)~{\rm fm}$~\cite{MeinelScale}.
The L\"uscher-Weisz gauge action~\cite{Luscher:1984xn}  was used with a clover-improved quark 
action~\cite{Sheikholeslami:1985ij} 
with one level of stout smearing ($\rho=0.125$)~\cite{Morningstar:2003gk}. 
The clover coefficient was set equal to its tree-level tadpole-improved value,  
a value that is consistent with an independent numerical study of the 
nonperturbative $c_{\rm SW} $
in the Schr\"odinger functional scheme~\cite{Hoffmann:2007nm,Edwards:2008ja,Brown:2014ena}, reducing discretization errors from
${\cal O}(b)$ to ${\cal O}(b^2)$. 
The $L=24,32$ and $48$ ensembles consist of 
$3.4\times 10^4$, 
$2.2\times 10^4$, and  
$1.5\times 10^4$
HMC evolution trajectories, respectively.
Calculations were performed on  gauge-field configurations taken at uniform intervals from these trajectories,
see Table~\ref{tab:gauageparams}.
The strange-quark mass was tuned to that of the physical strange quark, while the selected light-quark mass 
gave rise to a pion of mass  $m_\pi=\pionmassFULL$ and a kaon of mass $m_K=\kaonmassFULL$.
Many
details of the current study mirror those of our previous work at the SU(3) symmetric point, which can be found in
Refs.~\cite{Beane:2012vq,Beane:2013br}.
In each run on a given configuration, 48 quark propagators were generated 
from uniformly distributed Gaussian-smeared sources 
on a cubic grid with an origin randomly selected within the  volume. 
\textcolor{\revrevrevcolor}{
The parameters of the Gaussian smearing are the same as those used in Refs.~\cite{Beane:2012vq,Beane:2013br}.
}
Multiple runs were performed to increase statistical precision and the total number of measurements is recorded 
in Table~\ref{tab:gauageparams}.
Specifics of the ensembles and the number of sources used in each ensemble can also be found in 
Table~\ref{tab:gauageparams}.
\begin{table}
\begin{center}
\begin{minipage}[!ht]{16.5 cm}
  \caption{
  Parameters of the ensembles of gauge-field configurations and of the measurements used in this work.
    The lattices have dimension  $L^3\times T$, a lattice spacing $b$, and
    a bare quark mass $b\ m_q$ (in lattice units). 
$N_{\rm src}$ light-quark sources are used (as described in the text)
to perform measurements on $N_{\rm cfg}$ configurations in each ensemble.
  }  
\label{tab:gauageparams}
\end{minipage}
\setlength{\tabcolsep}{0.3em}
\begin{tabular}{c|cccccccccccc}
\hline
      Label & $L/b$ & $T/b$ & $\beta$ & $b\ m_l$ & $b\ m_s$ & $b$ [fm]  & $L$ [fm] & $T$
      [fm] & $m_\pi L$ & $m_\pi T$ & $N_{\rm cfg}$ & $N_{\rm src}$\\
\hline
      A & 24 & 64 & 6.1 & -0.2800 & -0.2450 & \bFULL  & \La & \Ta  &   6.390 & 17.04 & \NcfgA & \NscrA\\
      B & 32 & 96 & 6.1 & -0.2800 & -0.2450 & \bFULL  & \Lb & \Tb  & 8.514 & 25.54  & \NcfgB & \NscrB\\	
      C & 48 & 96 & 6.1 & -0.2800 & -0.2450 & \bFULL  & \Lc & \Tc  & 12.78 & 25.49 & \NcfgC & \NscrC\\
\hline
\end{tabular}
\begin{minipage}[t]{16.5 cm}
\vskip 0.0cm
\noindent
\end{minipage}
\end{center}
\end{table}     
Quark propagators were computed using the multigrid algorithm~\cite{Clark:2011}
or using GPUs~\cite{Clark:2010,Babich:2011}
with a tolerance of $10^{-12}$ in double precision. 
In the measurements performed on the $L=24$ and $32$ ensembles, 
the quark propagators, either unsmeared or smeared at the sink using the same parameters as used at the source, provided two sets of correlation functions for each combination of source and sink interpolating fields, labeled as SP and SS, respectively.  
In contrast, for the measurements performed on the $L=48$ ensemble only  SP correlation functions were produced.
The propagators were contracted into baryon blocks that were projected to a well-defined 
momentum at the sink, that were then used to form the one- and two-nucleon correlation functions.~\footnote{
\textcolor{\revrevrevcolor}{
As such, the same Gaussian smeared quark propagators were used to generate the single-nucleon and two-nucleon correlation functions.
We have employed a small number of different source and sink structures,
as described in Ref.~\cite{Detmold:2012eu}, 
and have presented optimal combinations for each hadron.
The ground state energies extracted from the correlation functions for a given species of hadron are consistent within uncertainties.
}
}
The  blocks are of the form
\begin{eqnarray}
{\cal B}_N^{ijk}({\bf p},t; x_0) & = & 
\sum_{\bf x} e^{i {\bf p}\cdot {\bf x}} 
S_i^{(f_1),i^\prime}({\bf x},t; x_0)
S_j^{(f_2),j^\prime}({\bf x},t; x_0)
S_k^{(f_3),k^\prime}({\bf x},t; x_0) 
b_{i^\prime j^\prime k^\prime}^{(N)}
\ \ \ ,
\label{eq:NucBlock}
\end{eqnarray}
where $S^{(f)}$ is a quark propagator of flavor $f=u,d$, and the indices are combined spin-color indices running 
over $i = 1, . . . , N_c N_s$, where $N_c=3$ is the number of colors and $N_s=4$ is the number of spin components. 
The choice of the $f_i$ and the tensor $b^{(N)}$ depend on the spin and flavor of the nucleon under consideration,
and the local interpolating fields constructed in Ref.~\cite{Detmold:2012eu}, restricted to those that contain only upper-spin components 
(in the Dirac spinor basis) are used.
This choice results in the simplest interpolating fields that also have good overlap with the nucleon ground states (from localized sources). 
Blocks are constructed for all lattice momenta $|{\bf p}|^2 < 5$ allowing for the study of two-nucleon systems with zero or nonzero total momentum.
In the production on the $L=32$ ensemble, correlation functions were produced for all of the spin states associated with each nuclear species.  
However, only one spin state per species was calculated on the $L=24$ and $L=48$ ensembles.

%%%%%%%%%%%%%%%%%%
\subsection{Robust Estimators: The Mean with Jacknife and the  Hodges-Lehmann Estimator with Bootstrap}
\label{sec:JNandHL}

The correlation functions are estimated from calculations performed from many source locations on many gauge-field configurations.
On any given configuration, these results are correlated 
and, because they
become translationally invariant after averaging,
they can be blocked together 
to generate one representative correlation function for each configuration.
More generally, because of the correlation between  nearby configurations 
produced in a Markov chain, 
the results obtained over multiple gauge-field configurations are  blocked together to produce 
one representative correlation function from any particular subsequence of the Markov chain.  
In this work, there are a large number of independent representative correlation functions which, by the central limit theorem,
tend to possess a Gaussian-distributed mean.
As computational resources are finite,  only a finite number of calculations of each correlation function can be performed.
The underlying distributions of the nuclear correlation functions are non-Gaussian with extended tails, and therefore 
outliers are typically present in any sample which
lead to slow convergence of the mean.
This can then lead to significant fluctuations in estimates of correlation functions when resampling methods, such as Bootstrap and 
Jacknife, are employed using the mean to estimate average values 
(for a discussion of the ``noise'' associated with these and other such calculations, see Ref.~\cite{Beane:2009kya,Endres:2011jm,Detmold:2014hla}).
Dealing with outliers of distributions is required in many areas beyond LQCD, and there is extensive literature on 
{\it robust estimators} that are resilient to the presence of outliers, such as the median or the Hodges-Lehmann (HL) estimator~\cite{Hodges:1963}.  
The vacuum expectation values  of interest in quantum field theory are defined by the mean value of a (generally non-Gaussian) distribution. 
Nevertheless, with sufficient blocking, the mean of the distribution will be Gaussian distributed, for which the mean, median, mode and HL estimator coincide. 
It therefore makes sense to consider such robust estimators for 
large sets of blocked LQCD correlation functions.

While the median 
of a sample $\{x_i\}$
is well known, the HL-estimator is less so.
It is a robust and unbiased estimator of the median of a sample, 
and is defined as~\cite{Hodges:1963}
\begin{eqnarray}
{\rm HL}(\{x_i\}) & = & {\rm Median}\left[ \{ (x_i+x_j)/2 \} \right]
\ \ \ ,
\end{eqnarray}
where the sample is summed over all $1<i,j <N$,
where $N$ is the sample size.
The  uncertainty associated with the HL-estimator is derived from the Median Absolute Deviation (MAD), defined as 
\begin{eqnarray}
{\rm MAD}(\{x_i\})  & = & {\rm Median}\left[  \{ |x_i - {\rm Median} \left[\{x_i \}\right] |\} \right]
\ \ \ .
\end{eqnarray}
For a Gaussian distributed sample, $1\sigma = 1.4826~{\rm MAD}$s.
The median, HL-estimator and other similar estimators 
cannot be computed straightforwardly under Jackknife, and instead such 
analyses are performed with  Bootstrap resampling.

In the present work, the correlation functions, and their ratios, are 
analyzed using both the mean under Jackknife  and HL under Bootstrap, 
from $\sim 100$  representative correlation functions constructed by blocking the full set of correlation functions.
In almost all cases, the HL with Bootstrap 
gives rise to 
smaller statistical fluctuations over the resampled ensembles
and, consequently, to smaller uncertainties in estimates of energies,
as seen in our previous investigation into robust estimators~\cite{Beane:2014oea}.
It is  found that 
outlying  blocked correlation functions 
cause a significant enlargement of the estimated variance of the mean, 
while the robust HL-estimator is  insensitive to them.

%%%%%%%%%%%%%%%%%%
\section{Single Mesons and Baryons}
\label{sec:SingleHadrons}

\begin{figure}[!ht]
  \centering
  \includegraphics[width=0.31 \columnwidth]{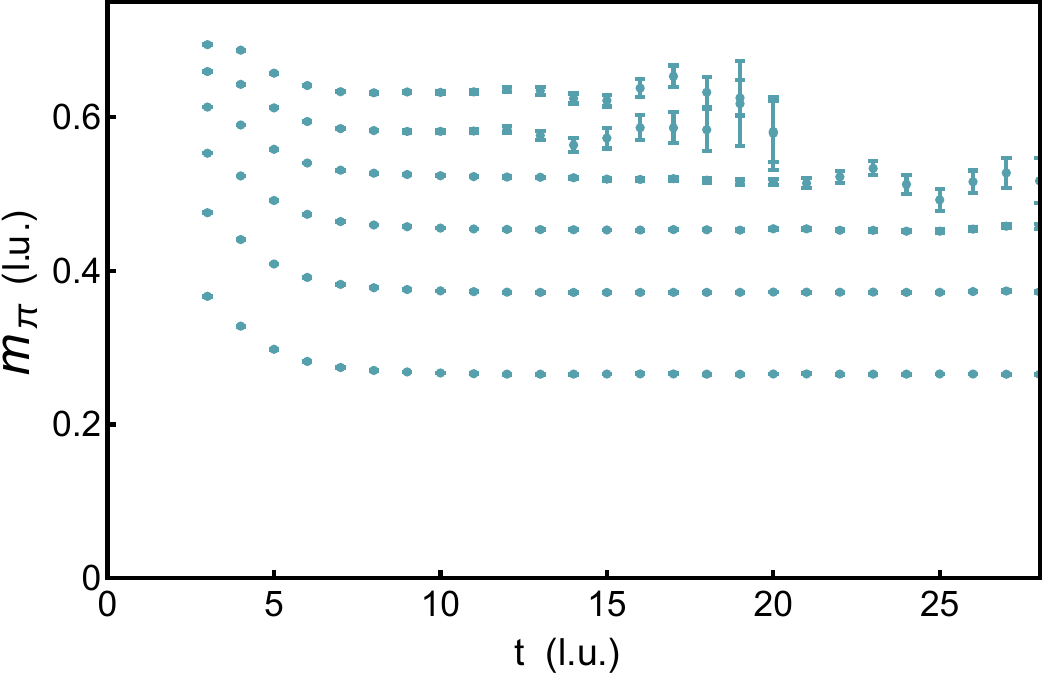}   \includegraphics[width=0.31 \columnwidth]{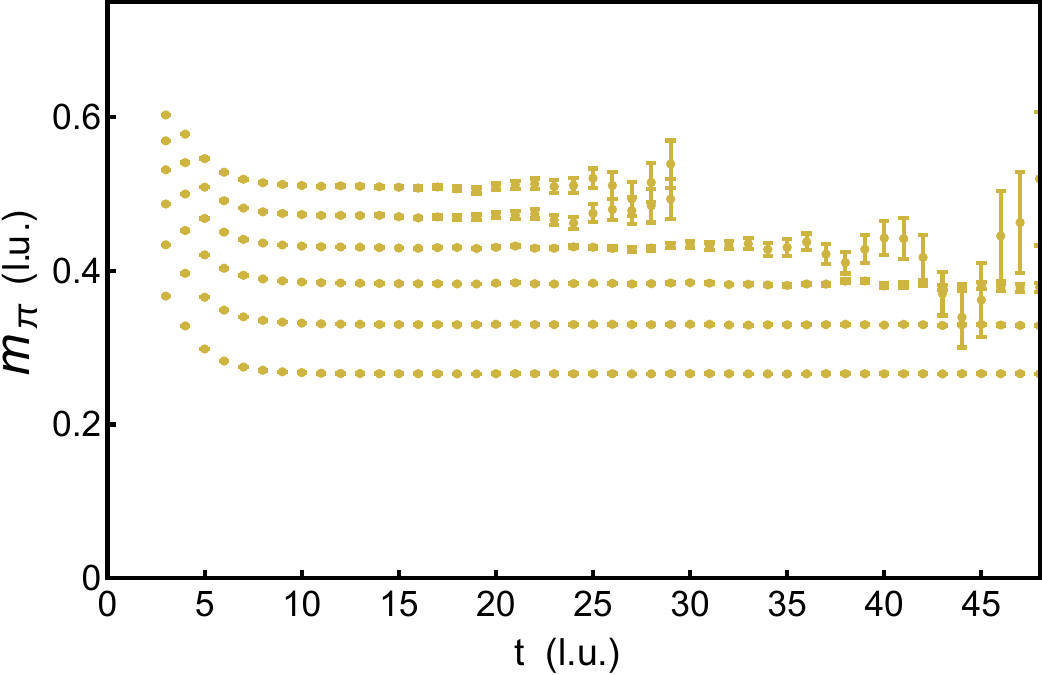}   \includegraphics[width=0.31 \columnwidth]{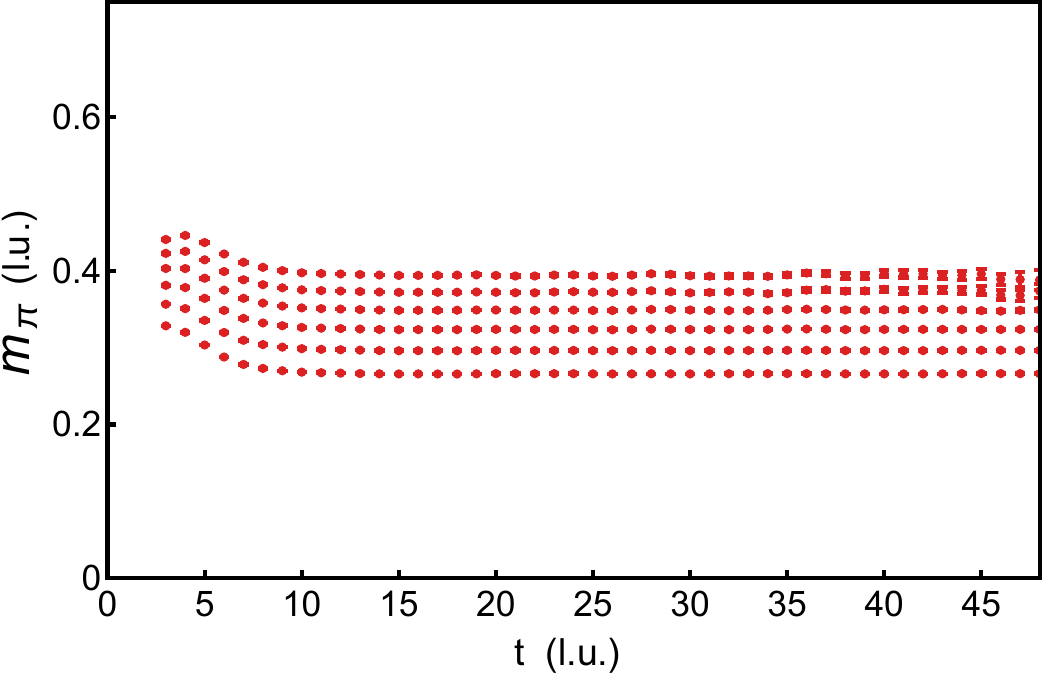} 
  \caption{
Cosh EMPs for the $\pi^\pm$ in the $L=24$ (left), $L=32$ (center), $L=48$ (right) lattice volumes, respectively.
In ascending order, the momenta are ${\bf P}=2\pi {\bf n} /L$ with $|{\bf n}|^2 = 0,1,2,3,4,5$.
    }
  \label{fig:pionEMPs}
\end{figure}

\begin{figure}[!ht]
  \centering
  \includegraphics[width=0.31 \columnwidth]{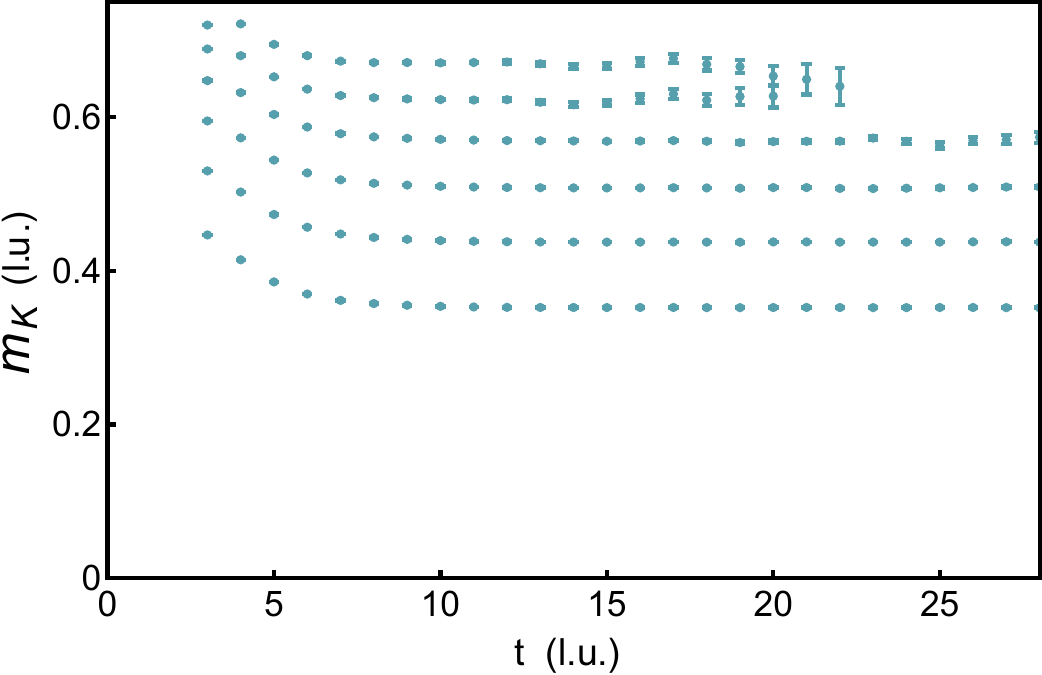}   \includegraphics[width=0.31 \columnwidth]{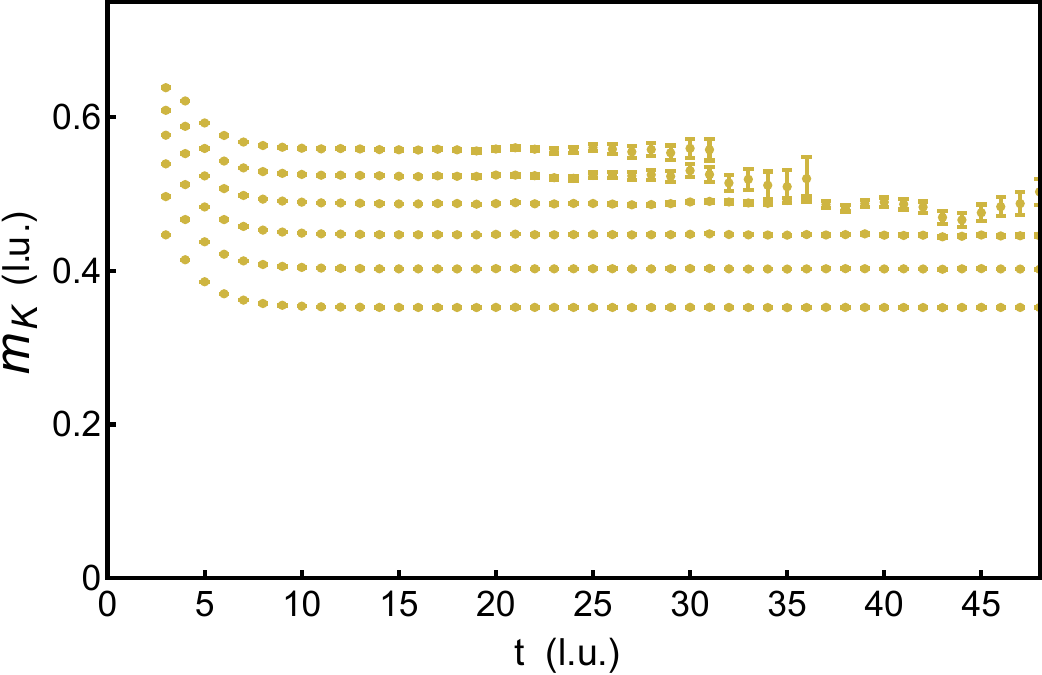}   \includegraphics[width=0.31 \columnwidth]{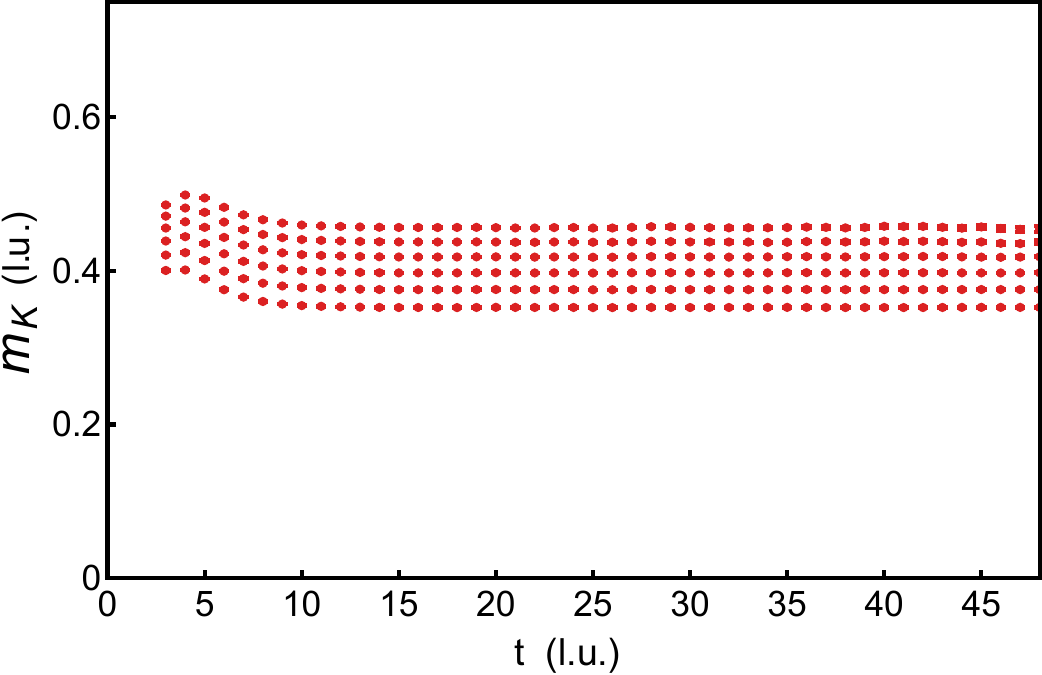} 
  \caption{
Cosh EMPs for the $K^\pm$ in the $L=24$ (left), $L=32$ (center), $L=48$ (right) lattice volumes, respectively.
In ascending order, the momenta are ${\bf P}=2\pi {\bf n} /L$ with $|{\bf n}|^2 = 0,1,2,3,4,5$.
    }
  \label{fig:kaonEMPs}
\end{figure}

\begin{figure}[!ht]
  \centering
  \includegraphics[width=0.31 \columnwidth]{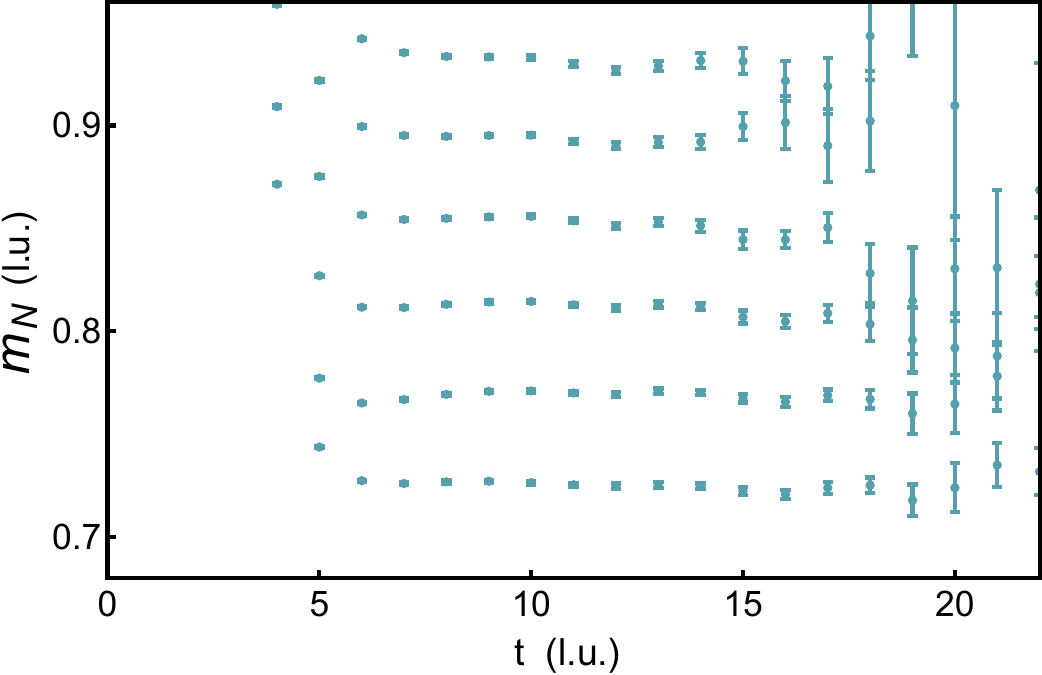}   \includegraphics[width=0.31 \columnwidth]{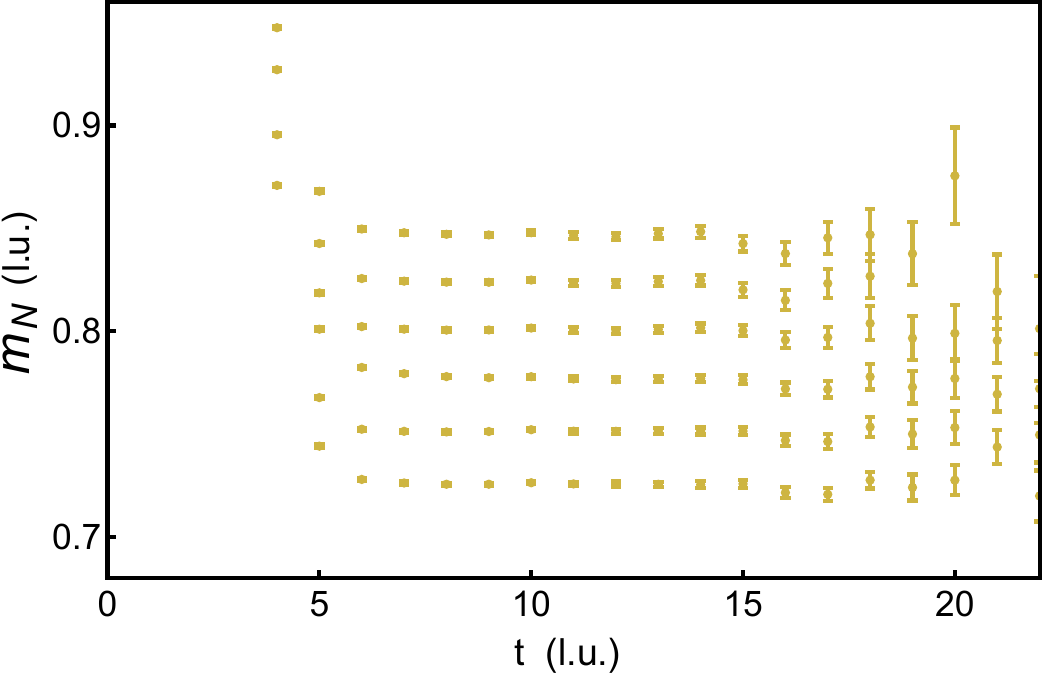}   \includegraphics[width=0.31 \columnwidth]{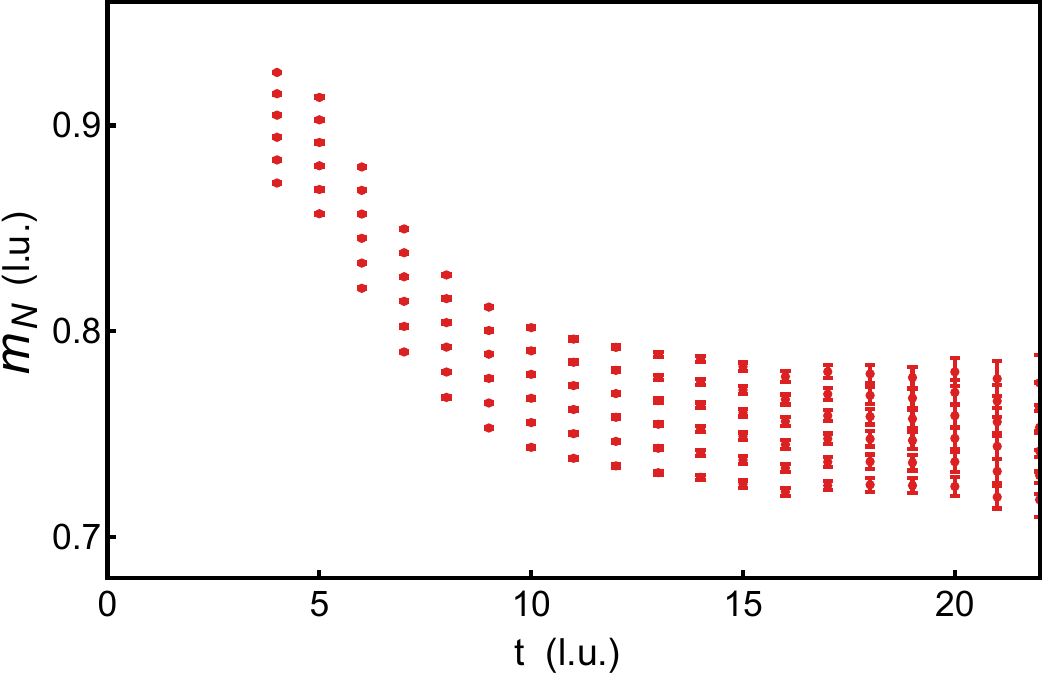} 
  \caption{
EMPs for the nucleon in the $L=24$ (left), $L=32$ (center), $L=48$ (right) lattice volumes, respectively.
In ascending order, the momenta are ${\bf P}=2\pi {\bf n} /L$ with $|{\bf n}|^2 = 0,1,2,3,4,5$.
    }
  \label{fig:NEMPs}
\end{figure}

\begin{figure}[!ht]
  \centering
  \includegraphics[width=0.31 \columnwidth]{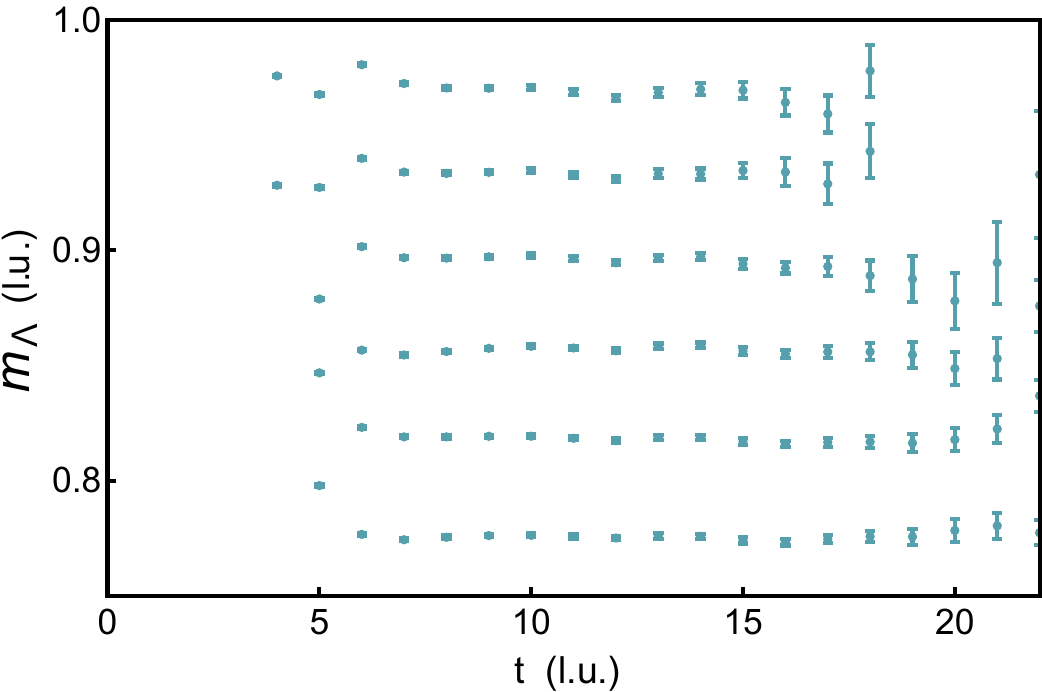}   \includegraphics[width=0.31 \columnwidth]{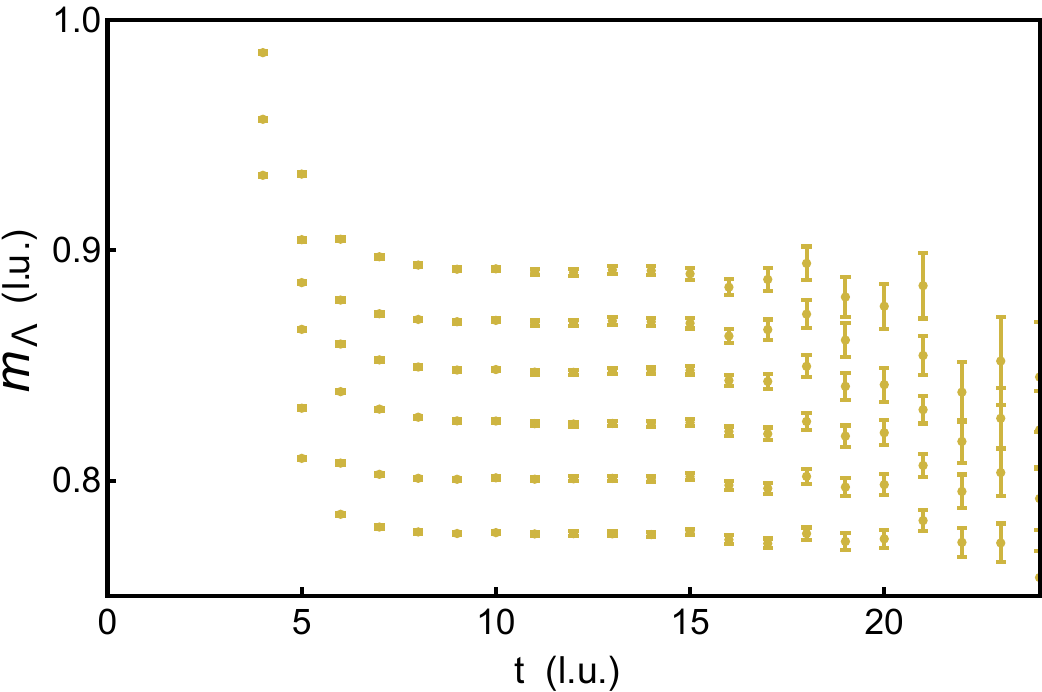}   \includegraphics[width=0.31 \columnwidth]{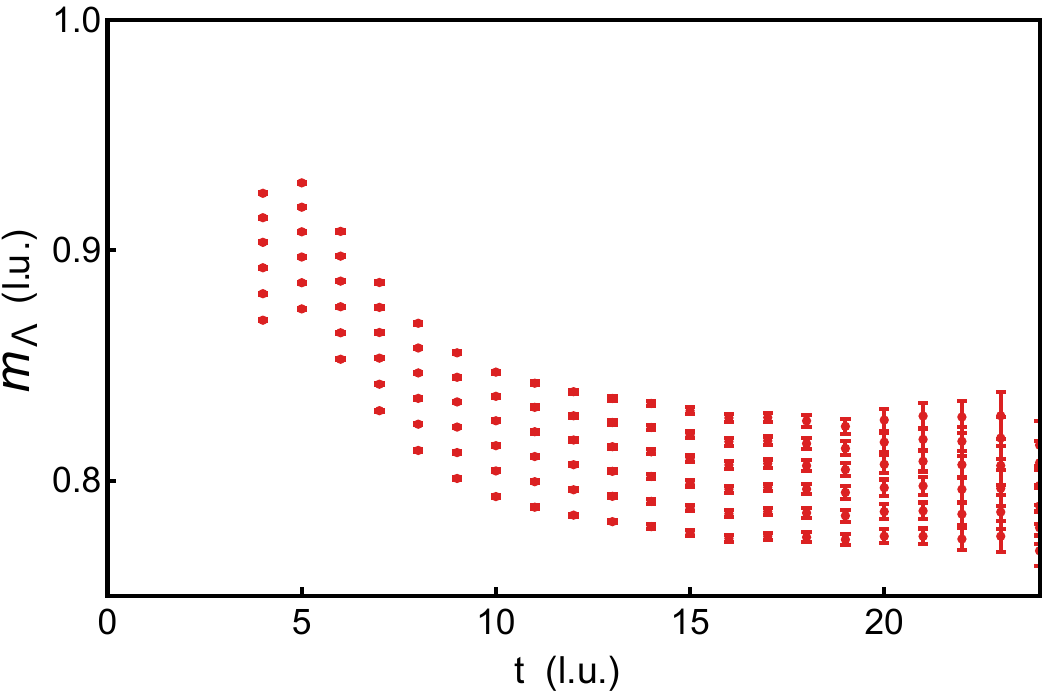} 
  \caption{
EMPs for the $\Lambda$ in the $L=24$ (left), $L=32$ (center), $L=48$ (right) lattice volumes, respectively.
In ascending order, the momenta are ${\bf P}=2\pi {\bf n} /L$ with $|{\bf n}|^2 = 0,1,2,3,4,5$.
    }
  \label{fig:LamEMPs}
\end{figure}

\begin{figure}[!ht]
  \centering
  \includegraphics[width=0.31 \columnwidth]{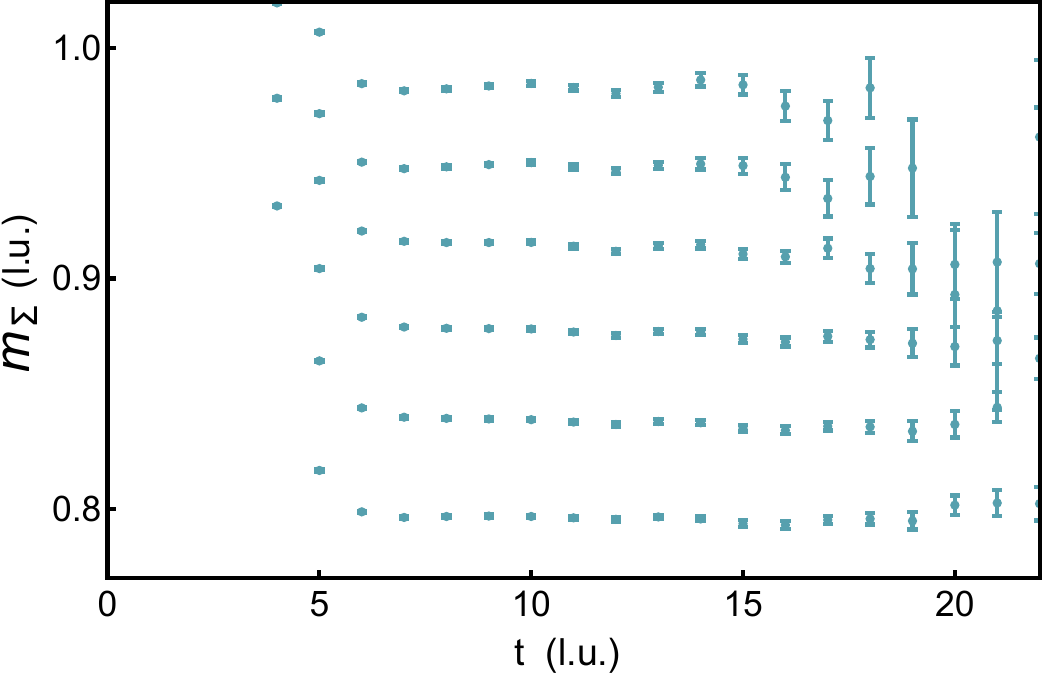}   \includegraphics[width=0.31 \columnwidth]{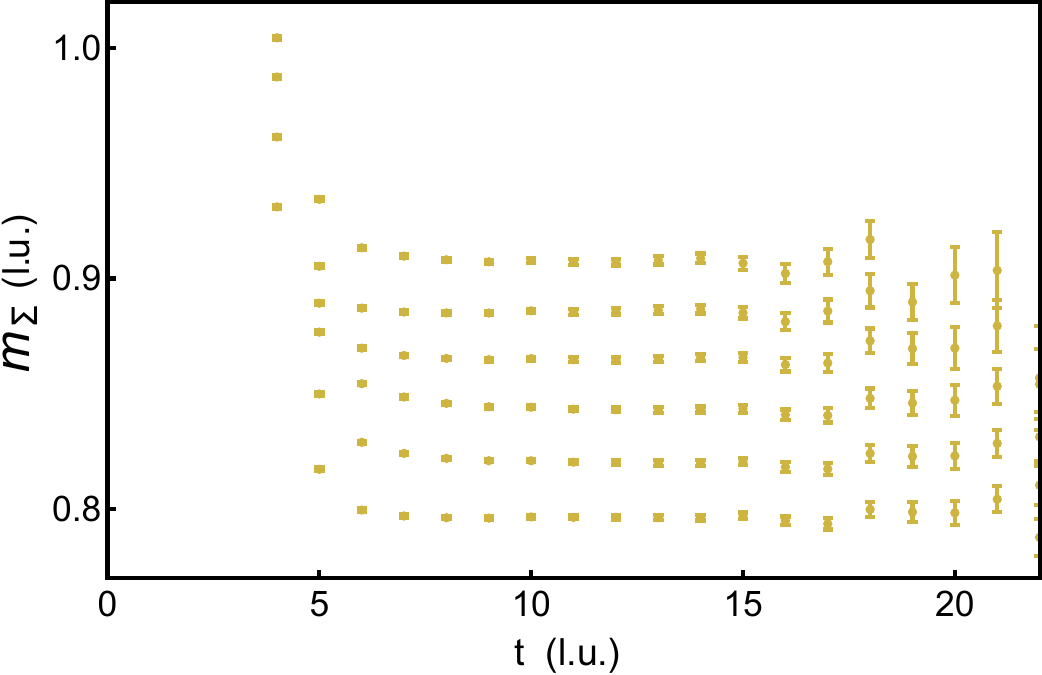}   \includegraphics[width=0.31 \columnwidth]{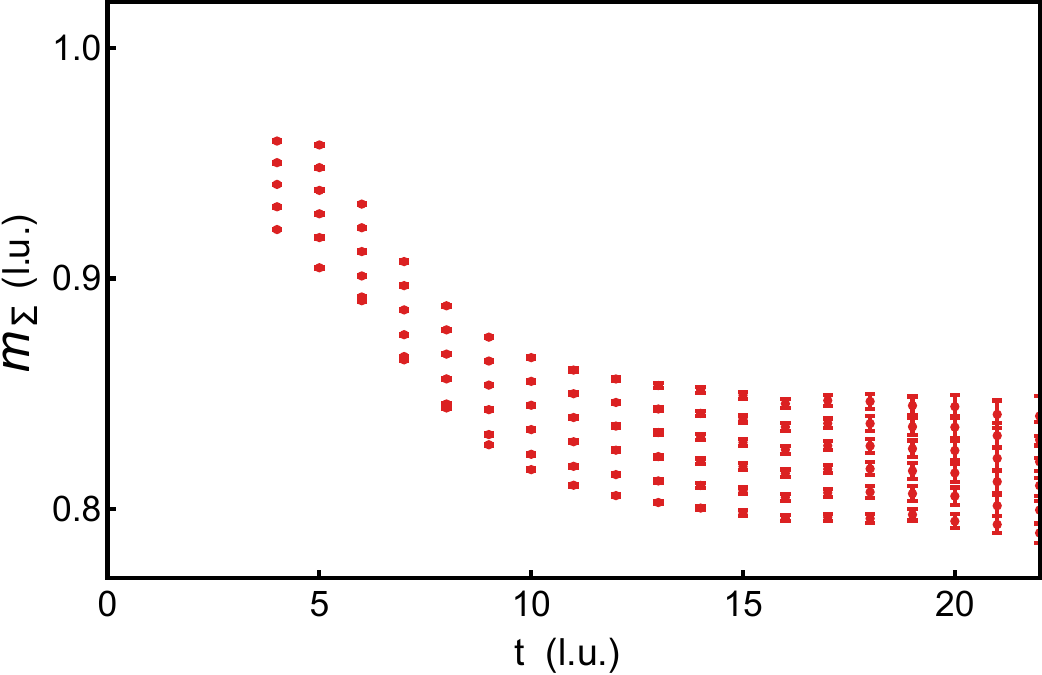} 
  \caption{
EMPs for the $\Sigma$ in the $L=24$ (left), $L=32$ (center), $L=48$ (right) lattice volumes, respectively.
In ascending order, the momenta are ${\bf P}=2\pi {\bf n} /L$ with $|{\bf n}|^2 = 0,1,2,3,4,5$.
    }
  \label{fig:SigEMPs}
\end{figure}

\begin{figure}[!ht]
  \centering
  \includegraphics[width=0.31 \columnwidth]{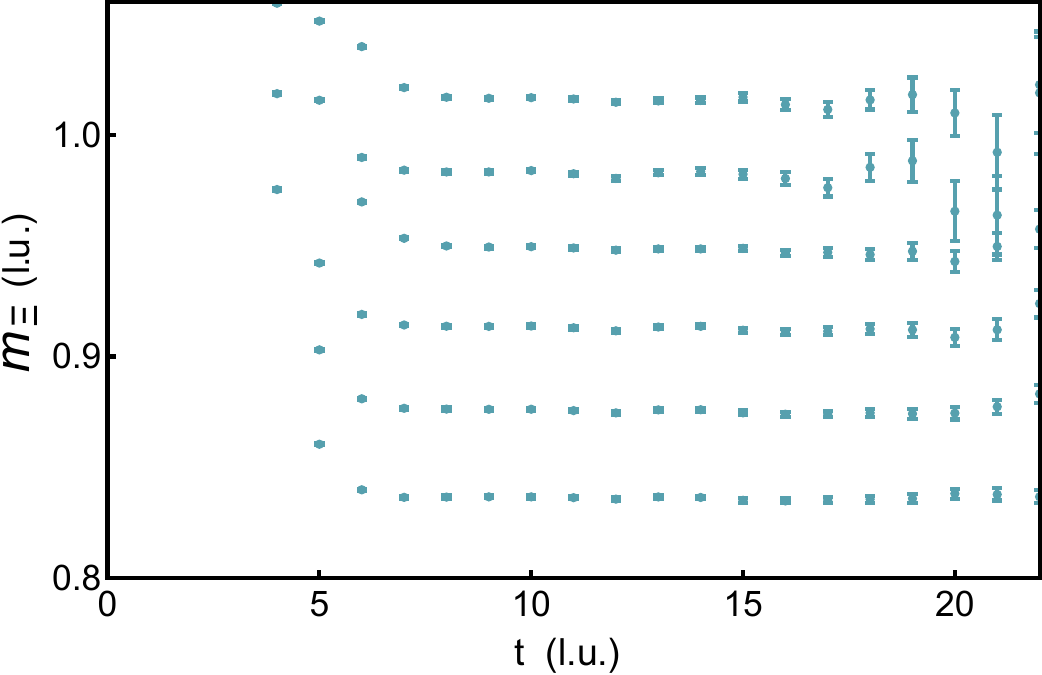}   \includegraphics[width=0.31 \columnwidth]{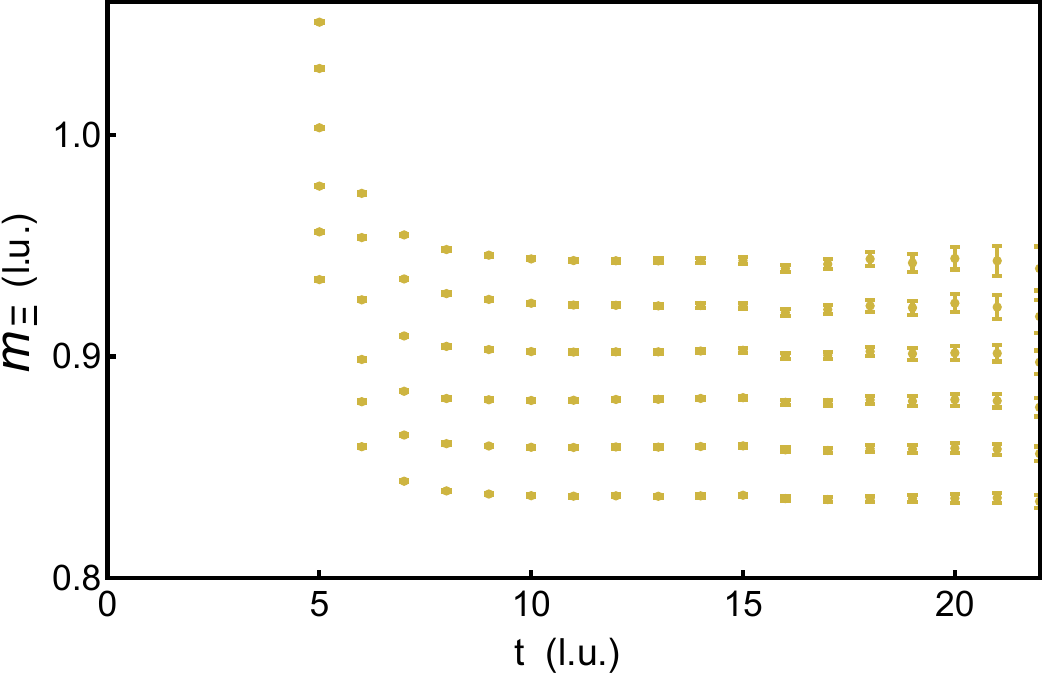}   \includegraphics[width=0.31 \columnwidth]{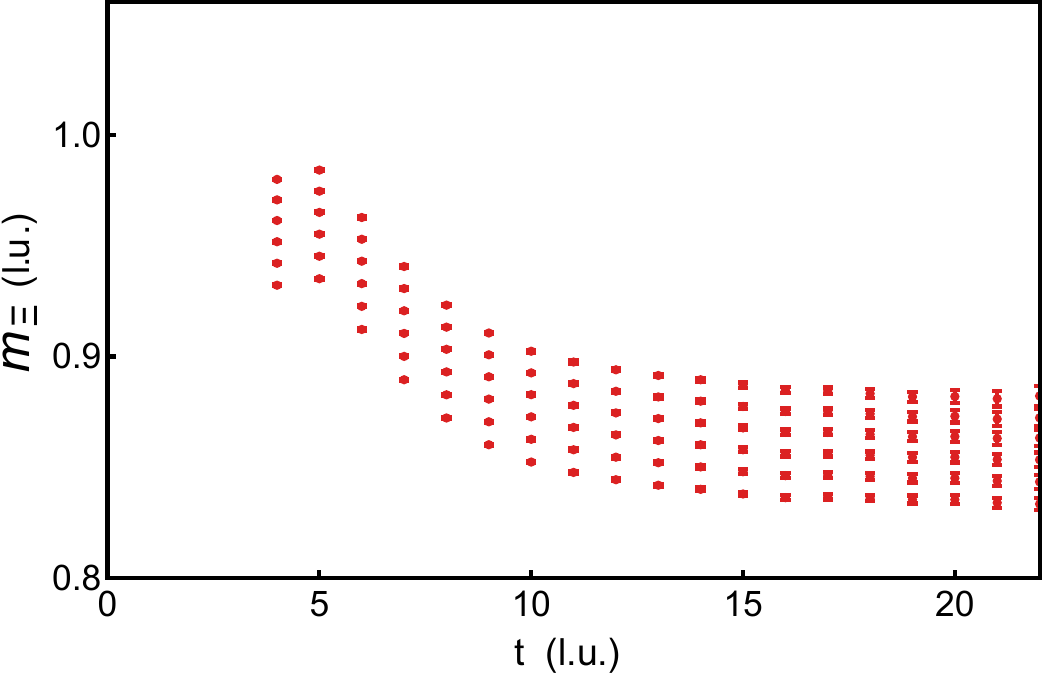} 
  \caption{
EMPs for the $\Xi$ in the $L=24$ (left), $L=32$ (center), $L=48$ (right) lattice volumes, respectively.
In ascending order, the momenta are ${\bf P}=2\pi {\bf n} /L$ with $|{\bf n}|^2 = 0,1,2,3,4,5$.
    }
  \label{fig:XiEMPs}
\end{figure}

Precision measurements of the single hadron masses, their dispersion relations and their 
volume dependence are essential for a complete analysis of multi-nucleon systems, in particular for a complete quantification of the uncertainties in binding energies and S-matrix elements.
Single hadron correlation functions for the 
$\pi^\pm$, $\rho^\pm$, $K^\pm$,  $K^{*,\pm}$,  the octet baryons and the decuplet baryons 
were calculated  in each of the three lattice volumes at 
six different momenta (in each volume),
from which ground-state energies 
for each momentum were extracted.
The hadron energies  were extracted from plateaus in the effective mass plots (EMPs)
derived from linear combinations (in the $L=24$ and $32$ ensembles) 
of the SP and SS correlation functions 
calculated at each lattice momentum.
The EMPs associated with the $\pi^\pm$ and $K^\pm$ are shown in Fig.~\ref{fig:pionEMPs} and Fig.~\ref{fig:kaonEMPs},
respectively,
while the EMPs for the octet baryons are shown in Figs.~\ref{fig:NEMPs},~\ref{fig:LamEMPs},~\ref{fig:SigEMPs} and \ref{fig:XiEMPs}.
It is clear from the EMPs that extended ground-state plateaus exists for all hadrons at all momenta, 
and as such relatively precise hadron masses and dispersion relations can be determined.
\textcolor{\revrevrevcolor}{
As only SP correlation functions were calculated for the L=48 ensemble, the ground state plateaus set in at larger times for each hadron in this ensemble than in the L=24 and L=32 ensembles.
}
A correlated $\chi^2$-minimization fit of the plateau region in 
combinations of correlation function to a constant energy was performed over a range of fit intervals to determine the energy, 
its statistical uncertainty
and the systematic uncertainty due to the selection of the fitting range.
The energies of the pseudoscalar mesons and octet baryons are
are shown in Table~\ref{tab:mesonMOM} and Table~\ref{tab:BaryonsMOM},
respectively.
\begin{table}[!ht]
\begin{center}
\begin{minipage}{16.5 cm}
  \caption{
  The pion and kaon energies (l.u.) as a function of momentum (l.u.),
    $|{\bf P}| = \left({2\pi\over L}\right) |{\bf n}|$,
    calculated on each ensemble of gauge-field configurations.
The infinite-volume meson masses, determined by fitting  expressions of the form in
eq.~(\ref{eq:FVforms}), are also given.
The first uncertainty associated with each extraction 
is statistical and the second is the fitting systematic.
In the case of the extrapolated values, the systematic uncertainty also contains the estimated uncertainty 
due to the extrapolation (which is small in both cases).
  }
\label{tab:mesonMOM}
\end{minipage}
\setlength{\tabcolsep}{0.2em}
\resizebox{\linewidth}{!}{%
\begin{tabular}{c|c|cccccc}
\hline
meson 
& ensemble  
& $|{\bf n}|=0$ 
& $|{\bf n}|^2=1$ 
& $|{\bf n}|^2=2$
& $|{\bf n}|^2=3$ 
& $|{\bf n}|^2=4$ 
& $|{\bf n}|^2=5$  \\
\hline
& \cfga 
& \textcolor{\revcolor}{0.26626(36)(14)} 
& \textcolor{\revcolor}{0.37184(28)(34)} 
& \textcolor{\revcolor}{0.45341(30)(45)} 
& \textcolor{\revcolor}{0.5204(07)(13)}
& \textcolor{\revcolor}{0.5812(10)(17)} 
& \textcolor{\revcolor}{0.6329(09)(12)} \\
$\pi^\pm $
&\cfgb 
&  \textcolor{\revcolor}{0.26607(23)(09)} 
&  \textcolor{\revcolor}{0.33006(20)(14)} 
&  \textcolor{\revcolor}{0.38330(21)(16)} 
&  \textcolor{\revcolor}{0.43042(26)(28)} 
&  \textcolor{\revcolor}{0.47156(43)(93)} 
&  \textcolor{\revcolor}{0.5093(05)(12)} \\
&\cfgc 
&  \textcolor{\revcolor}{0.26607(17)(11)} 
&  \textcolor{\revcolor}{0.29624(14)(05)} 
&  \textcolor{\revcolor}{0.32365(13)(10)} 
& \textcolor{\revcolor}{0.34895(16)(10)} 
& \textcolor{\revcolor}{0.37221(22)(18)} 
& \textcolor{\revcolor}{0.39404(31)(35)} \\
\hline
 &     $L=\infty$ &  \textcolor{\revcolor}{0.26606(14)(08) } & & & & &  \\
\hline
\hline
& \cfga 
& \textcolor{\revcolor}{0.35239(30)(16)} 
& \textcolor{\revcolor}{0.43749(24)(25)} 
& \textcolor{\revcolor}{0.50810(22)(25)} 
& \textcolor{\revcolor}{0.56947(35)(50)}
& \textcolor{\revcolor}{0.6224(07)(13)} 
& \textcolor{\revcolor}{0.67109(52)(55)} \\
$K^\pm$ 
& \cfgb 
&  \textcolor{\revcolor}{0.35248(18)(08)} 
&  \textcolor{\revcolor}{0.40259(16)(17)} 
&  \textcolor{\revcolor}{0.44725(17)(09)} 
&  \textcolor{\revcolor}{0.48782(24)(49)} 
&  \textcolor{\revcolor}{0.52357(45)(60)} 
&  \textcolor{\revcolor}{0.55727(46)(88)} \\
& \cfgc 
&  \textcolor{\revcolor}{0.35236(16)(25)} 
&  \textcolor{\revcolor}{0.37559(13)(06)} 
&  \textcolor{\revcolor}{0.39744(13)(06)} 
& \textcolor{\revcolor}{0.41814(13)(06)} 
& \textcolor{\revcolor}{0.43760(17)(05)} 
& \textcolor{\revcolor}{0.45628(21)(09)} \\
\hline
&      $L=\infty$ &  \textcolor{\revcolor}{0.35240(11)(03)} & & & & &  \\
\hline
\end{tabular}
}
\begin{minipage}[t]{16.5 cm}
\vskip 0.0cm
\noindent
\end{minipage}
\end{center}
\end{table}     
\begin{table}[!ht]
\begin{center}
\begin{minipage}[!ht]{16.5 cm}
  \caption{
  The baryon energies (l.u.) as a function of momentum (l.u.),
    $|{\bf P}| = \left({2\pi\over L}\right) |{\bf n}|$,
    calculated on each ensemble of gauge-field configurations.
The infinite-volume  masses, determined by fitting the expression in
eq.~(\ref{eq:FVforms}), are also given.
The first uncertainty is statistical and the second is the fitting systematic.
In the case of the extrapolated values, the systematic uncertainty also contains the estimated uncertainty 
due to the extrapolation (which is small in all cases).
  }
\label{tab:BaryonsMOM}
\end{minipage}
\setlength{\tabcolsep}{0.2em}
\resizebox{\linewidth}{!}{%
\begin{tabular}{c|c|cccccc}
\hline
baryon 
& ensemble  
& $|{\bf n}|=0$ 
& $|{\bf n}|^2=1$ 
& $|{\bf n}|^2=2$
& $|{\bf n}|^2=3$ 
& $|{\bf n}|^2=4$ 
& $|{\bf n}|^2=5$  \\
\hline
& \cfga 
& \textcolor{\revcolor}{0.7251(04)(11)} 
& \textcolor{\revcolor}{0.7699(10(13)} 
& \textcolor{\revcolor}{0.8108(10)(13)} 
& \textcolor{\revcolor}{0.8497(13)(21)}
& \textcolor{\revcolor}{0.8944(16)(23)} 
& \textcolor{\revcolor}{0.9311(17)(23)} \\
N
& \cfgb 
&  \textcolor{\revcolor}{0.72546(47)(31)} 
&  \textcolor{\revcolor}{0.75160(60)(47)} 
&  \textcolor{\revcolor}{0.77657(75)(89)} 
&  \textcolor{\revcolor}{0.80098(62)(81)} 
&  \textcolor{\revcolor}{0.8238(07)(11)} 
&  \textcolor{\revcolor}{0.8467(07)(10)} \\
& \cfgc 
&  \textcolor{\revcolor}{0.7245(10)(13)} 
&  \textcolor{\revcolor}{0.7359(21)(34)} 
&  \textcolor{\revcolor}{0.7471(23)(35)} 
& \textcolor{\revcolor}{0.7556(20)(36)} 
& \textcolor{\revcolor}{0.7661(21)(40)} 
& \textcolor{\revcolor}{0.7771(22)(42)} \\
\hline
&       $L=\infty$ &  \textcolor{\revcolor}{0.72524(46)(35) } & & & & &  \\
\hline
\hline
& \cfga 
& \textcolor{\revcolor}{0.77609(42)(66)} 
& \textcolor{\revcolor}{0.8165(14)(18)} 
& \textcolor{\revcolor}{0.8533(14)(21)} 
& \textcolor{\revcolor}{0.8918(23)(34)}
& \textcolor{\revcolor}{0.9336(14)(22)} 
& \textcolor{\revcolor}{0.9709(12)(16)} \\
$\Lambda$
& \cfgb 
&  \textcolor{\revcolor}{0.77633(45)(48)} 
&  \textcolor{\revcolor}{0.80059(60)(48)} 
&  \textcolor{\revcolor}{0.82435(75)(51)} 
&  \textcolor{\revcolor}{0.84687(78)(54)} 
&  \textcolor{\revcolor}{0.8680(10(14)} 
&  \textcolor{\revcolor}{0.8900(08)(10)} \\
& \cfgc 
&  \textcolor{\revcolor}{0.77650(94)(80)} 
&  \textcolor{\revcolor}{0.7858(14)(20)} 
&  \textcolor{\revcolor}{0.7963(14)(21)} 
& \textcolor{\revcolor}{0.8066(15)(23)} 
& \textcolor{\revcolor}{0.8166(16)(27)} 
& \textcolor{\revcolor}{0.8268(16)(29)} \\
\hline
 &     $L=\infty$ &  \textcolor{\revcolor}{ 0.77638(42)(48)} & & & & &  \\
\hline
\hline
& \cfga 
& \textcolor{\revcolor}{0.79520(70)(65)} 
& \textcolor{\revcolor}{0.83608(73)(62)} 
& \textcolor{\revcolor}{0.87550(75)(87)} 
& \textcolor{\revcolor}{0.9147(07)(13)}
& \textcolor{\revcolor}{0.9485(07)(10)} 
& \textcolor{\revcolor}{0.9855(10)(24)} \\
$\Sigma$ 
& \cfgb 
&  \textcolor{\revcolor}{0.79634(31)(49)} 
&  \textcolor{\revcolor}{0.82033(60)(61)} 
&  \textcolor{\revcolor}{0.84320(63)(75)} 
&  \textcolor{\revcolor}{0.86502(71)(51)} 
&  \textcolor{\revcolor}{0.88575(60)(57)} 
&  \textcolor{\revcolor}{0.90755(65)(63)} \\
& \cfgc 
&  \textcolor{\revcolor}{0.7958(12)(13)} 
&  \textcolor{\revcolor}{0.8050(14)(23)} 
&  \textcolor{\revcolor}{0.8152(15)(24)} 
& \textcolor{\revcolor}{0.8253(16)(26)} 
& \textcolor{\revcolor}{0.8351(16)(28)} 
& \textcolor{\revcolor}{0.8451(17)(30)} \\
\hline
&       $L=\infty$ &  \textcolor{\revcolor}{ 0.79638(33)(54)} & & & & &  \\
\hline
\hline
& \cfga 
& \textcolor{\revcolor}{0.83646(63)(49)} 
& \textcolor{\revcolor}{0.87594(60)(58)} 
& \textcolor{\revcolor}{0.91318(58)(54)} 
& \textcolor{\revcolor}{0.9487(06)(10)}
& \textcolor{\revcolor}{0.9828(06)(11)} 
& \textcolor{\revcolor}{1.01668(60)(95)} \\
$\Xi$
&\cfgb 
&  \textcolor{\revcolor}{0.83715(53)(58)} 
&  \textcolor{\revcolor}{0.85886(49)(59)} 
&  \textcolor{\revcolor}{0.88044(50)(57)} 
&  \textcolor{\revcolor}{0.90201(51)(36)} 
&  \textcolor{\revcolor}{0.92261(62)(89)} 
&  \textcolor{\revcolor}{0.94276(66)(89)} \\
&\cfgc 
&  \textcolor{\revcolor}{0.83643(68)(72)} 
&  \textcolor{\revcolor}{0.8460(11)(10)} 
&  \textcolor{\revcolor}{0.8557(12)(11)} 
& \textcolor{\revcolor}{0.8652(12)(13)} 
& \textcolor{\revcolor}{0.8744(13)(14)} 
& \textcolor{\revcolor}{0.8837(14)(17)} \\
\hline
&      $L=\infty$ &  \textcolor{\revcolor}{ 0.83690(45)(50)} & & & & &  \\
\hline
\hline
\end{tabular}
}
%noalign{\smallskip\hrule}\cr}
\begin{minipage}[t]{16.5 cm}
\vskip 0.0cm
\noindent
\end{minipage}
\end{center}
\end{table}     

The energies determined at zero momentum are used to extrapolate the hadron masses to infinite volume,
and are combined with the other energies to determine their dispersion relations.
With the large values of $m_\pi L$ in the ensembles of gauge configurations, 
it is sufficient to use the leading-order (LO) finite-volume (FV) corrections to the hadron masses to extrapolate from the 
volumes of the calculations to infinite volume.  
The LO modifications to the pseudoscalar masses, $m_M$, and 
baryon masses, $M_B$, are given by 
\begin{eqnarray}
m_M^{(V)}(m_\pi L)  & = & 
m_M^{(\infty)} + c_M\ { e^{-m_\pi L}\over (m_\pi L)^{3/2}}\ +\ ...
\nonumber\\
M_B^{(V)}(m_\pi L)  & = & 
M_B^{(\infty)} + c_B\ { e^{-m_\pi L}\over m_\pi L}\ +\ ...
\ \ \ ,
\label{eq:FVforms}
\end{eqnarray}
where the forms are those of p-regime chiral perturbation theory ($\chi$PT) and 
heavy-baryon $\chi$PT (HB$\chi$PT~\cite{Jenkins:1990jv}).  The infinite-volume masses, 
$m_M^{(\infty)}$ and $M_B^{(\infty)}$, and the coefficients of the LO volume dependence,
$c_M$ and $c_B$, are quantities determined by fits to the LQCD calculations, and will, in general, be different for each hadron.

\begin{figure}[!ht]
  \centering
  \includegraphics[width=0.48 \columnwidth]{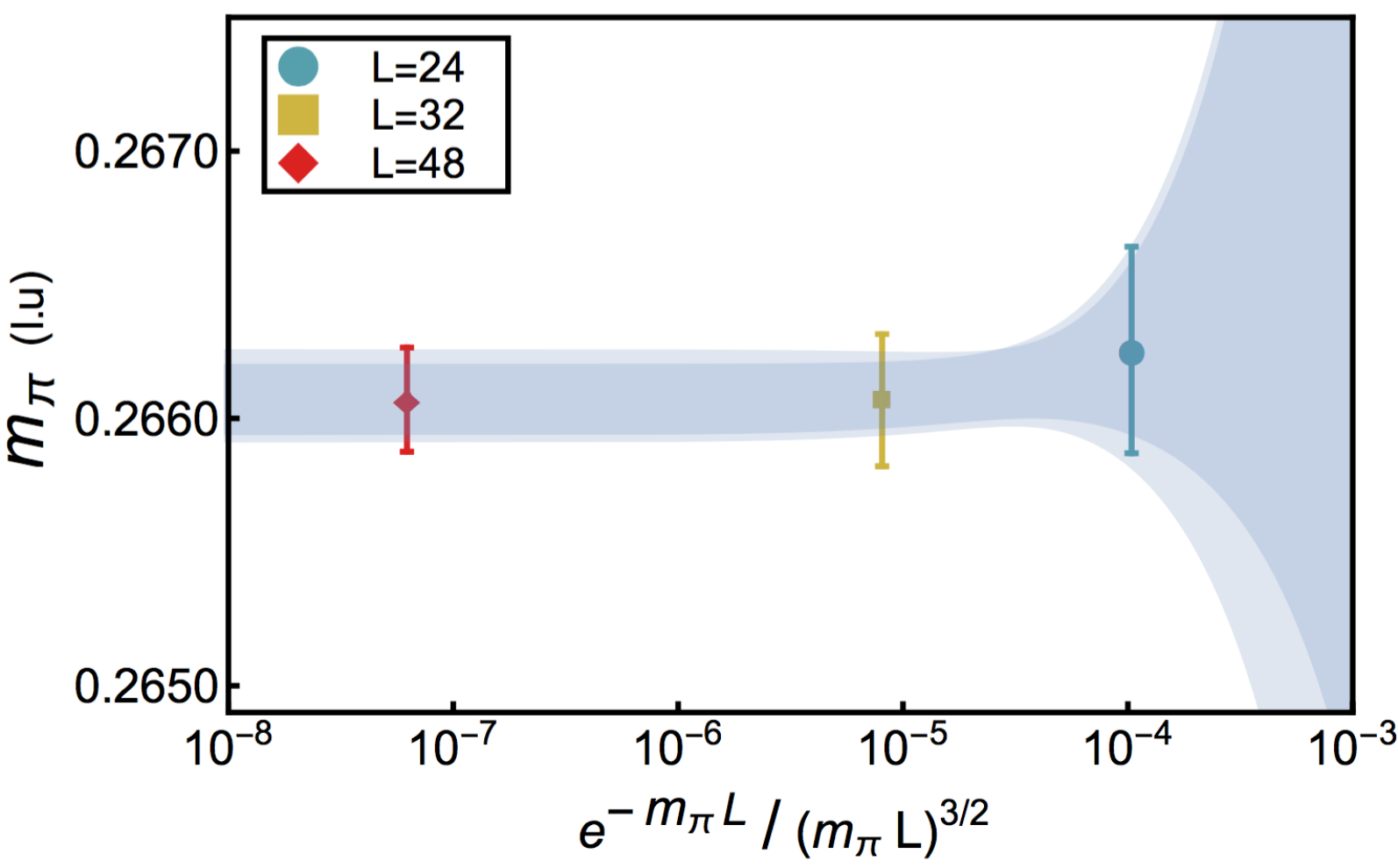}  \includegraphics[width=0.48 \columnwidth]{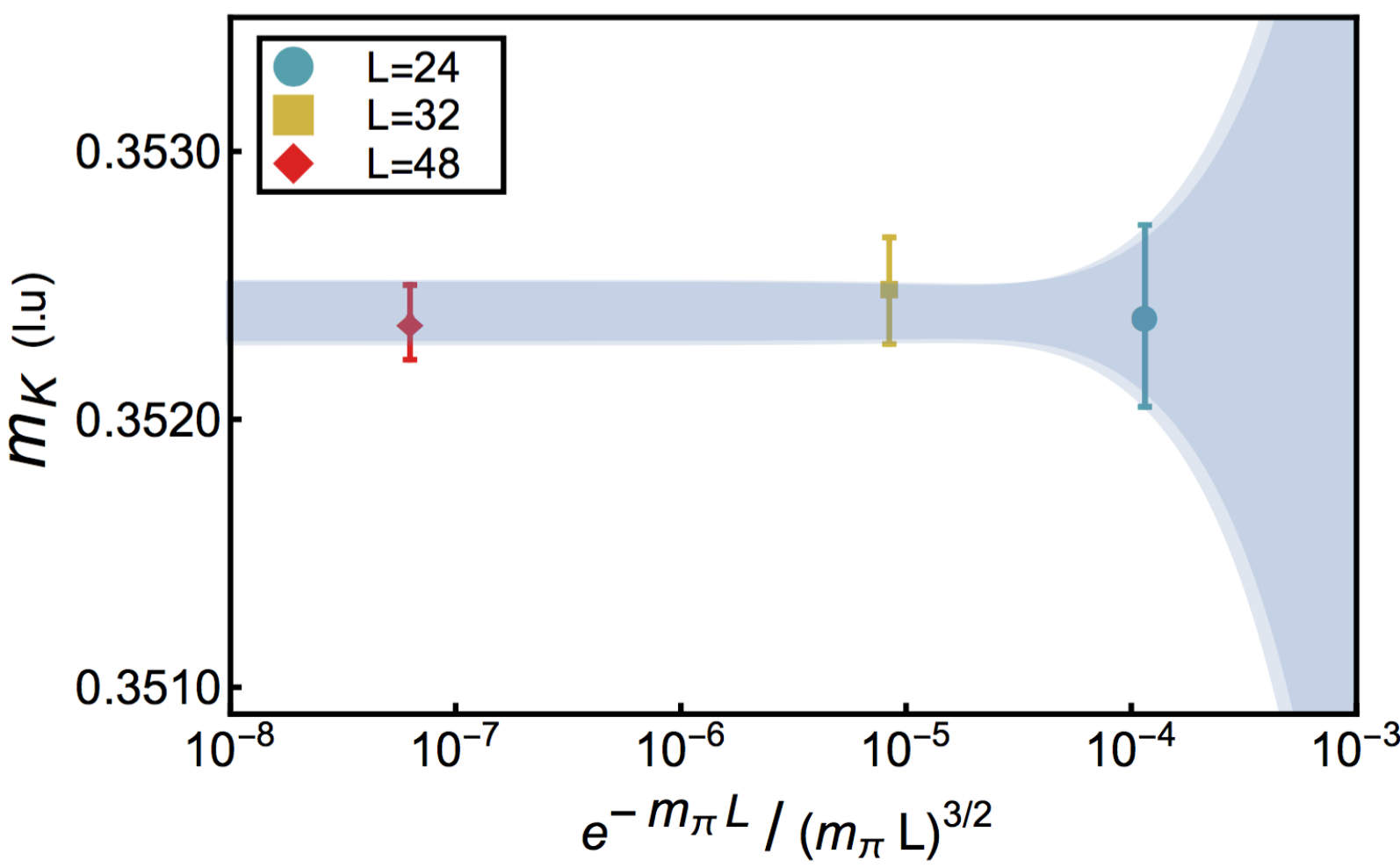}\ \
  \caption{
The volume dependence of the $\pi^+$ (left panel) and $K^\pm$ (right panel)  masses.
Energies (l.u.) in the $L=24, 32$ and $48$ lattice volumes are shown as the blue, yellow and red points, respectively,
while the results of fits to these results of the form given in eq.~(\protect\ref{eq:FVforms})
are  shown by the shaded regions with the inner (outer) band denoting the statistical (statistical and systematic combined in quadrature) uncertainties.   
    }
  \label{fig:mesonVOL}
\end{figure}

\begin{figure}[!ht]
  \centering
  \includegraphics[width=0.48 \columnwidth]{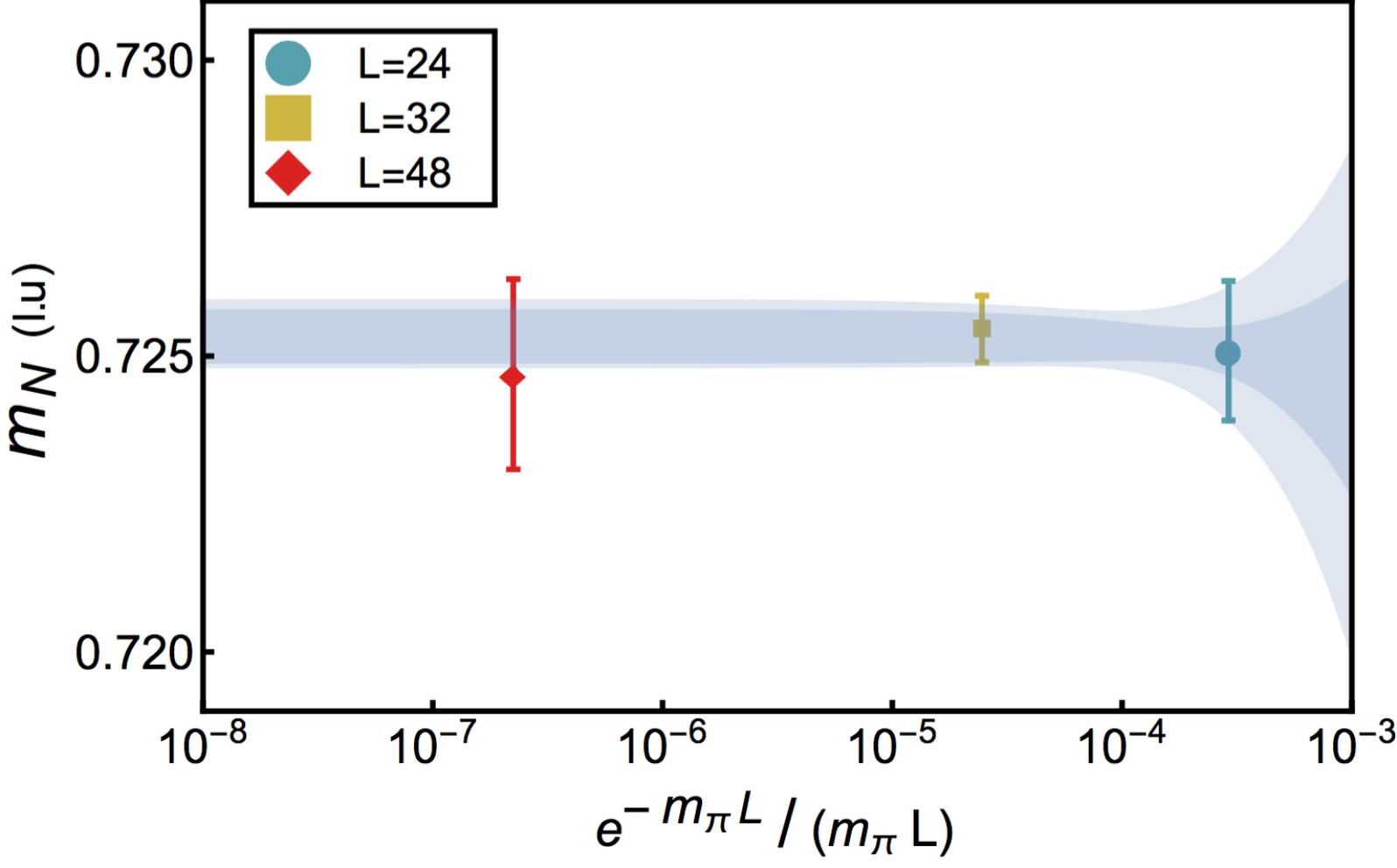}  \includegraphics[width=0.48 \columnwidth]{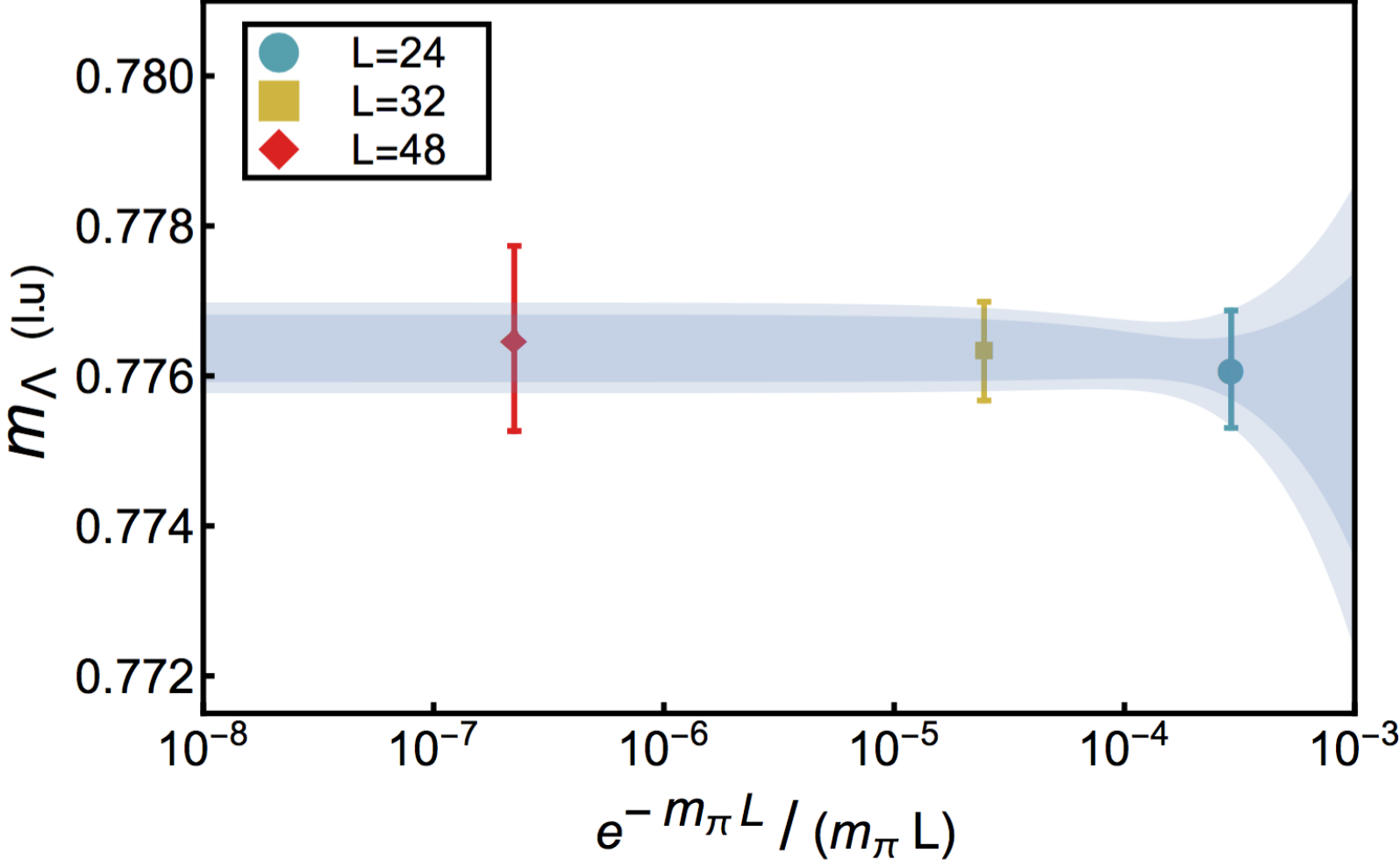}\ \
  \includegraphics[width=0.48 \columnwidth]{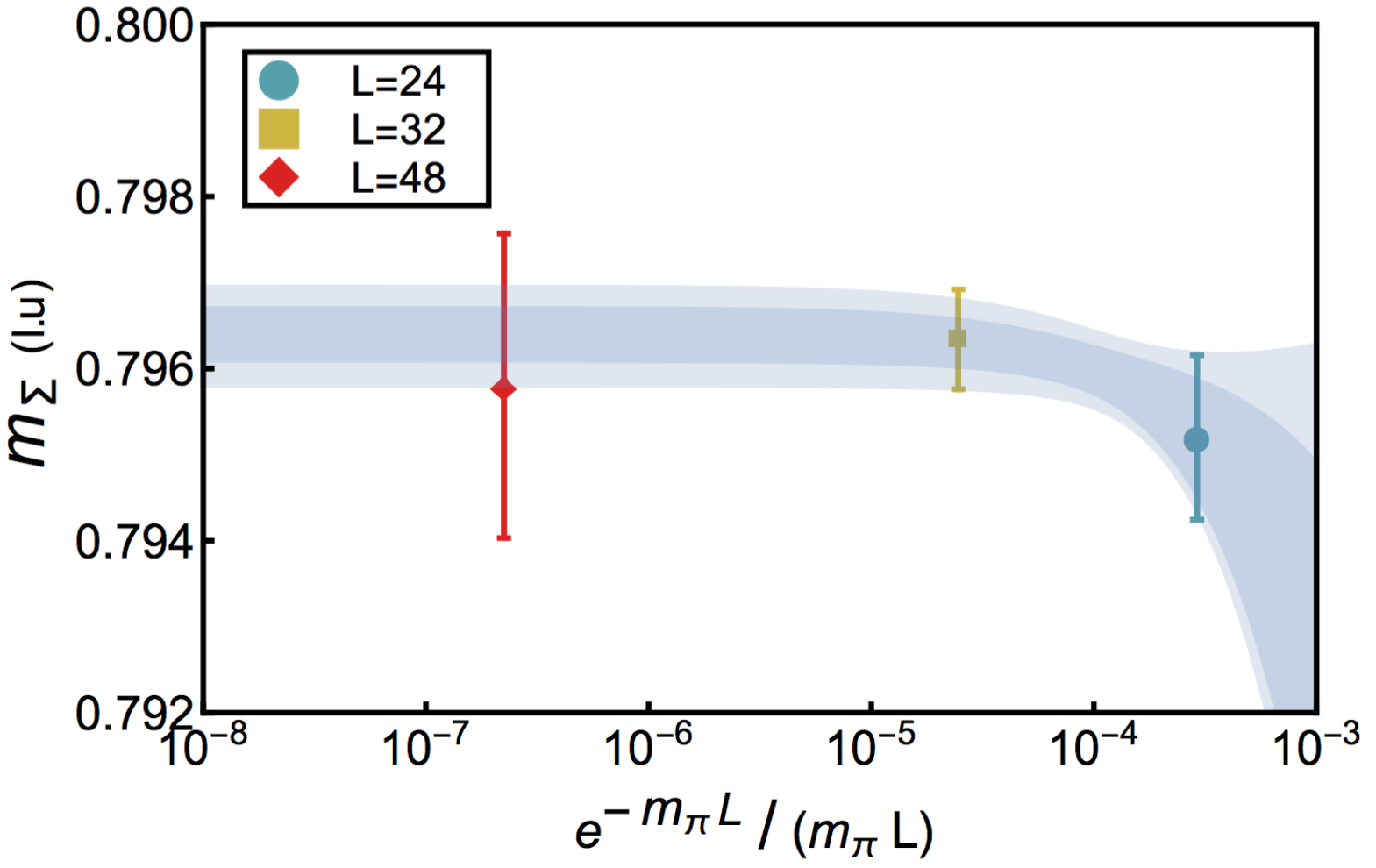}  \includegraphics[width=0.48 \columnwidth]{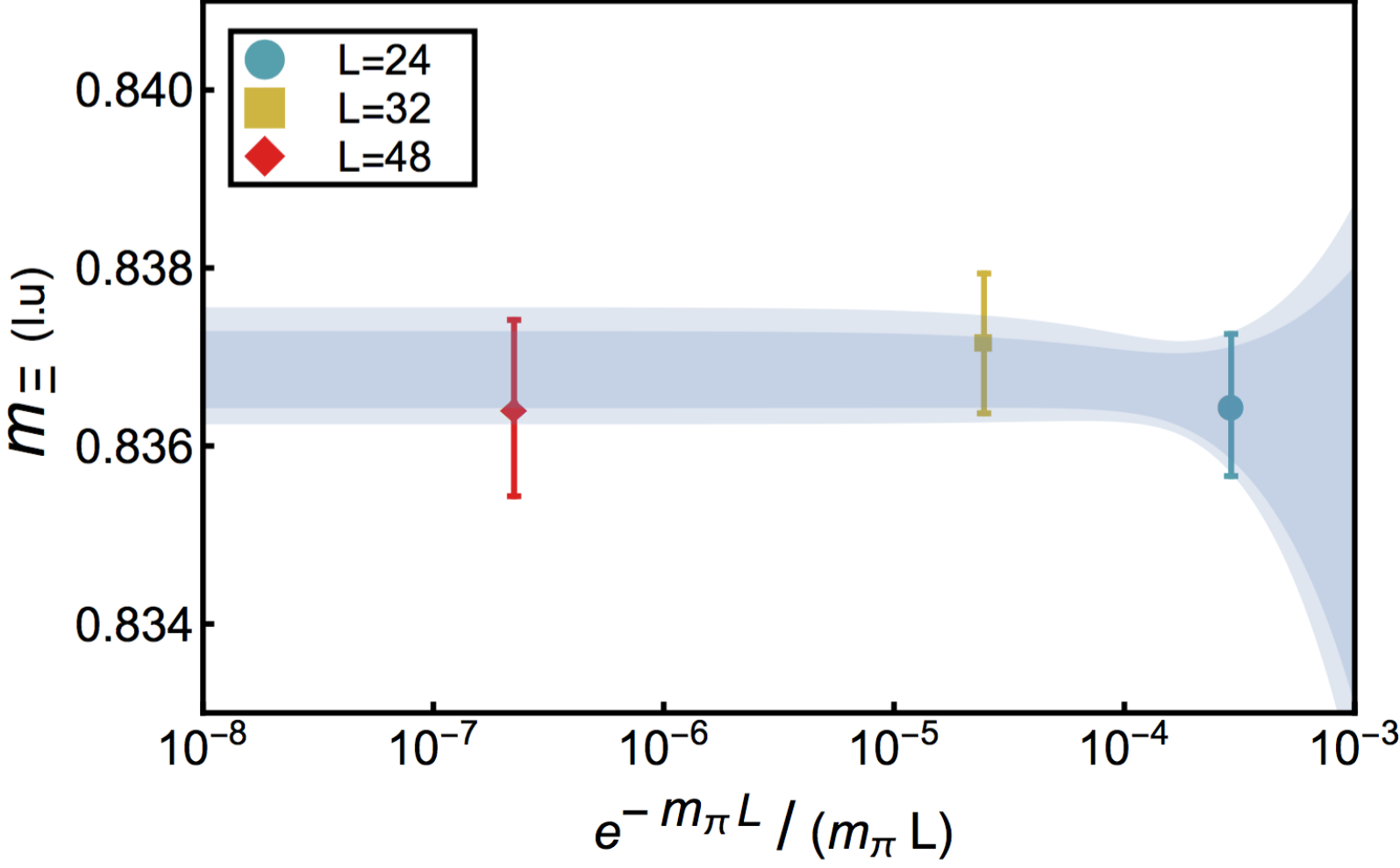}\ \
  \caption{
The volume dependence of the $N$, $\Lambda$, $\Sigma$ and $\Xi$  masses.
Energies (l.u.) in the $L=24, 32$ and $48$ lattice volumes are shown as the blue, yellow and red points, respectively,
  while fits to these results are  shown by the gray, shaded regions with the inner (outer) band denoting the statistical (statistical and systematic combined in quadrature) uncertainties.   
    }
  \label{fig:BaryonVOL}
\end{figure}

The zero-momentum energies of the pseudoscalar mesons 
and their infinite-volume extrapolation
are
given in Table~\ref{tab:mesonMOM} and shown in Fig.~\ref{fig:mesonVOL}.
The energies of both mesons are found to be independent of the lattice volume within the uncertainties of the calculations.
Despite the larger number of correlation functions in the $L=24$ ensemble, the uncertainties in the meson masses are larger than those extracted from the $L=32$ ensemble.
The zero-momentum energies of the octet baryons 
and their infinite-volume extrapolation
are given in Table~\ref{tab:BaryonsMOM} and shown in Fig.~\ref{fig:BaryonVOL}.
As with the mesons, there is no statistically significant volume dependence observed for any of the octet-baryon masses.
Two-parameter  $\chi^2$-minimization fits of the form given in eq.~(\ref{eq:FVforms})
were performed to the volume-dependence of each 
hadron to extract its infinite-volume mass.
Because of the negligible volume dependence in the LQCD results,  limited constraints can be placed on the $c_{M,B}$ coefficients.
In addition to the $\pi^\pm$, $K^\pm$ and octet baryons, analogous extrapolations were performed with the results obtained for the $\rho^\pm$, $K^{*\pm}$ and decuplet baryons, the combined results of which are shown in Table~\ref{tab:IVmasses}
(in both l.u. and MeV).
\renewcommand{\arraystretch}{1.5}% Wider
\begin{table}[!ht]
\begin{center}
\begin{minipage}[!ht]{16.5 cm}
  \caption{
The infinite-volume hadron masses obtained by extrapolating zero-momentum ground-state energies 
with the volume dependence given in eq.~(\protect\ref{eq:FVforms}).
The first and second uncertainties are the statistical and systematic, respectively, while the third for values in units of MeV results from the uncertainty in the scale setting.
  } 
\label{tab:IVmasses}
\end{minipage}
\setlength{\tabcolsep}{0.3em}
\begin{tabular}{c|cc||c|cc}
\hline
      hadron & 
      Mass  (l.u.) & 
      Mass  (MeV)  &
      hadron & 
      Mass  (l.u.) & 
      Mass  (MeV)  
      \\
\hline
$\pi^\pm$    & 0.26614(15)(15)  & \pionmassFULLa & $K^\pm$  & 0.35241(12)(11) & \kaonmassFULLa\\
$\rho^\pm$  & 0.5248(14)(15) &  \rhomassFULLa & $K^{*\pm}$  &  0.56923(89)(51)& \kstarmassFULLa \\
$N$  & 0.72524(46)(35) & \MNFULLa & $\Lambda$  & 0.77638(42)(48) & \lammassFULLa  \\
$\Sigma$  & 0.79638(33)(54) & \sigmassFULLa & $\Xi$  & 0.83690(45)(50) & \ximassFULLa\\
$\Delta$  & 0.8791(14)(17) & \deltamassFULLa & $\Sigma^*$  & 0.9211(17)(19) &  \sigstarmassFULLa \\
$\Xi^*$  & 0.9637(09)(17) & \xistarmassFULLa & $\Omega$  & 1.0059(06)(12) & \omegamassFULLa  \\
\hline
\end{tabular}
\begin{minipage}[t]{16.5 cm}
\vskip 0.0cm
\noindent
\end{minipage}
\end{center}
\end{table}     

Deviations of the single hadron dispersion relations from that of special relativity lead to modifications to 
L\"uscher's quantization conditions (QCs) in two-body systems.
To address this, the dispersion relations have been precisely determined, and the deviations from special relativity are propagated through 
the extraction of S-matrix elements using the QCs.
In each of the ensembles,  single hadron correlation functions were calculated for each of the hadrons of interest with 
momenta  $|{\bf p}|\le  \sqrt{5} \left(2\pi/L\right)$, the results of which are given in Table~\ref{tab:mesonMOM} and 
Table~\ref{tab:BaryonsMOM}.
Energy-momentum relations that are fit to the results obtained for each hadron, $h$,  are of the form
\begin{eqnarray}
E_h^2 = M_h^2 + v_h^2 |{\bf p}|^2 + \eta_h \left( |{\bf p}|^2\right)^2
\ \ \ ,
\label{eq:dispfitform}
\end{eqnarray}
where the hadron speed of light, 
$v_h$, and the higher-order deviation from special relativity, parameterized by $\eta_h$, are 
determined by fits to the results of the LQCD calculations.
With this parameterization, the $v_h$ are consistent with unity and the 
$\eta_h$ are  consistent with zero (for all hadrons).  
There is a Lorentz-breaking term that could be considered at this order in a momentum expansion,
$\sum\limits_j {\bf p}_j^4$, but this is also found to be consistent with zero.
\begin{figure}[!ht]
  \centering
  \includegraphics[width=0.48 \columnwidth]{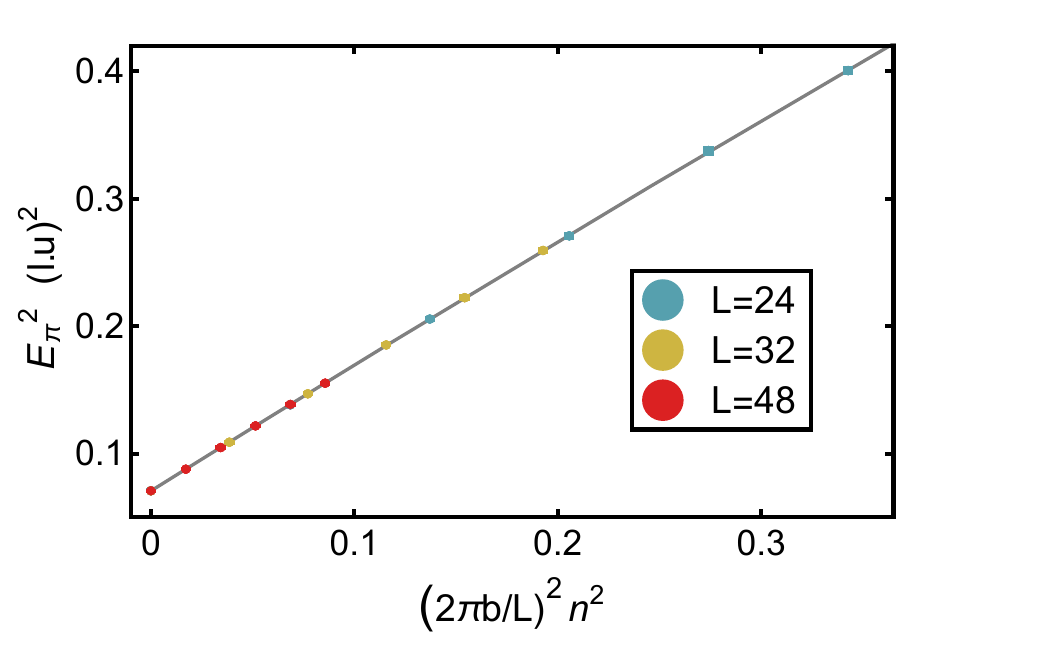}  \includegraphics[width=0.48 \columnwidth]{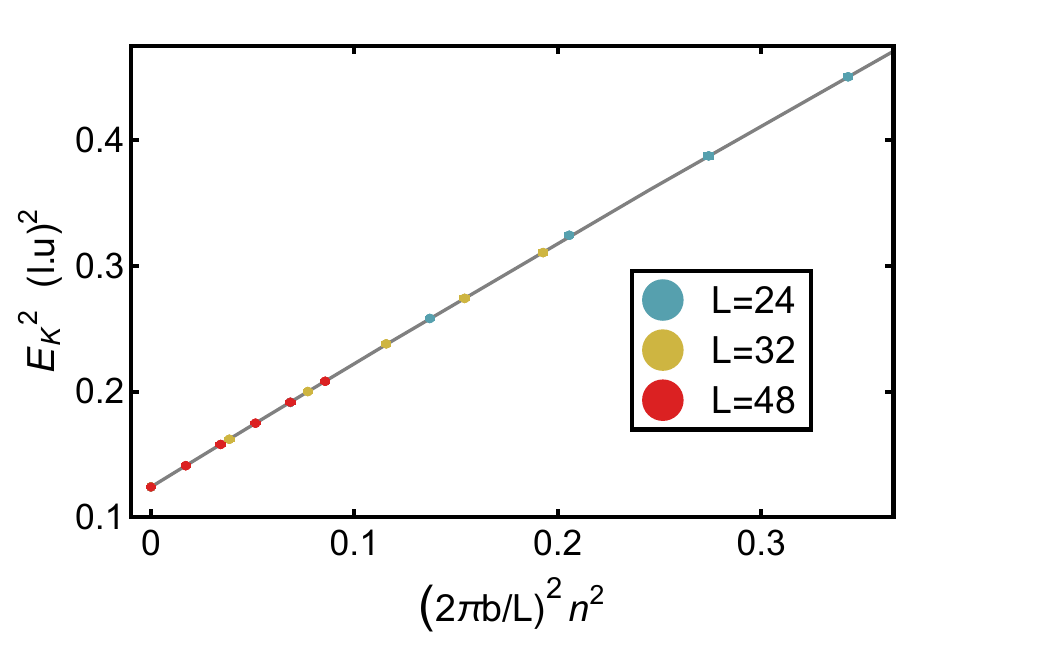}\ \
  \caption{
Dispersion relations of the $\pi^\pm$, $K^\pm$ .
  The results in the $L=24, 32$ and $48$ lattice volumes are shown as the blue, yellow and red points, respectively,
  while fits to these results are  shown by the gray curves.   
    }
  \label{fig:pionDISP}
\end{figure}
\begin{figure}[!ht]
  \centering
  \includegraphics[width=0.48 \columnwidth]{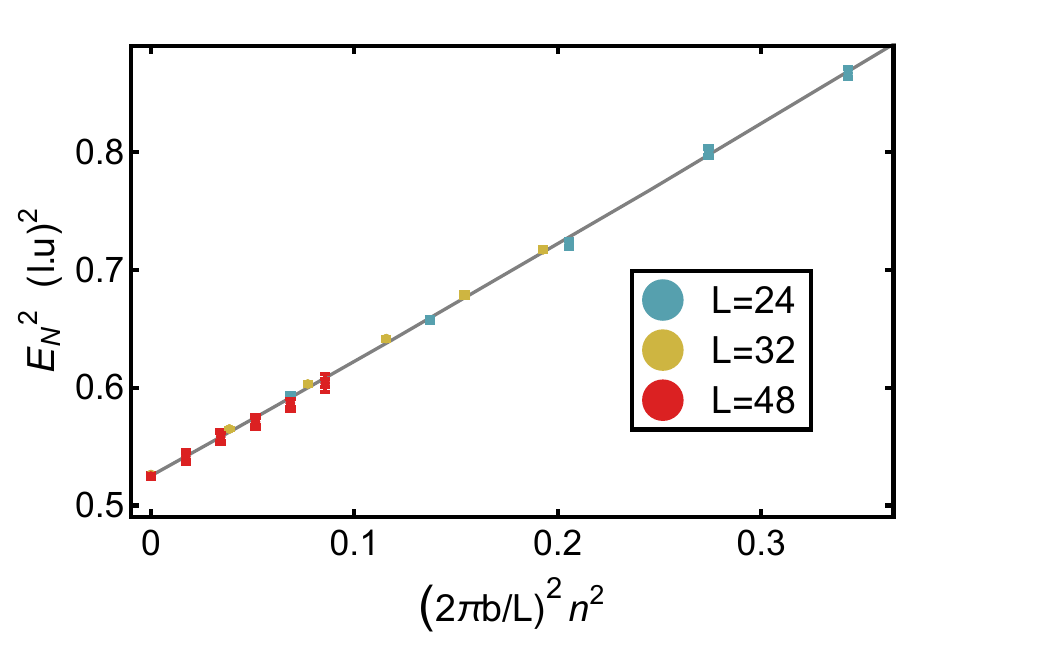}   \includegraphics[width=0.48 \columnwidth]{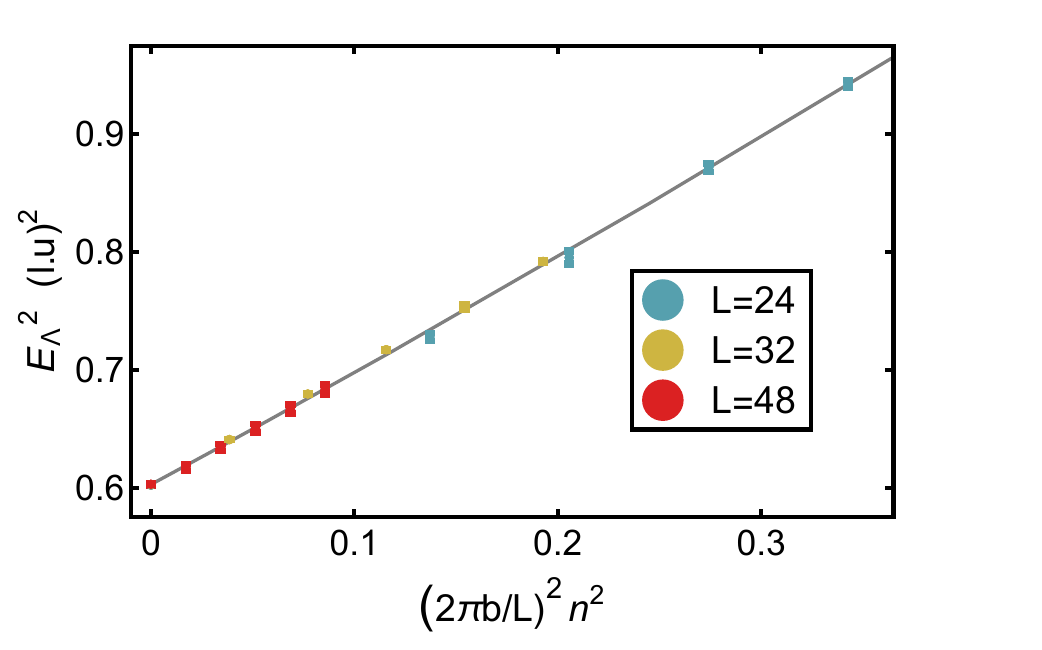}   \
  \includegraphics[width=0.48 \columnwidth]{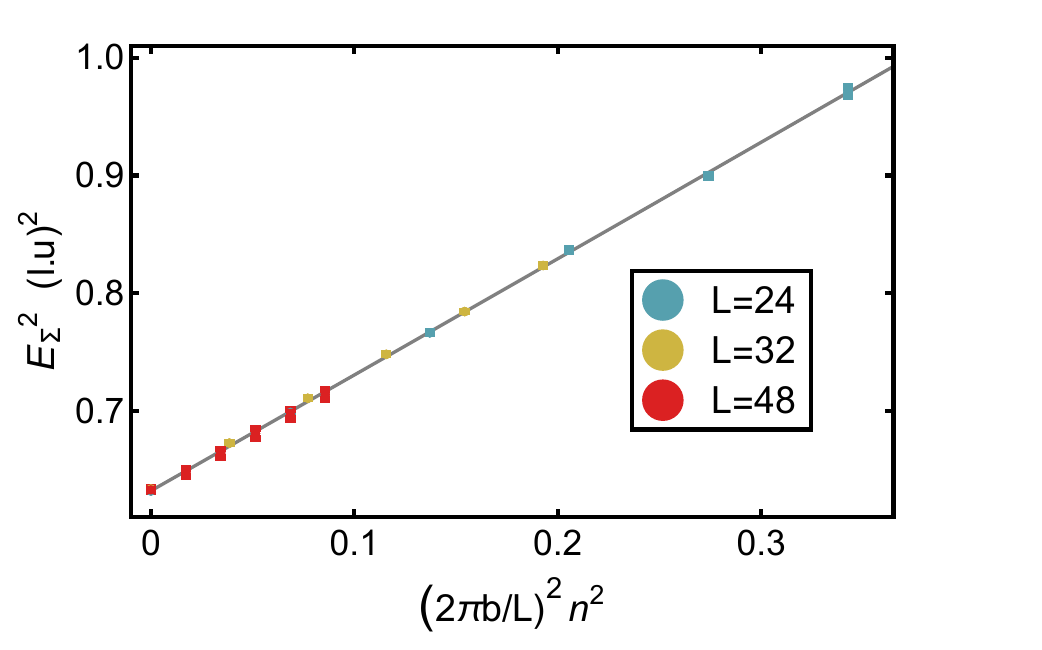}   \includegraphics[width=0.48 \columnwidth]{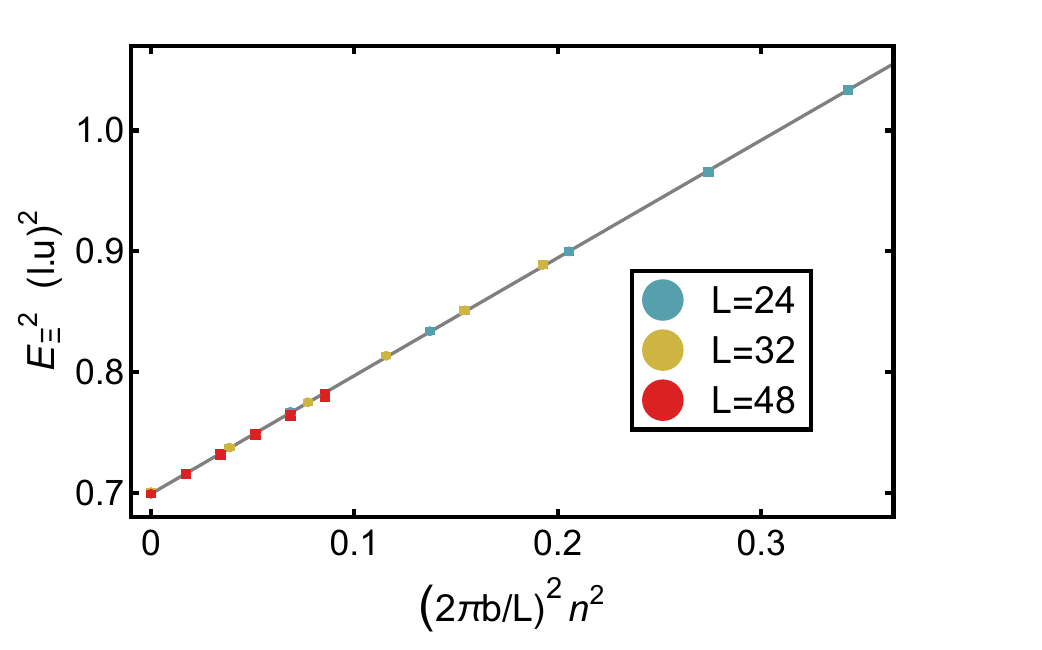}   \
  \caption{
Dispersion relations of the octet baryons.
 The results in the $L=24, 32$ and $48$ lattice volumes are shown as the blue, yellow and red points, respectively,
  while fits to these results are  shown by the gray curves.   
    }
  \label{fig:baryDISP}
\end{figure}
The energies of the $\pi^\pm$, $K^\pm$ and  octet baryons as a function of momentum
are given in Table~\ref{tab:mesonMOM} and Table~\ref{tab:BaryonsMOM} and
shown in Fig.~\ref{fig:pionDISP}  and Fig.~\ref{fig:baryDISP}.
$\chi^2$-minimization fits to the energy-momentum 
dispersion relation are performed to extract the speed of light for each hadron,
the results of which are shown in Table~\ref{tab:velocitiesFIT}.
\renewcommand{\arraystretch}{1.5}% Wider
\begin{table}[!ht]
\begin{center}
\begin{minipage}[!ht]{16.5 cm}
  \caption{
The   speed of light of each hadron determined from fits to the energy-momentum results.
} 
\label{tab:velocitiesFIT}
\end{minipage}
\setlength{\tabcolsep}{0.3em}
\begin{tabular}{c|c||c|c}
\hline
      hadron & 
      $v_h$ & 
      hadron & 
      $v_h$
      \\
\hline
$\pi^\pm$    &   1.0025(18)(08) & $K^\pm$      &  1.0038(20)(12)   \\
 $N$            &   1.010(16)(07)   & $\Lambda$  &   1.018(15)(01)    \\
$\Sigma$    &   1.010(12)(03)   & $\Xi$            & 1.0102(61)(13)    \\
 \hline
\end{tabular}
\begin{minipage}[t]{16.5 cm}
\vskip 0.0cm
\noindent
\end{minipage}
\end{center}
\end{table}     
In the low-energy regime relevant to the two-nucleon systems, the dispersion relation of special relativity is found to hold at the 
$\sim 1\%$ level.

%%%%%%%%%%%%%%%
\subsection{The Gell-Mann-Okubo Mass Relation}
\label{sec:GMO}
Given the precise determinations of the single hadron spectrum, it is important to  test relations between 
baryon masses that are predicted to hold in particular limits of QCD. 
The Gell-Mann--Okubo mass relation~\cite{GellMann:1962xb,Okubo:1961jc} 
arises from SU(3) flavor symmetry and its violation, quantified by 
\begin{eqnarray}
T_{\rm GMO} & = &  M_\Lambda + {1\over 3} M_\Sigma - {2\over 3} M_N - {2\over 3} M_\Xi
\ \ \ ,
\label{eq:gmo}
\end{eqnarray}
results from SU(3) breaking transforming 
in the {\bf 27}-plet irreducible representation
(irrep) of flavor SU(3) 
which can only arise from multiple insertions of the light-quark mass matrix 
or from non-analytic meson-mass dependence induced  by  loops in $\chi$PT.
Further, it has been shown that $T_{\rm GMO} $ vanishes in the large-$N_c$ limit as $1/N_c$~\cite{Dashen:1994qi}.
In previous work~\cite{Beane:2006pt}, 
we performed the first LQCD determination of this quantity, after which more precise LQCD determinations~\cite{WalkerLoud:2008bp} 
were performed.  
In this work, by far the most precise determination of $T_{\rm GMO}$ was obtained from the $L=32$ ensemble, where we find 
$T_{\rm GMO} = +0.000546(51)(81)~{\rm l.u.} = +0.92(09)(14)(01)~{\rm MeV}$
(compared with $T_{\rm GMO} = +0.00056(19)(38)~{\rm l.u.} = +0.96(33)(64)(01)~{\rm MeV}$ and 
$T_{\rm GMO} = +0.00104(27)(29)~{\rm l.u.} = +1.76(46)(49)(02)~{\rm MeV}$, 
from the $L=24$ and $48$ ensembles, respectively).
It is conventional to form the dimensionless quantity $\delta_{\rm GMO} = T_{\rm GMO}/M_0$, 
where $M_0$ is the centroid of the octet baryons masses.
In the present calculations, 
the centroid is found to be  
$M_0=0.78658(51)(36)~{\rm l.u.} =1329(01)(01)(14)~{\rm MeV}$, 
from which $\delta_{\rm GMO} = 0.00069(06)(10)$. 
This value is consistent with our previously published result close to this pion mass 
and is also consistent with other subsequent determinations~\cite{WalkerLoud:2008bp},
but far more precise.
\textcolor{\revrevrevcolor}{
However, as the present calculations have been performed at only one lattice spacing, there is a systematic 
uncertainty associated with extrapolating to the continuum that is not directly quantified, but which we have estimated to be small.
}  
It is worth noting that the experimental value, 
$T_{\rm GMO}^{\rm expt} =+8.76(08)~{\rm MeV}$, is
an order of magnitude larger than the value we have determined at this heavier pion mass.

%%%%%%%%%%%%%%%%%%%%%%%%%%%%%%%%%%%%%%%%%%%%%%%%
\section{The $\siii$-$\diii$ Coupled Channels and the Deuteron}
\label{sec:3s1}

The phenomenology of the $\siii$-$\diii$ coupled $J=1$ channels in finite volumes has been explored recently using the
experimentally constrained phase shifts and mixing angles in an effort to understand what might be 
expected in future LQCD calculations~\cite{Briceno:2013bda}.
One goal of that study was to estimate the lattice volumes, and identify the 
correlation functions, required to extract the phase shifts
and mixing parameter describing these channels in infinite volume.
It was found to be convenient in those FV studies~\cite{Briceno:2013bda}
to use the Blatt-Biedenharn (BB)~\cite{Blatt:1952zza} parameterization of the $2\times 2$ S-matrix 
(below the inelastic threshold),
\begin{eqnarray}
S_{(J=1)} & = & 
\left(
\begin{array}{cc}
\cos\epsilon_1 & -\sin\epsilon_1\\
\sin\epsilon_1 & \cos\epsilon_1
\end{array}
\right)
\ 
\left(
\begin{array}{cc}
e^{2i\delta_{1\alpha}} & 0 \\
0 & e^{2i\delta_{1\beta}} 
\end{array}
\right)
\ 
\left(
\begin{array}{cc}
\cos\epsilon_1 & \sin\epsilon_1\\
-\sin\epsilon_1 & \cos\epsilon_1
\end{array}
\right)
\ \ \ ,
\label{eq:BBSmat}
\end{eqnarray}
from which the QCs associated with these channels can be determined.
For the two-nucleon system at rest in a cubic volume, embedded in the 
even parity 
$\mathbb{T}_1$ irrep 
of the cubic group,  
the QC in the limit of vanishing $\delta_{1\beta}$,  D-waves and higher phase shifts
becomes~\cite{Briceno:2013bda}
\begin{eqnarray}
k^*_{\mathbb{T}_1}\cot\delta_{1\alpha}(k^*_{\mathbb{T}_1}) & = & 4\pi c_{00}^{(0,0,0)}(k^*_{\mathbb{T}_1};L)
\ \ \ ,
\label{eq:Tone}
\end{eqnarray}
where $k^*_{\mathbb{T}_1}$ is the magnitude of the momentum in the center-of-momentum (CoM) 
frame, and 
the function $c_{00}^{(0,0,0)}(k^*_{\mathbb{T}_1};L)$ is proportional to the L\"uscher ${\cal Z}_{00}$ function, 
as given in Ref.~\cite{Luscher:1986pf,Luscher:1990ux}.  
The phase shift $\delta_{1\alpha}$ is evaluated at $k^*_{\mathbb{T}_1}$.  
The three $j_z$-substates are degenerate and their energies are insensitive to the mixing parameter 
$\epsilon_1$.

In contrast, for the two-nucleon system carrying one unit of lattice  momentum along the z-axis,
${\bf P}_{\rm tot.}= {2\pi\over L}{\bf d}$ with
${\bf d}=(0,0,1)$,
the three substates are embedded into two distinct 
even-parity 
 irreps of the cubic group - the one-dimensional $\mathbb{A}_2$ representation
 and the 
two-dimensional $\mathbb{E}$ representation, containing the $j_z=0$ and $j_z=\pm 1$ states, respectively.
In the same limit as taken to derive
eq.~(\ref{eq:Tone}), the QCs for these two irreps are~\cite{Briceno:2013bda}
\begin{eqnarray}
k^*_{\mathbb{A}_2}\cot\delta_{1\alpha}(k^*_{\mathbb{A}_2}) & = & 4\pi c_{00}^{(0,0,1)}(k^*_{\mathbb{A}_2};L)
\ -\ {1\over\sqrt{5}} {4\pi\over k^{*2}_{\mathbb{A}_2}} 
\ c_{20}^{(0,0,1)}(k^*_{\mathbb{A}_2};L)
\ s_{\epsilon_1}(k^*_{\mathbb{A}_2})
\ \ ,
\nonumber\\
k^*_{\mathbb{E}}\cot\delta_{1\alpha}(k^*_{\mathbb{E}}) & = & 4\pi c_{00}^{(0,0,1)}(k^*_{\mathbb{E}};L)
\ +\ {1\over 2 \sqrt{5}} {4\pi\over k^{*2}_{\mathbb{E}}} 
\ c_{20}^{(0,0,1)}(k^*_{\mathbb{E}};L)
\ s_{\epsilon_1}(k^*_{\mathbb{E}})
\ \ \ ,
\label{eq:A2E}
\end{eqnarray}
where
\begin{eqnarray}
s_{\epsilon_1}(k^*) & = & \sqrt{2}\sin2\epsilon_1(k^*) - \sin^2\epsilon_1(k^*)
\ \ \ .
\label{eq:mixdef}
\end{eqnarray}
The difference in energy between the $\mathbb{A}_2$ and $\mathbb{E}$ FV eigenstates provides a measure of $\epsilon_1$,
but this is complicated by the fact that 
they are  evaluated at two slightly different energies.
This analysis can be extended to other lattice momenta~\cite{Briceno:2013bda}, but the QCs in eq.~(\ref{eq:Tone}) and 
eq.~(\ref{eq:A2E}) are sufficient for the present purposes.

Correlation functions for two nucleons in the $\mathbb{T}_1$, $\mathbb{A}_2$ and $\mathbb{E}$ irreps are straightforwardly 
constructed from the nucleon blocks we have described previously.
In fact, multiple correlation functions are constructed in each irrep.
In the $L=24$ and $48$ ensembles, 
the spin projections were not performed to permit construction of the $\mathbb{A}_2$ irrep,
and so only  $L=32$ correlation functions can be used to constrain $\epsilon_1$.

%%%%%%%%%%%%%%%%%%%%%
\subsection{The Deuteron}
\label{subsec:deut}

\begin{figure}[!ht]
  \centering
  \includegraphics[width=0.32 \columnwidth]{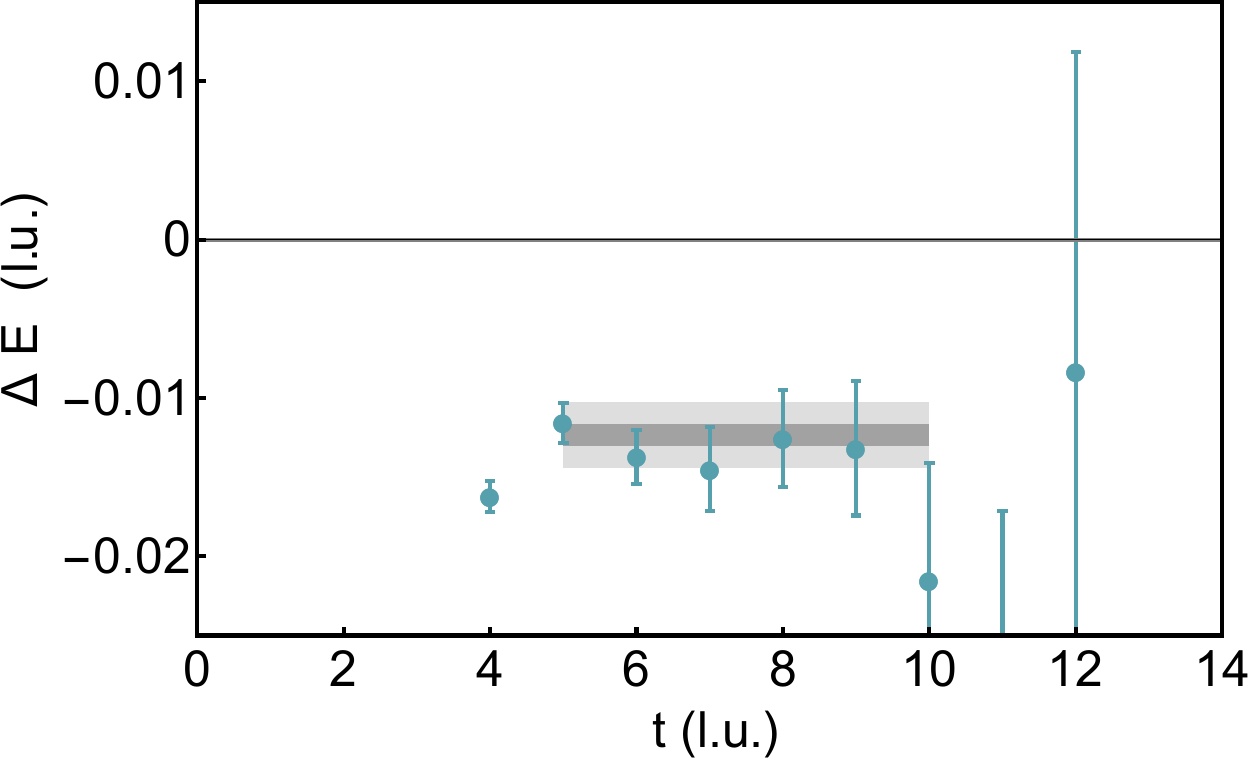}   \includegraphics[width=0.32 \columnwidth]{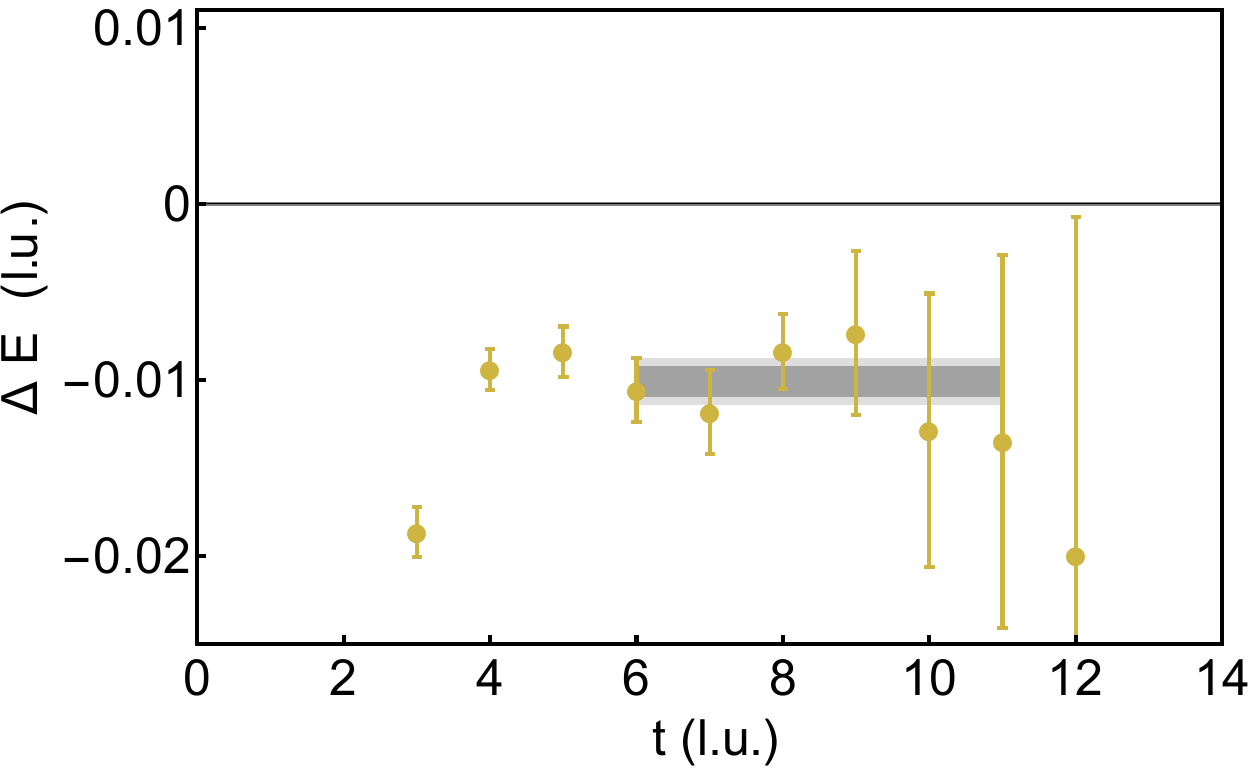}     \includegraphics[width=0.32 \columnwidth]{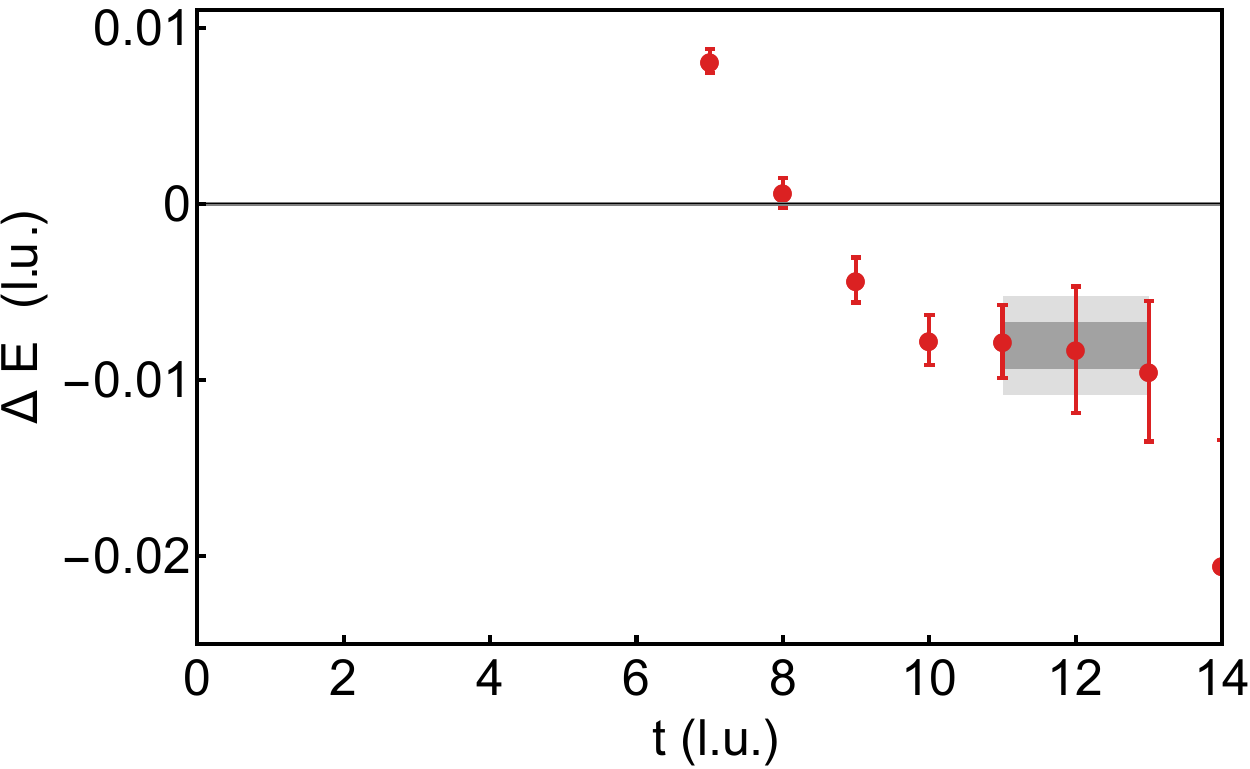}    
  \caption{
EMPs for the energy difference between the deuteron and twice the nucleon in the $L=24$ (left), $L=32$ (center) and $L=48$ (right) ensembles,
along with fits to the plateau regions.
The extracted binding energies are given in Table~\protect\ref{tab:deutbinding}.
  }
  \label{fig:DeutEMPs}
\end{figure}

In nature, the deuteron is the only bound state in the two-nucleon systems, residing in the $\siii$-$\diii$ coupled channels,
and it has a special position in nuclear physics.
The deuteron has always provided a benchmmark when deriving phenomenological interactions between nucleons, and it
will play a critical role in verifying LQCD as a viable calculational tool.
Correlation functions for two nucleons in the even-parity $\mathbb{T}_1$ irrep of the cubic group 
were constructed,
from which,
after a correlated subtraction of  twice  
the energy of a single nucleon,
the EMPs shown in Fig.~\ref{fig:DeutEMPs} were derived~\footnote{
The single nucleon correlation function, the square of which  is divided out of  two-nucleon correlation functions to yield a plateau on the energy difference,
have been temporally displaced, in some instances, to enhance the plateau region in the difference.
Further, due to the nature of the HL estimator, the first few time-slices in the difference correlation function have been removed, leading to temporal displacements of the EMPs.
The EMPs  defining  energy differences in this work correspond to both the one-nucleon and two-nucleon correlation functions being in their respective ground states
(as defined by plateaus in their respective individual EMPs).
}.
As  with the single hadrons,  correlated $\chi^2$-minimization fits of a constant to the plateau regions were performed 
to estimate the deuteron binding energy, $B_d$, and associated uncertainties.
The deuteron binding energies extracted from each ensemble are given in Table~\ref{tab:deutbinding},
along with
the values of $e^{-\kappa L}$, where $\kappa = \sqrt{ M_N B_d}$ is the binding momentum
of the deuteron.
As $e^{-\kappa L}$ is seen to  change from $\sim 3\%$ in the largest volume to $\sim 11\%$ in the smallest, 
an extrapolation in volume is desirable.
\renewcommand{\arraystretch}{1.5}
\begin{table}
\begin{center}
\begin{minipage}[!ht]{16.5 cm}
  \caption{
The deuteron binding energies extracted from plateaus in the EMPs 
shown in Fig.~\protect\ref{fig:DeutEMPs},
along with the infinite-volume extrapolated value.
The size of the FV effects is characterized by $e^{-\kappa L}$, shown in the last column.
The first uncertainty corresponds to the statistical uncertainty associated with the fit, the second corresponds 
to the systematic uncertainty associated with the selection of the fitting interval (determined by varying this range).  
In the case of dimensionful quantities, the third uncertainty is associated with scale setting.
For the infinite-volume values of the binding energy, the last uncertainty 
is introduced by the finite-volume extrapolation in eq.~(\protect\ref{eq:Bvol}), 
and is estimated by considering the effect of omitted terms scaling as 
$e^{-2 \kappa_0 L}/ L$.
  }  
\label{tab:deutbinding}
\end{minipage}
\setlength{\tabcolsep}{0.3em}
\begin{tabular}{c|cc|c}
\hline
      Ensemble & 
      $ \Delta E $\ (l.u.) & 
      $B_d $\ (MeV) &
      $e^{-\kappa L}$
      \\
\hline
\cfga &     -0.01157(73)(96) & 19.6(1.2)(1.6)(0.2) & 0.111\\
\cfgb &     -0.01037(89)(96) & 17.5(1.5)(1.6)(0.2) & 0.063\\
\cfgc &     -0.0078(12)(19) & 13.3(2.0)(3.2)(0.2) & 0.027\\
\hline 
$L=\infty$
& $-0.0085^{+(10)(16)(01)}_{-(10)(11)(01)}$ 
& $\Bd$  
& \\
\hline
\end{tabular}
\begin{minipage}[t]{16.5 cm}
\vskip 0.0cm
\noindent
\end{minipage}
\end{center}
\end{table}     

\begin{figure}[!ht]
  \centering
  \includegraphics[width=0.48 \columnwidth]{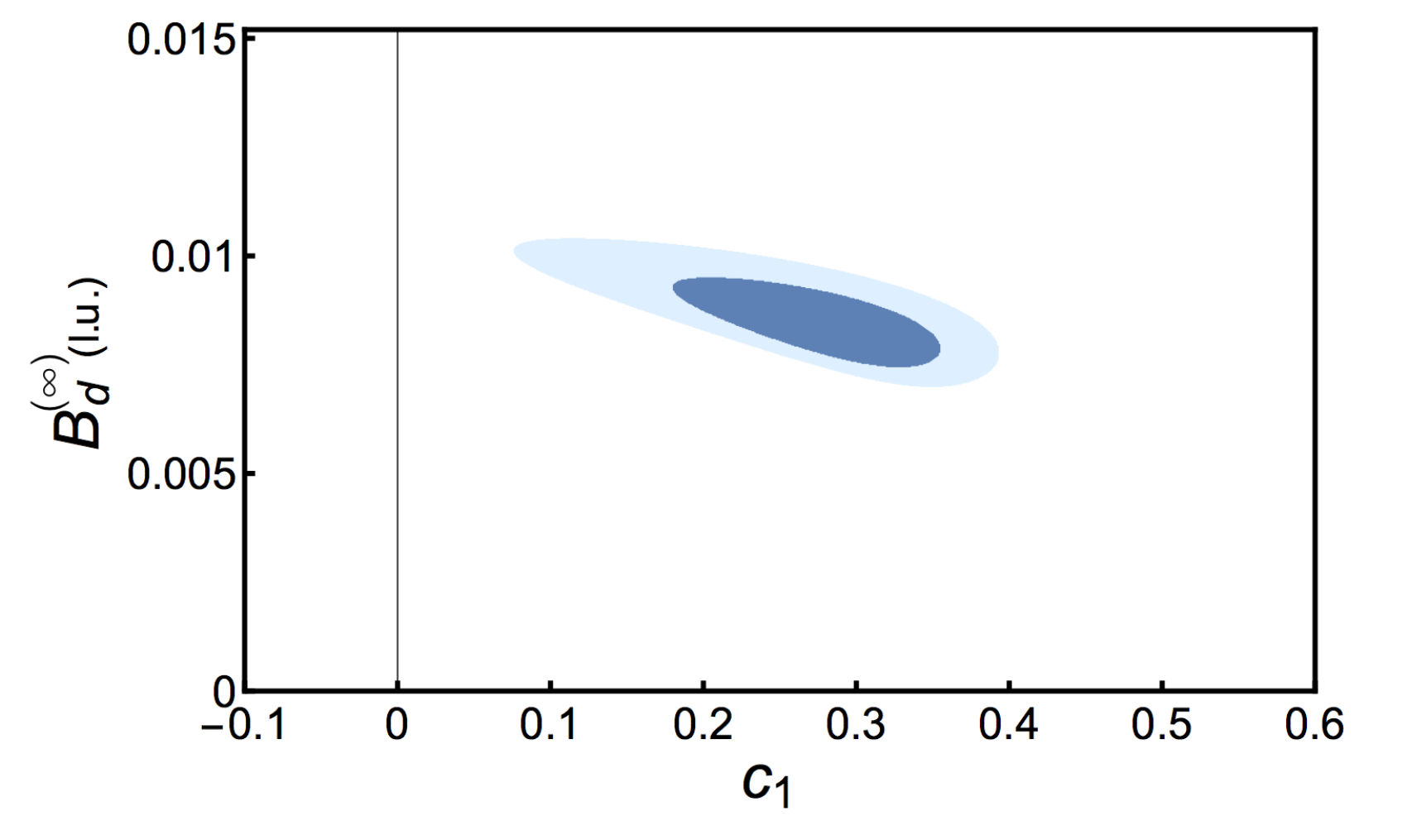}    
  \caption{
 The region in $c_1$-$B_d^{(\infty)}$ parameter space 
 defined by $\chi^2\rightarrow\chi^2_{\rm min}+1$.
 The inner region is defined by the statistical uncertainty, while the outer region is defined by the statistical and 
 systematic uncertainties combined in quadrature.
  }
  \label{fig:DeutBc}
\end{figure}

Inspired by the FV contributions to the binding of a shallow bound state resulting from 
short-range interactions~\cite{Beane:2003da,Bour:2011ef,Davoudi:2011md},
the extrapolation to infinite volume was performed by fitting a function of the form,
\begin{eqnarray}
B_d(L) & = & B_d^{(\infty)} + c_1 \left[ 
{ e^{-\kappa_0 L}\over L} \ + \ \sqrt{2} { e^{-\sqrt{2} \kappa_0 L}\over L} 
\ +\ {4\over 3\sqrt{3}} { e^{-\sqrt{3} \kappa_0 L}\over L} 
\right]\ +\ ...
\ \ \ ,
\label{eq:Bvol}
\end{eqnarray}
to the results obtained in the three lattice volumes,
where $\kappa_0 = \sqrt{ M_N B_d^{(\infty)} }$ 
(with $B_d^{(\infty)}$ the deuteron binding energy in infinite volume)
and $c_1$ are the fit parameters.
The ellipsis denote terms that are ${\cal O}( e^{- 2 \kappa_0 L})$ and higher.
A  $\chi^2$-minimization fit to the deuteron binding energies in Table~\ref{tab:deutbinding}
generates the region in $c_1$-$B_d^{(\infty)}$ parameter space shown in Fig.~\ref{fig:DeutBc}, 
defined by $\chi^2\rightarrow\chi^2_{\rm min}+1$.
The deuteron binding energy found from  extrapolating to infinite volume is
\begin{eqnarray}
B_d^{(\infty)} & = & \Bd~{\rm MeV}
\ \ \ .
\label{eq:Bd}
\end{eqnarray}
The first uncertainty corresponds to the statistical uncertainty,
the second corresponds to the fitting systematic uncertainty, 
the third  is associated with scale setting, and the last uncertainty 
is introduced by the finite-volume extrapolation in eq.~(\protect\ref{eq:Bvol}), 
and is estimated by considering the effect of terms scaling as 
$~\sim e^{-2 \kappa_0 L}/ L$.
Combining the errors in eq.~(\ref{eq:Bd}) 
in quadrature leads to 
$B_d^{(\infty)} = \BdSUMMARY~{\rm MeV}$.

%%%%%%%%%%%%%%%%%%%%%
\subsubsection{The Mixing Parameter, $\epsilon_1$}
\label{subsec:mixingparam}

For  a deuteron that is moving in the lattice volume,
the energy eigenvalues are sensitive to the mixing parameter $\epsilon_1$,
as expected from the QCs given in eq.~(\ref{eq:A2E}) for the specific boost  ${\bf d}=(0,0,1)$.
Explicitly evaluating the $c_{lm}^{\bf d}$ functions that appear in eq.~(\ref{eq:A2E}) 
for the two irreps containing the deuteron gives the QCs,
\begin{eqnarray}
k^*_{\mathbb{A}_2}\cot\delta_{1\alpha} (i \kappa_{\mathbb{A}_2})
+ \kappa_{\mathbb{A}_2} 
& = & {2 e^{-\kappa_{\mathbb{A}_2} L}\over L} 
\left[1 + 2 \left(1 + {3\over\kappa_{\mathbb{A}_2} L} + {3\over(\kappa_{\mathbb{A}_2} L)^2} \right)
s_{\epsilon_1}(i\kappa_{\mathbb{A}_2})
\right] ,
\nonumber\\
k^*_{\mathbb{E}}\cot\delta_{1\alpha}(i\kappa_{\mathbb{E}}) + \kappa_{\mathbb{E}} 
& = & {2 e^{-\kappa_{\mathbb{E}} L}\over L} 
\left[ 1 - \left(1 + {3\over\kappa_{\mathbb{E}} L} + {3\over(\kappa_{\mathbb{E}} L)^2} \right)
s_{\epsilon_1}(i\kappa_{\mathbb{E}})
\right] 
\ ,
\end{eqnarray}
where $s_{\epsilon_1}(k^*)$ is defined in eq.~(\ref{eq:mixdef}).
For both irreps, the functions $k^* \cot\delta$  and $s_{\epsilon_1}$
are evaluated at $k^* = i\kappa$.
Iteratively solving these QCs in terms of the infinite-volume binding momentum, $\kappa_0$ 
($\kappa_{\mathbb{A}_2} , \kappa_{\mathbb{E}} \rightarrow \kappa_0$ in the infinite-volume limit),
the spin-averaged binding energy of the $\mathbb{A}_2$ and $\mathbb{E} $ irreps  is
\begin{eqnarray}
\overline{B}_d^{(0,0,1)}
& = & B_d^{(\infty)}\ +\ 
{4\kappa_0\over M} {Z_\psi^2\over L} e^{-\kappa_0 L}
\ +\ ...
\ \ \ ,
\end{eqnarray}
where the ellipses denote terms ${\cal O}(e^{-\sqrt{2} \kappa_0 L})$ and higher, 
which is consistent,
at this order,
with the binding energy extracted from the $\mathbb{T}_1$ irrep for the deuteron at rest.
In the above expression,
$Z^2_\psi$ is the residue of the deuteron pole.
The difference in energies is
\begin{eqnarray}
\delta B_d^{(0,0,1)} & = & - 
{12\kappa_0\over M} {Z_\psi^2\over L} e^{-\kappa_0 L}
  \left(1 + {3\over\kappa_0 L} + {3\over(\kappa_0 L)^2} \right) 
s_{\epsilon_1}(i\kappa_0)
\ +\ ...
\ \ \ .
\end{eqnarray}
Calculating the exponentially small difference between the energies of these two states provides a direct measure of 
$\epsilon_1$ evaluated at the deuteron pole.
In order to extract a meaningful constraint on $\epsilon_1$, 
the FV corrections must be statistically different from zero, otherwise the coefficient of the leading contribution to the energy difference vanishes.

In the present production, it has been only possible to decompose the ${\bf d}=(0,0,1)$
boosted deuteron correlation functions 
into the $\mathbb{E}$ ($j_z=\pm 1$) and $\mathbb{A}_2$ ($j_z=0$) irreps 
in calculations performed with the $L=32$ ensemble.  
\begin{figure}[!ht]
  \centering
  \includegraphics[width=0.48 \columnwidth]{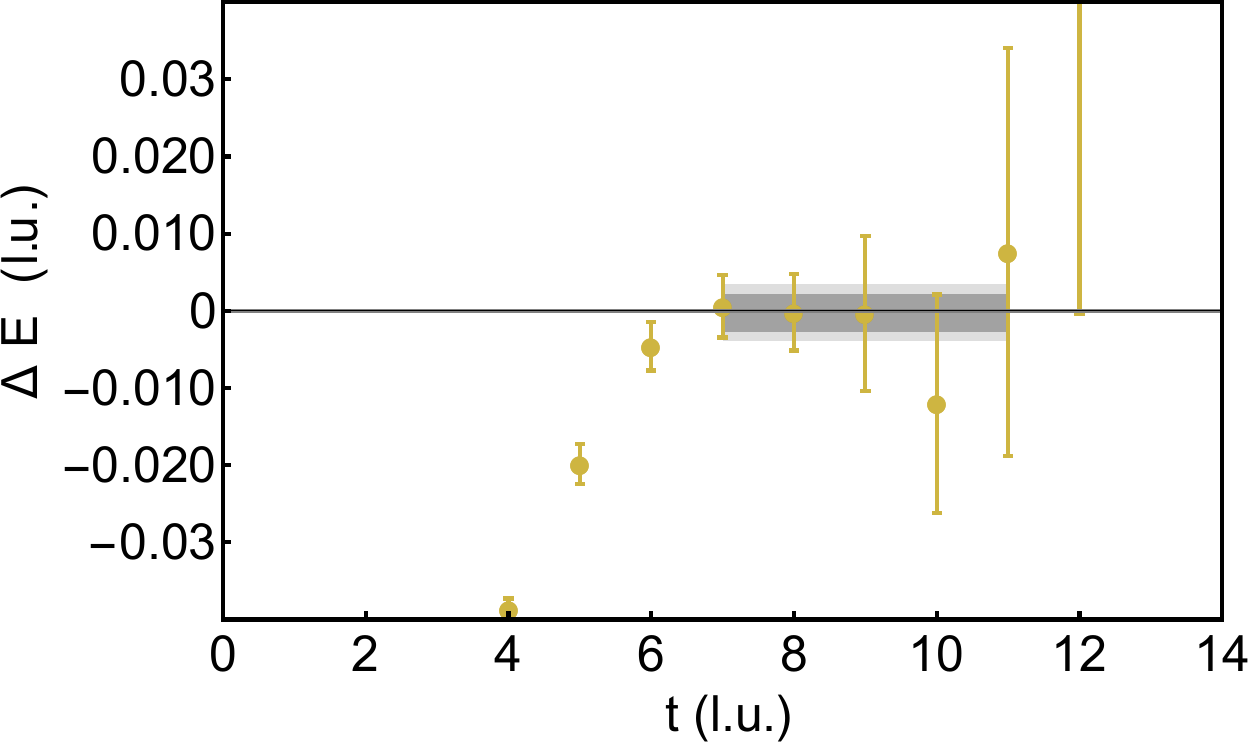}     \caption{
EMP associated with the energy
difference between the 
$\mathbb{E}$ ($j_z=\pm 1$) and $\mathbb{A}_2$ ($j_z=0$) deuteron states
with boost vector ${\bf d}=(0,0,1)$
in the  $L=32$ ensemble, along with fits to the plateau region.
The energy difference depends upon the mixing parameter $\epsilon_1$.
  }
  \label{fig:BostedDeutEMPs}
\end{figure}
The EMP associated with the difference in energies between these irreps is shown in Fig.~\ref{fig:BostedDeutEMPs},
and 
the energy difference extracted from fitting the plateau region is consistent with zero,
$\delta B_d^{(0,0,1)} (L=32) = -0.4(4.1)(4.6)~{\rm MeV}$.
While this energy difference  is bounded in magnitude, the fact that the FV contributions 
to the deuteron binding energy
are consistent with zero in this lattice volume means that no useful bound can be placed upon $\epsilon_1$.

%%%%%%%%%%%%%%%%%%%%%
\subsubsection{A Compilation of Deuteron Binding Energies from LQCD}
\label{subsec:deutBindLQCD}

\begin{figure}[!ht]
  \centering
  \includegraphics[width=0.48 \columnwidth]{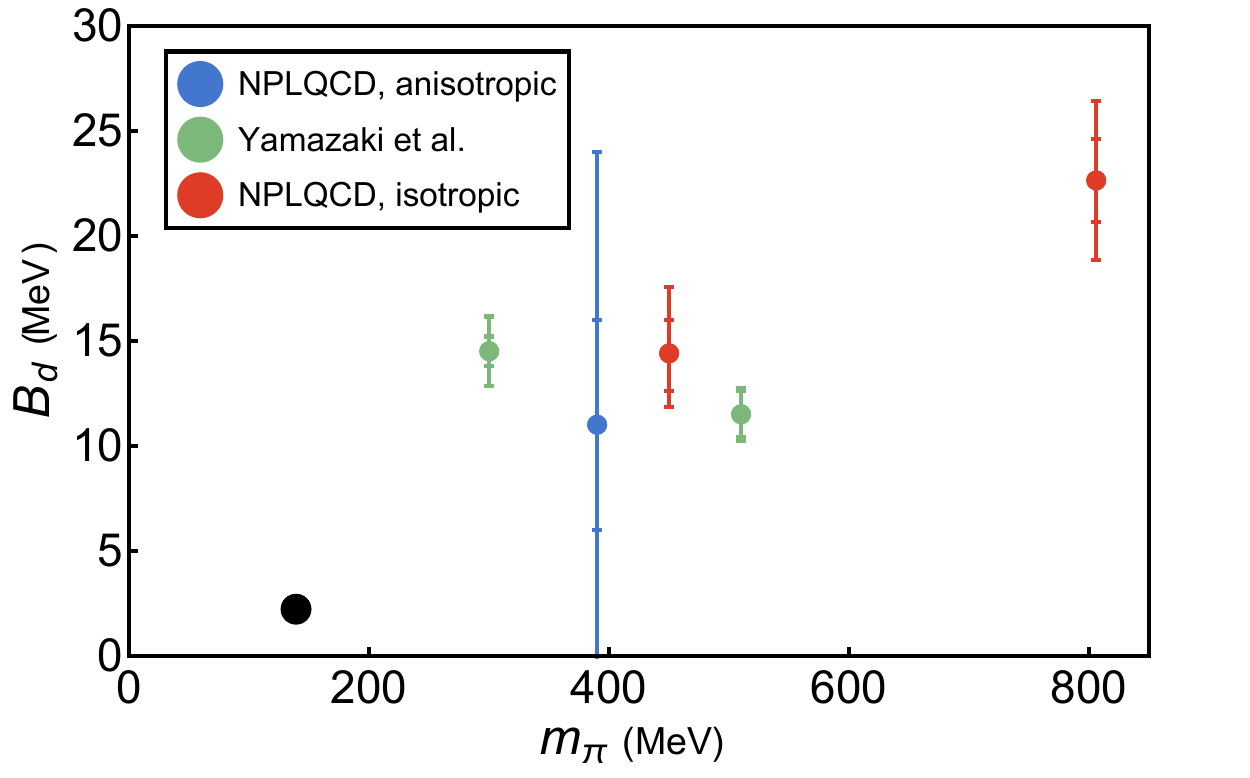}\ \
  \caption{
  The pion mass dependence of the deuteron binding energy calculated with LQCD.
  The NPLQCD anisotropic-clover  
  result is from Ref.~\protect\cite{Beane:2011iw}, the Yamazaki {\it et al}. results are from Refs.~\protect\cite{Yamazaki:2012hi,Yamazaki:2015asa}
  and the NPLQCD isotropic-clover results are from this work and  Ref.~\protect\cite{Beane:2012vq}.
  The black disk corresponds to the experimental binding energy.
    }
  \label{fig:DeutALLlqcd}
\end{figure}

The current calculation of the deuteron binding energy adds to a small number of previous calculations 
over a range of pion masses above 
$\sim 300~{\rm MeV}$~\protect\cite{Beane:2011iw,Beane:2012vq,Yamazaki:2012hi,Yamazaki:2015asa},~\footnote{
The deuteron and dineutron binding energies at $m_\pi\sim 800~{\rm MeV}$ in the $L=24$ and $L=32$ ensembles
presented in Ref.~\cite{Beane:2012vq} 
have been reproduced in Ref.~\cite{Berkowitz:2015eaa}, within uncertainties, on the same gauge ensembles.}
as  shown in Fig.~\ref{fig:DeutALLlqcd}.~\footnote{
The results of quenched calculations, and of calculations that have not been extrapolated to infinite volume~\cite{Beane:2006mx}, 
have not been shown.
The results  from Ref.~\cite{Yamazaki:2012hi,Yamazaki:2015asa} were obtained  with a power-law  extrapolation to infinite volume. 
This is not the correct form for a loosely bound state, and tends to lead to significantly smaller uncertainties than from extrapolations performed with 
the known
exponential form.
}
The present result is consistent, within uncertainties, 
with the results at $m_\pi\sim 300~{\rm MeV}$ and $m_\pi\sim 500~{\rm MeV}$ 
from Refs.~\cite{Yamazaki:2012hi,Yamazaki:2015asa}.
Further LQCD calculations at lighter quark masses are required to quantify the approach to the physical deuteron binding 
(for related NNEFT work see Ref.~\cite{Baru:2015ira}).

%%%%%%%%%%%%%%%%%%%%%
\subsection{Scattering in the $\siii$-$\diii$ Coupled Channels}
\label{subsec:tripletscattering}

To recover the S-matrix in the  $\siii$-$\diii$ coupled channels, 
calculations must be performed that isolate the phase shifts and mixing angle,
$\delta_{1\alpha}$, $\epsilon_1$ and $\delta_{1\beta}$, defined in eq.~(\ref{eq:BBSmat}),
from the FV  observables accessible to LQCD calculations.
The formalism with which to perform this analysis~\cite{He:2005ey,Briceno:2012yi,Briceno:2013lba,Briceno:2013bda}
is an extension of the seminal work of  L\"uscher~\cite{Luscher:1986pf,Luscher:1990ux}.
For vanishing total momentum, 
assuming that the contribution from  $\delta_{1\beta}$, $D$-waves and higher 
are negligible, 
the energies of  the $\mathbb{T}_1$ irreps are insensitive to $\epsilon_1$, 
as demonstrated  in eq.~(\ref{eq:Tone}).
Therefore, the shifts in energies of the two nucleon states in the $\mathbb{T}_1$
irrep for various total momentum from the energy of two free nucleons 
can be used to 
extract $\delta_{1\alpha}$ below the inelastic threshold.

\begin{figure}[!ht]
  \centering
  \includegraphics[width=0.32 \columnwidth]{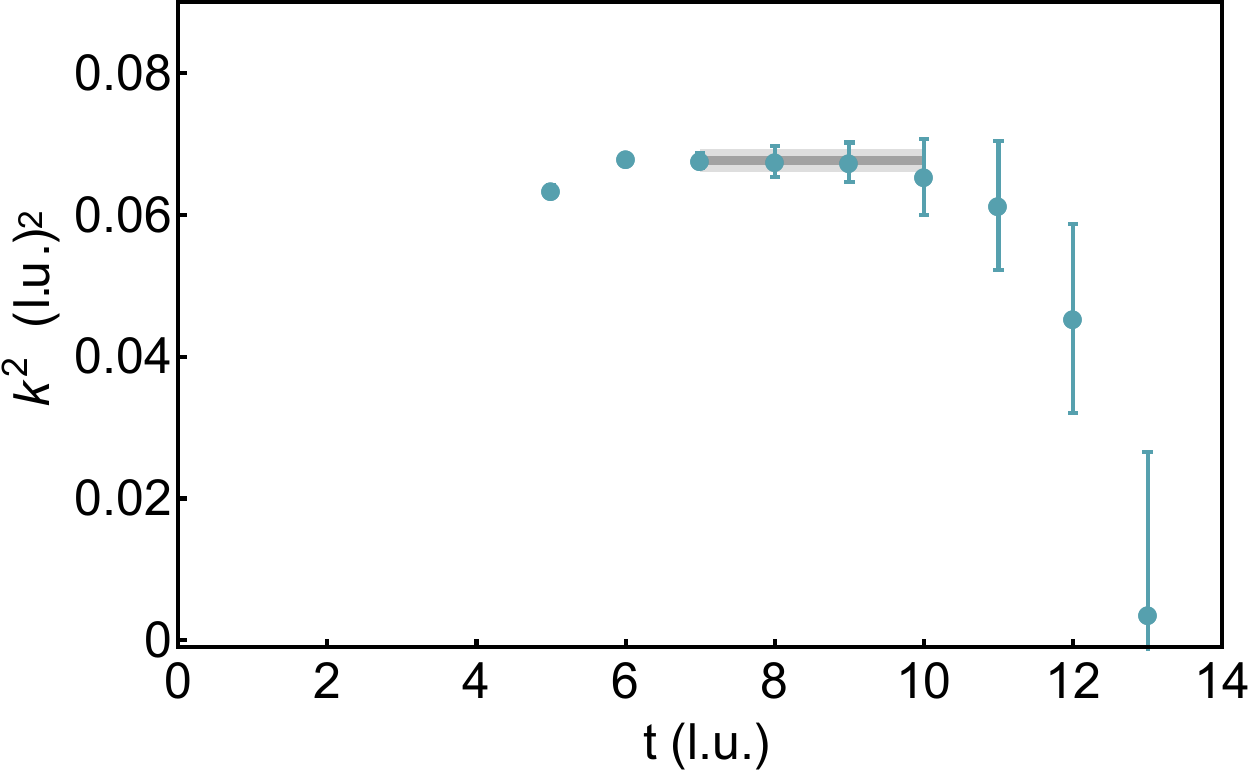}   
  \includegraphics[width=0.32 \columnwidth]{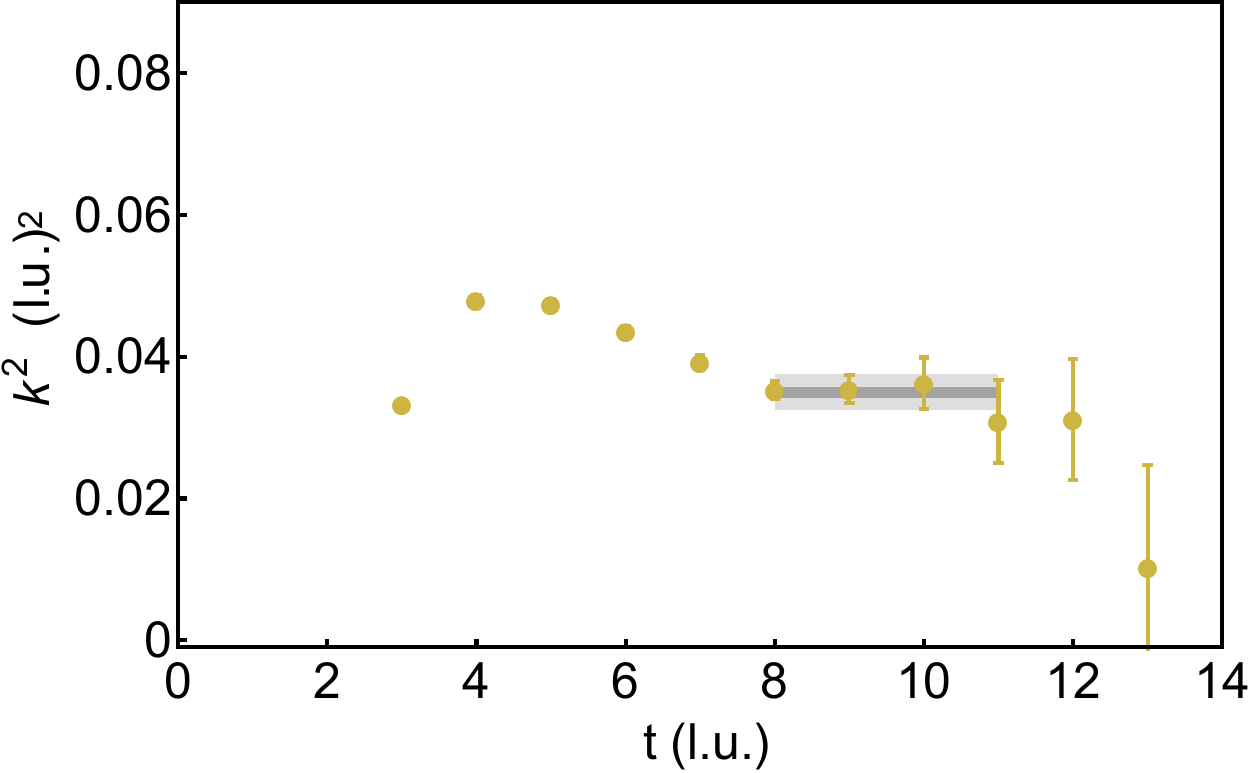}     
  \includegraphics[width=0.32 \columnwidth]{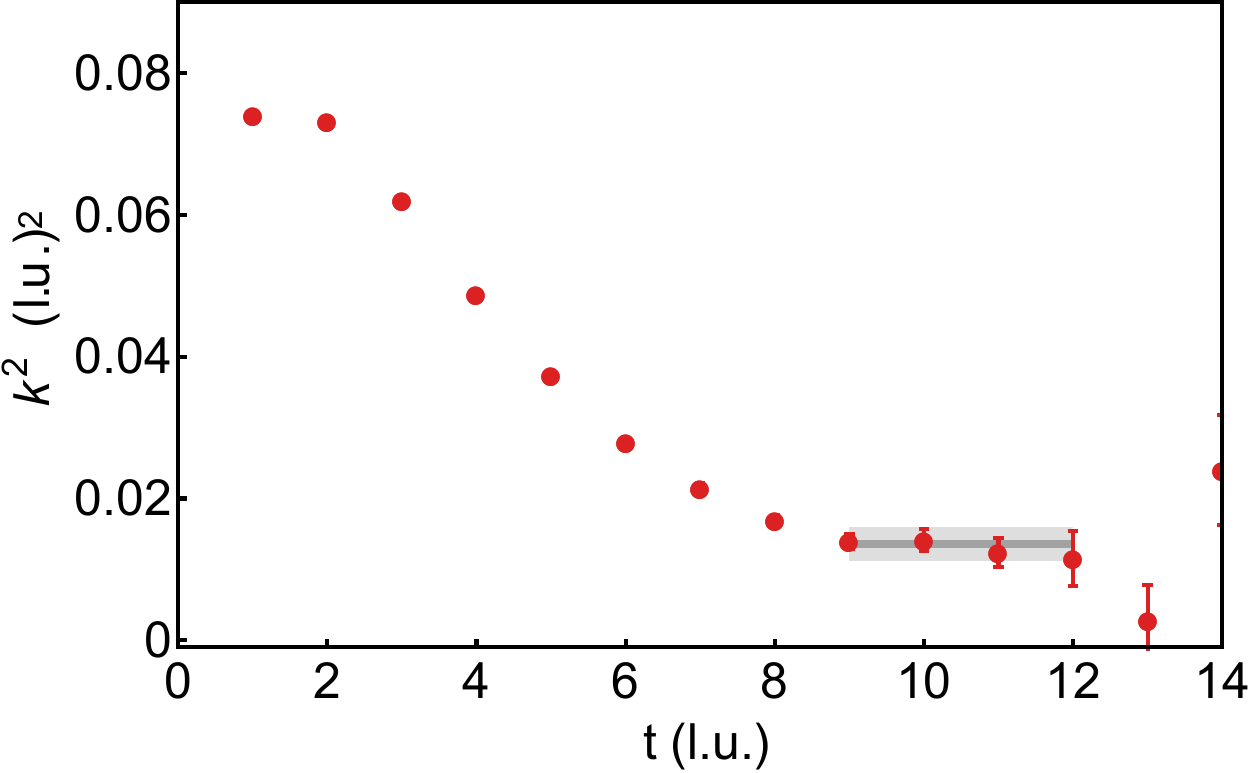}    
  \caption{
Ek2Ps for the lowest lying continuum
$\siii$-$\diii$ NN states near $k^*=2\pi/L$
 in the $L=24$ (left), $L=32$ (center) and $L=48$ (right) ensembles,
along with fits to the plateau regions.
  }
  \label{fig:neq1EMPs}
\end{figure}

\begin{figure}[!ht]
  \centering
 \includegraphics[width=0.32 \columnwidth]{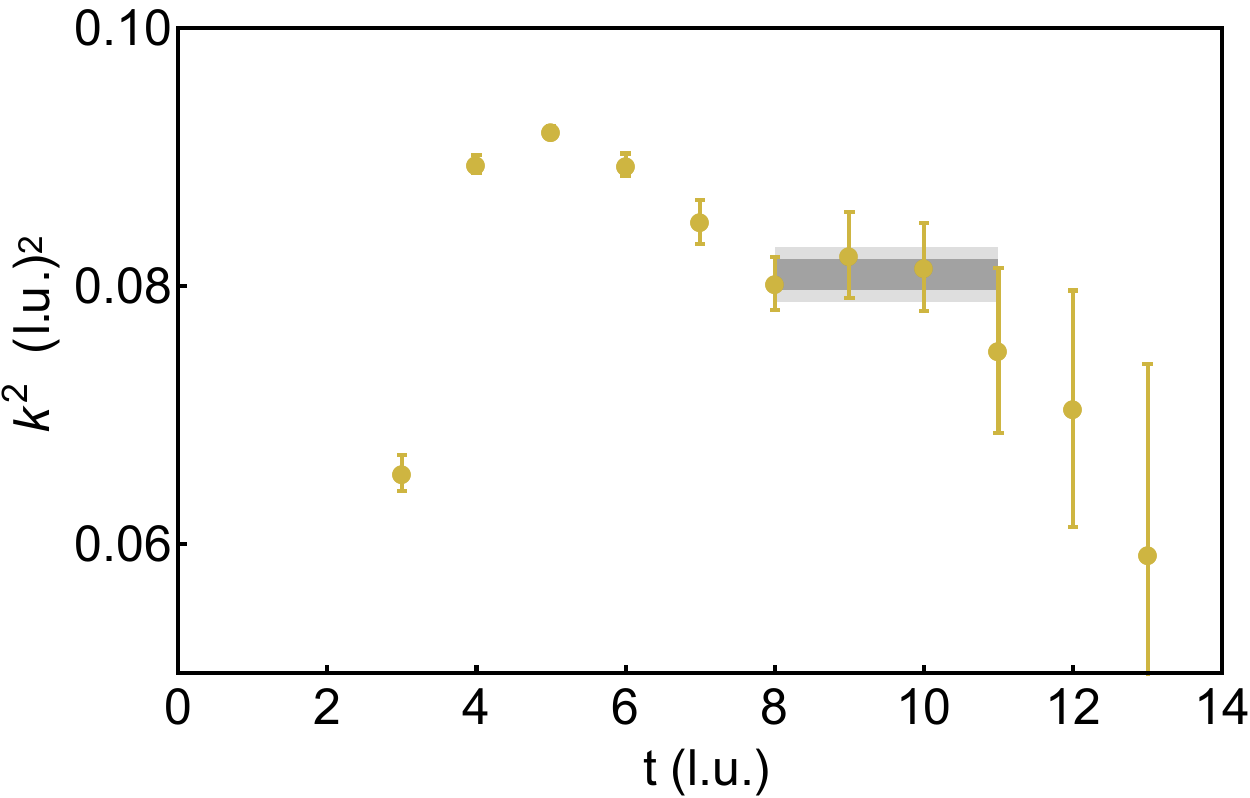}     \includegraphics[width=0.32 \columnwidth]{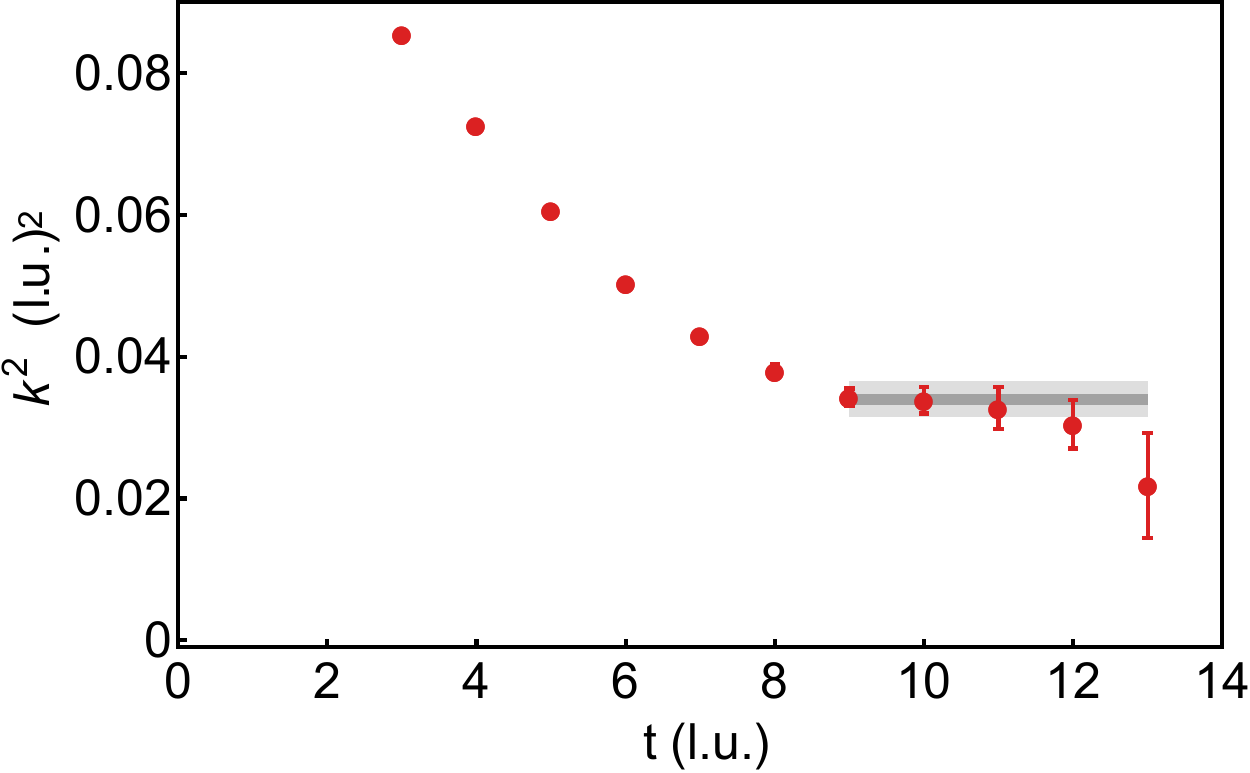}    
  \caption{
Ek2Ps for the continuum
$\siii$-$\diii$ NN states near 
\textcolor{\revrevcolor}{$k^*=2\sqrt{2}\pi/L$}
 in the $L=32$ (left) and $L=48$ (right) ensembles,
along with fits to the plateau regions.
  }
  \label{fig:neq2EMPs}
\end{figure}

\begin{figure}[!ht]
  \centering
 \includegraphics[width=0.32 \columnwidth]{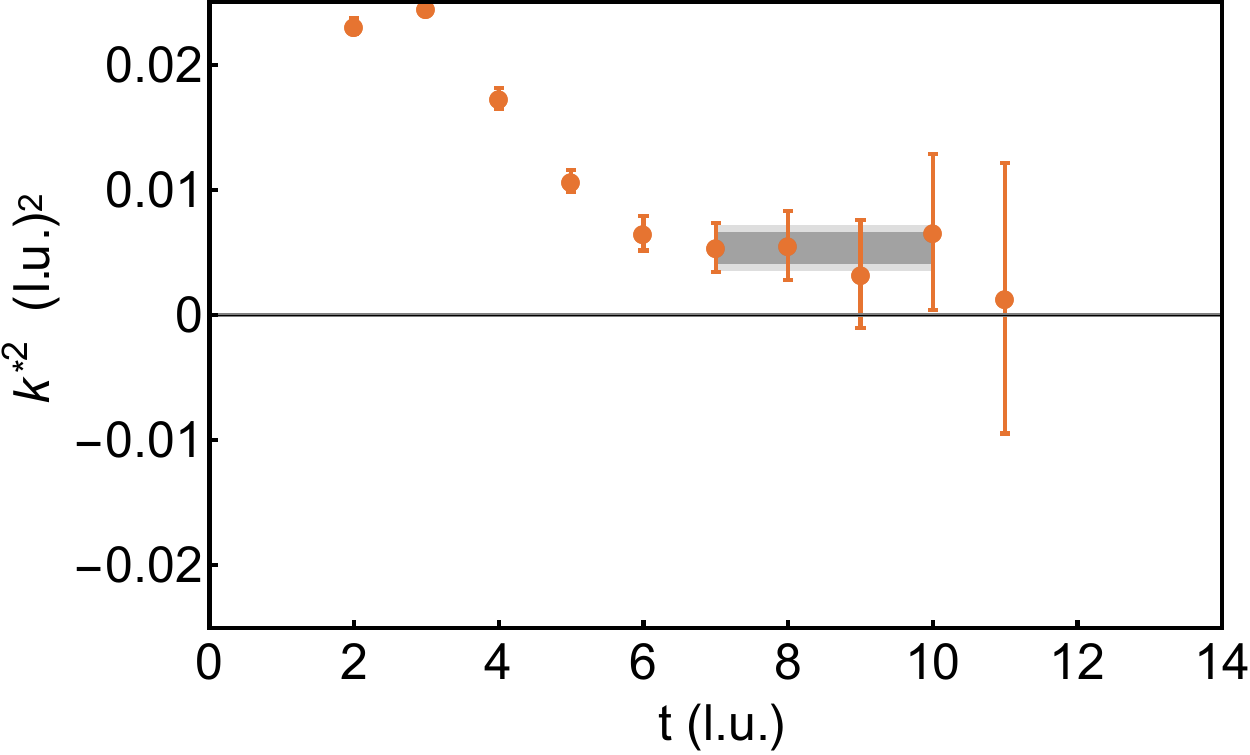}      
  \caption{
Ek2Ps for the 
spin-averaged 
continuum
$\siii$-$\diii$ NN states 
with ${\bf d}=(0,0,1)$ near  
\textcolor{\revrevcolor}{$k^*=0$}
 in the $L=32$ ensemble,
along with the fit to the plateau region.
  }
  \label{fig:deutPeq1EMPs}
\end{figure}

\begin{figure}[!ht]
  \centering
  \includegraphics[width=0.48 \columnwidth]{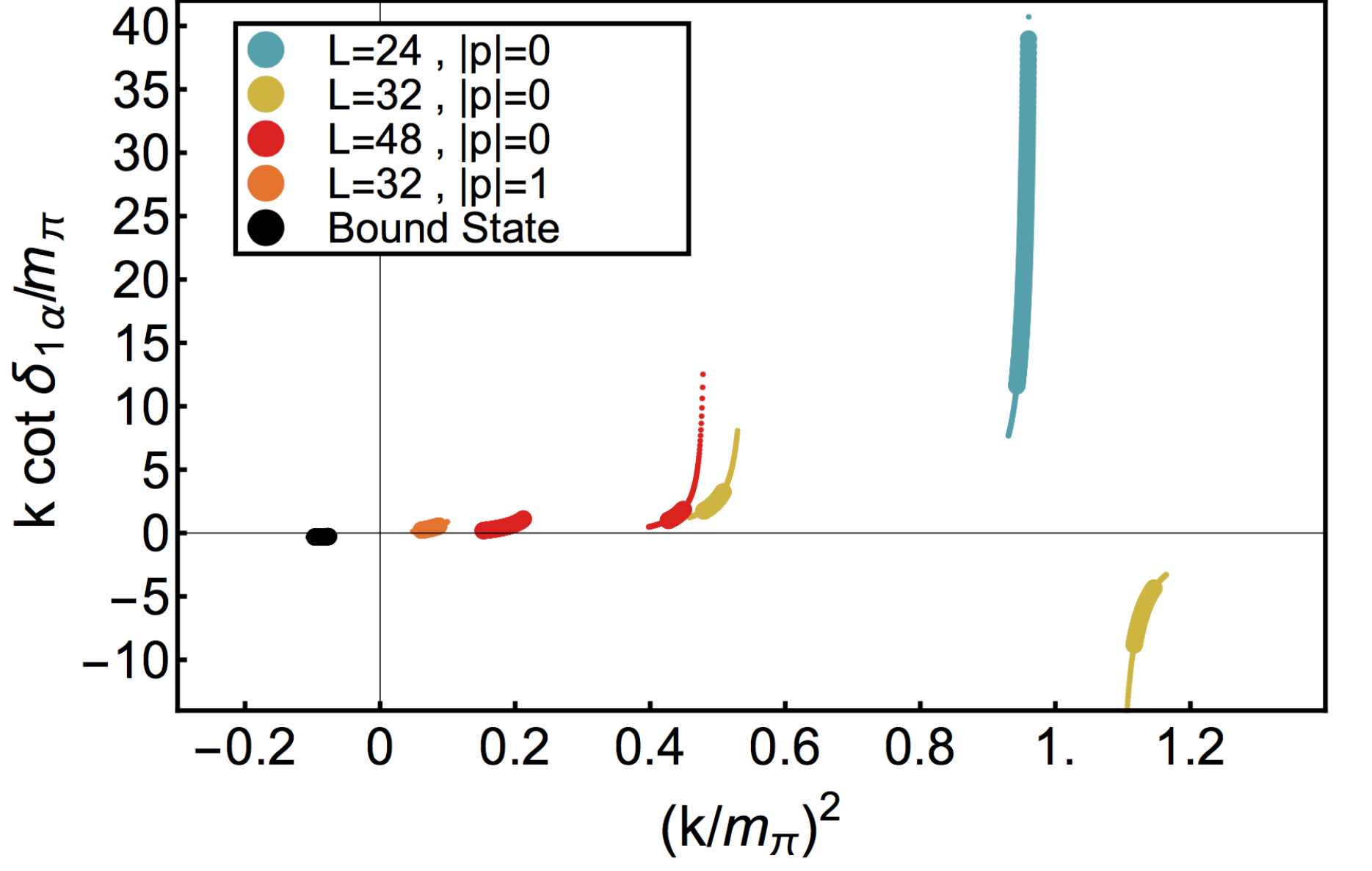}      
  \includegraphics[width=0.48 \columnwidth]{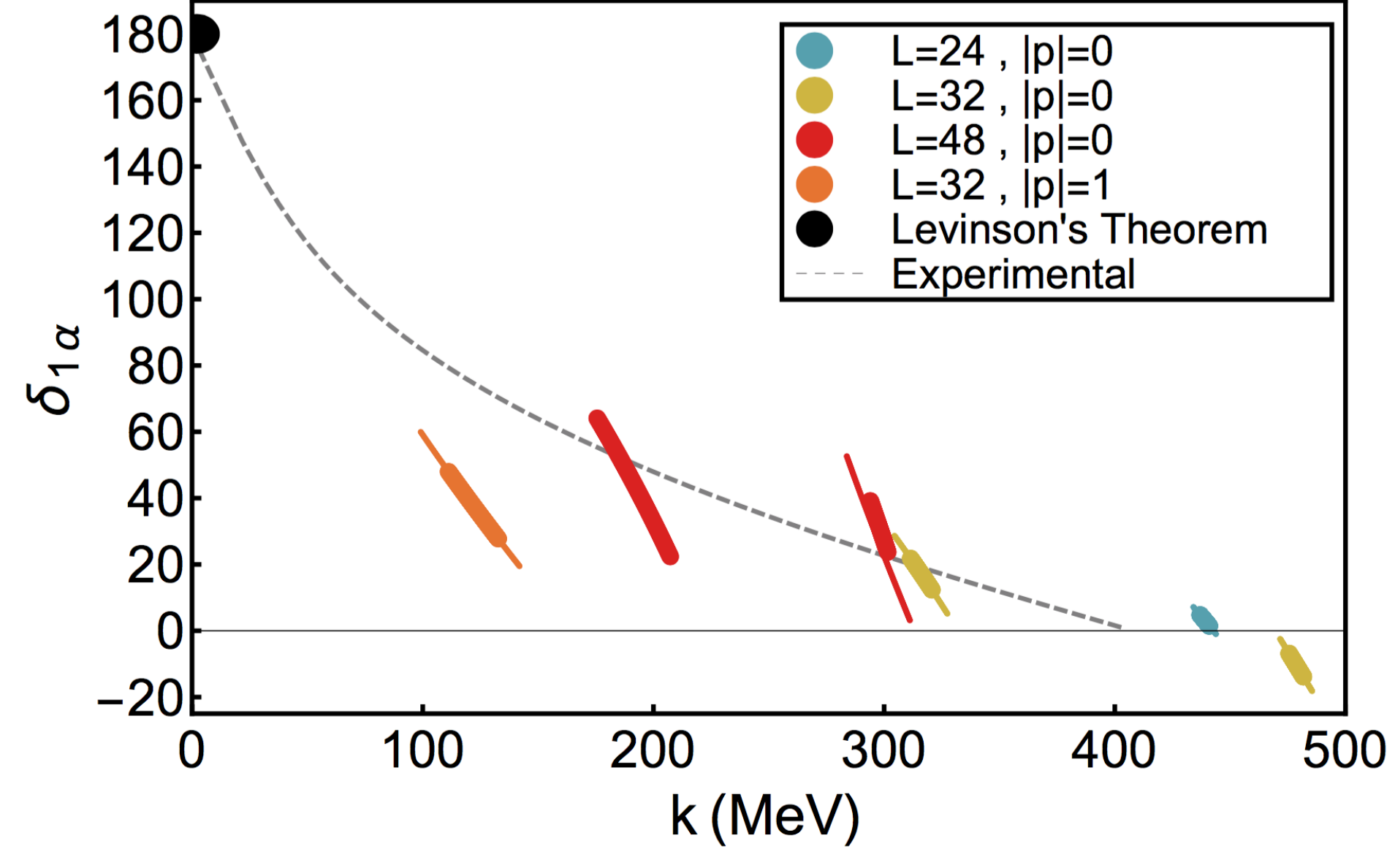}    
  \caption{
Scattering in the $\siii$-$\diii$ coupled channels.  
The left panel shows $k^*\cot\delta_{1\alpha}/m_\pi$ as a function of $k^{*2}/m_\pi^2$,
while the right panel shows the phase shift as a function of momentum in MeV,
assuming that $\delta_{1\beta}$ and the D-wave and higher partial-wave phase shifts vanish.  
The thick (thin) region of each result correspond to the statistical uncertainty (statistical and systematic uncertainties combined in quadrature).
The black circle in the right panel corresponds to the known result from Levinson's theorem, while
the dashed-gray curve corresponds to the  phase shift extracted from the Nijmegen partial-wave analysis of experimental data~\cite{NijmegenPWA}.
  }
  \label{fig:NN3s1kcotdelta}
\end{figure}

\begin{figure}[!ht]
  \centering
  \includegraphics[width=0.48 \columnwidth]{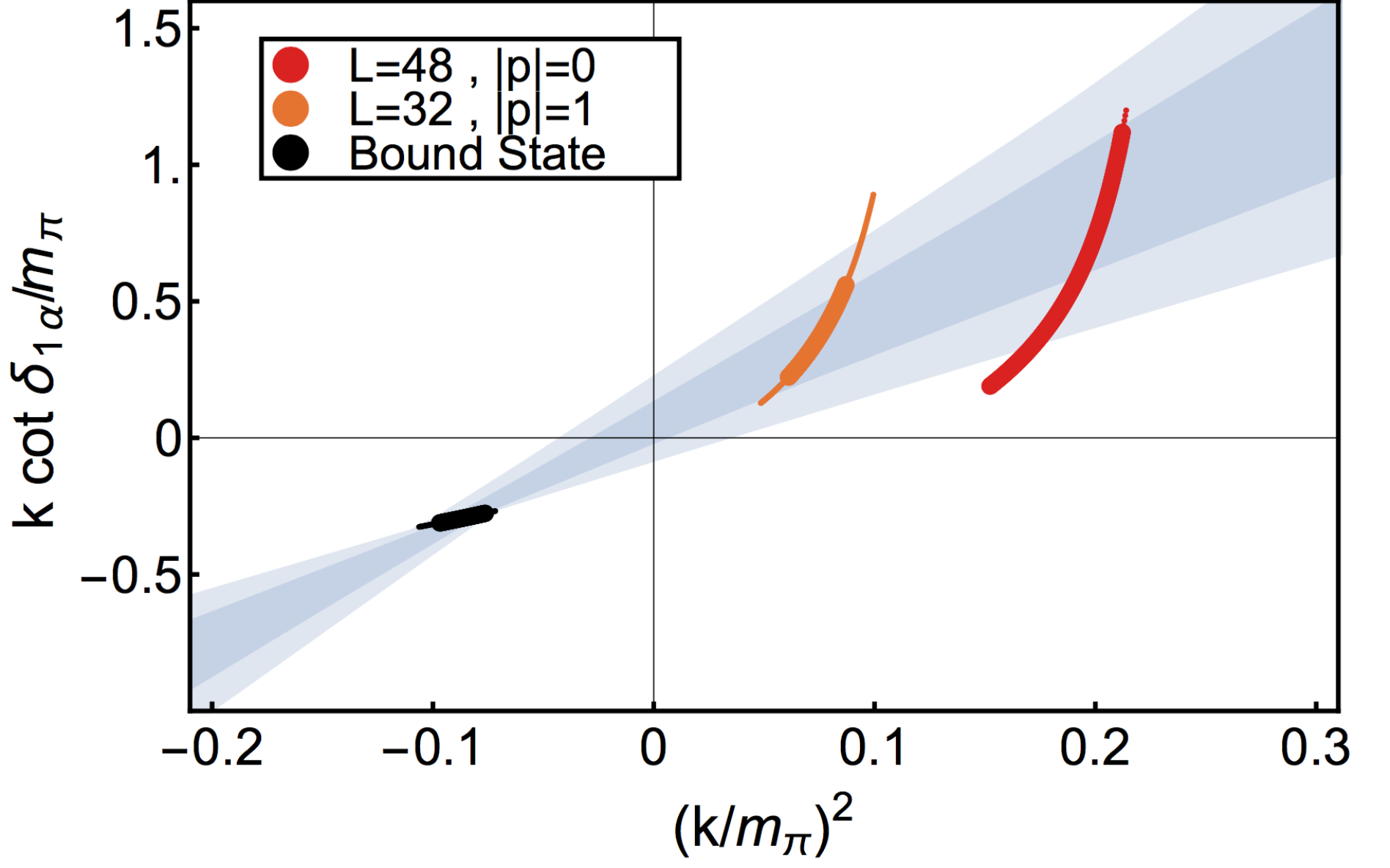}    
  \includegraphics[width=0.44 \columnwidth]{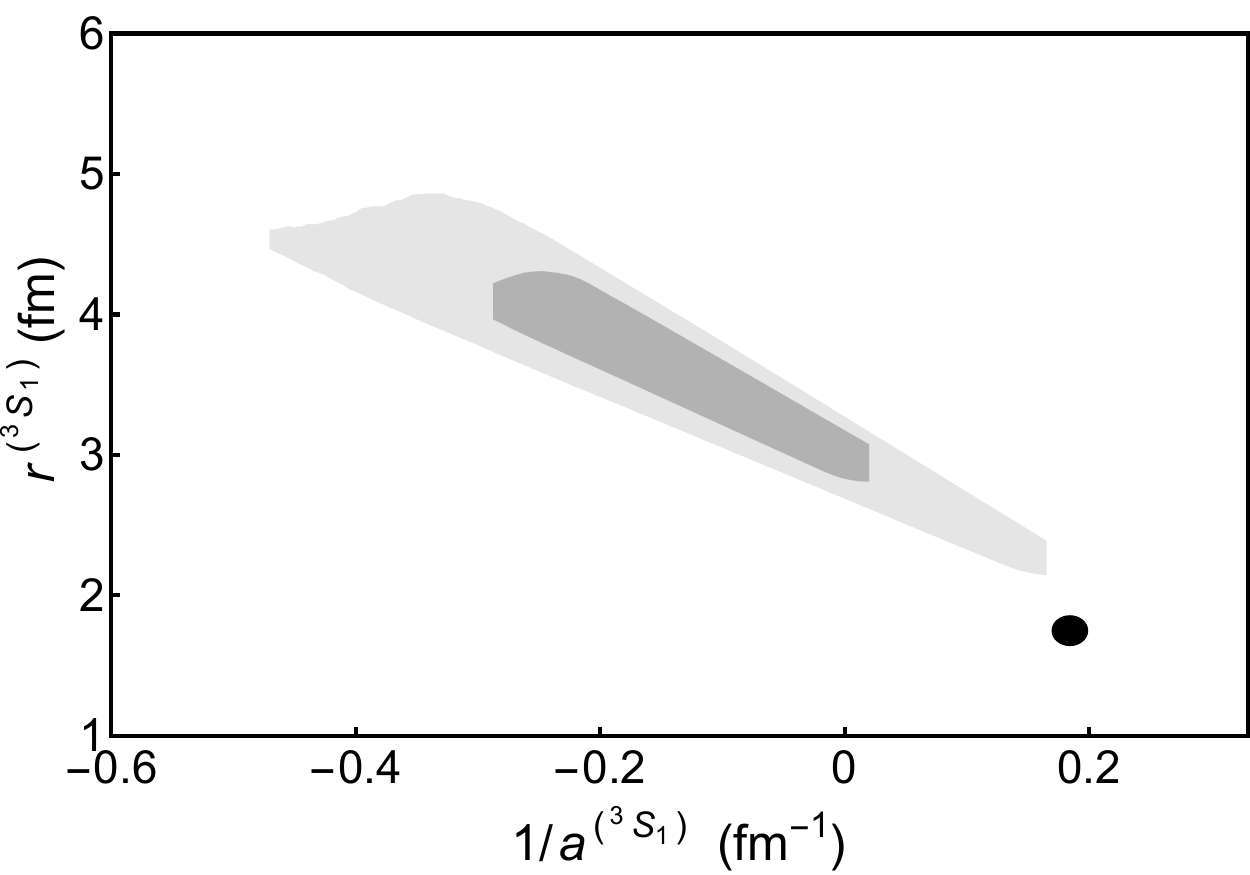}    
  \caption{
Scattering in the $\siii$-$\diii$ coupled channels below the start of the t-channel cut,
$k^{*2} < m_\pi^2/4$,
assuming that $\delta_{1\beta}$ and the D-wave and higher partial-wave phase shifts vanish.  
The left panel shows solid region corresponding to linear fits associated with the statistical uncertainty
and the statistical and systematic uncertainties combined in quadrature.
The right panel shows the scattering parameters, $1/a^{(\siii)}$ and $r^{(\siii)}$ 
determined from fits to scattering results below the t-channel cut.
The solid circle corresponds to the experimental values.
}
  \label{fig:NN3s1kcotdeltaER}
\end{figure}

Figure~\ref{fig:neq1EMPs} show the effective-$k^{*2}$ plots (Ek2Ps) associated with the first continuum 
$\mathbb{T}_1$ states in each ensemble, with momentum near \textcolor{\revrevcolor}{$k^*=2\pi/L$}.
These show the values of the interaction momentum $k^{*2}$ extracted from the LQCD correlation functions as a function of 
Euclidean time. 
As with the EMPs, plateau behavior indicates the dominance of a single state.
\textcolor{\revrevcolor}{
Because of the $\mathbb{E}$ irrep of the cubic group that is present in the $k^*=2\pi/L$ shell,
the spectrum is expected to have a predominantly D-wave state  that is close by~\cite{Briceno:2013bda}.
The overlap of our sources and sinks onto this state will be small, dictated  by the small mixing between the 
S-waves and D-waves. Analogous states are also present in higher-$k^*$ shells and in boosted systems.
}
For an arbitrary two-body system, comprised of particles with masses $m_1$ and $m_2$,
with zero CoM momentum, 
the interaction momentum
$k^{*2}$, is defined through
\begin{eqnarray}
\delta E^* & = & E^*- m_1 - m_2 
\ =\ 
\sqrt{k^{*2} + m_1^2} + \sqrt{k^{*2} + m_2^2} - m_1 - m_2 
\ \ \ ,
\label{eq:k2def}
\end{eqnarray}
where 
$E^*$ is the energy in the CoM frame, defined by 
$E^*=\sqrt{E^2-|{\bf P}_{\rm tot.}|^2}$ where $E$ is the total energy, 
and ${\bf P}_{\rm tot.}$ is the total momentum, of the system.
Figure~\ref{fig:neq2EMPs} shows the Ek2Ps for states with momentum near \textcolor{\revrevcolor}{$k^*=2\sqrt{2}\pi/L$},
while Fig.~\ref{fig:deutPeq1EMPs} shows the Ek2P for the system with  ${\bf d}=(0,0,1)$
on the $L=32$ ensemble.
Inserting the values of $k^*$ extracted from the plateau regions of the Ek2Ps in  Fig.~\ref{fig:neq1EMPs} and
Fig.~\ref{fig:neq2EMPs} 
into the QC in eq.~(\ref{eq:Tone}) gives rise to the values  of 
$k^*\cot\delta_{1\alpha}$  and $\delta_{1\alpha}$
given in Table~\ref{tab:deutquantities} and shown in Fig.~\ref{fig:NN3s1kcotdelta}.
Additionally, the result of inserting the value of $k^*$ extracted from the plateau in
Fig.~\ref{fig:deutPeq1EMPs} into the QC for the 
 $\mathbb{A}_2$ and  $\mathbb{E}$ irreps 
 in eq.~(\ref{eq:A2E})
 is  shown in Table~\ref{tab:deutquantities} and Fig.~\ref{fig:NN3s1kcotdelta}.
\renewcommand{\arraystretch}{1.5}% Wider
\begin{table}
\begin{center}
\begin{minipage}[!ht]{16.5 cm}
  \caption{
Scattering information in the $\siii$-$\diii$ coupled channels.
A ``-''  indicates that the uncertainty extends across a singularity of 
the L\"uscher function, or that it is associated with the bound state. 
The  uncertainties in these quantities are highly correlated, as can be seen from 
Fig.~\protect\ref{fig:NN3s1kcotdelta}.
}  
\label{tab:deutquantities}
\end{minipage}
\setlength{\tabcolsep}{0.3em}
\begin{tabular}{c|c|ccc}
\hline
      Ensemble &
      $|{\bf P}_{\rm tot}|$ (l.u.)& 
      $ k^*/m_\pi $  & 
      $ k^* \cot\delta_{1\alpha} / m_\pi$  &
      $\delta_{1\alpha}$ (degrees)        \\
\hline
All & 0 & $i 0.294^{+(17)(27)}_{-(18)(24)}  $ & $- 0.294^{+(17)(27)}_{-(18)(24)}  $ & - \\
\hline
\cfga & 0 &    $0.9754^{+(44)(98)}_{-(45)(99)}$ & - & $3.1(1.7)(3.7)$ \\
\hline
\cfgb &  0 & $0.702^{+(10)(23)}_{-(10)(24)}$ &    $2.3^{+(1.0)(5.7)}_{-(0.55)(0.89)} $ & $17(5)(11) $\\
\cfgb &  0 & $1.065^{+(07)(16)}_{-(08)(17)}$ &    $-5.4^{+(1.4)(2.1)}_{-(2.9)(29.5)} $ & $-11.1(3.8)(8.5) $ \\
\cfgb &  1 & $0.270^{+(26)(29)}_{-(40)(51)}$ &    $+0.35^{+(24)(15)}_{-(59)(20)} $ & $+38^{+(13)(23)}_{-(11)(16)}  $ \\
\hline
\cfgc &  0 &  $0.426(03)(12)$ & $0.45^{+(67)(34)}_{-(26)(08)}$ & $44^{+(21)(07)}_{-(21)(08)}$ \\
\cfgc &  0 &  $0.662(08)(29)$ & $0.35^{+(0.14)(3.0)}_{-(0.09)(0.21)}$ & $26^{+(07)(25)}_{-(07)(22)}$  \\
\hline
\end{tabular}
%noalign{\smallskip\hrule}\cr}
\begin{minipage}[t]{16.5 cm}
\vskip 0.0cm
\noindent
\end{minipage}
\end{center}
\end{table}     
The uncertainties in each of the extractions are relatively large,
magnified by their close proximity to a singularity in the kinematic functions $c_{00}^{{\bf d}}$.  
Even subject to these issues, a zero in the phase shift is  visible near  $k^*\sim m_\pi\sim 450~{\rm MeV}$,
indicative of an attractive interaction with a repulsive core.
It is interesting to compare this phase shift, at   a pion mass of $m_\pi\sim\pionmass$, with that of nature, illustrated
by the dashed curve in 
Fig.~\ref{fig:NN3s1kcotdelta}.
The phase shift resulting from a partial-wave analysis of experimental data is consistent, 
within uncertainties, with the phase shift calculated at $m_\pi\sim\pionmass$ over a large range of momenta.
The zeros of the phase shift occur at different momenta, but they are nearby.
Without results at smaller $k^*$, a precise extraction of the scattering parameters, such as the scattering length and effective range, is not feasible,
and additional calculations are required in order to accomplish this.
However, the   determination of the binding energy and the two continuum states 
that lie below the threshold of the t-channel cut (set by the pion mass, $k^*=m_\pi/2$)
can be used to perform an approximate determination
of the inverse scattering length and effective range. 
A linear fit was performed, $k^*\cot\delta = -1/a + {1\over 2} r k^{*2}$, as shown in Fig.~\ref{fig:NN3s1kcotdeltaER}.
The range of linear fits straddle  $k^*\cot\delta =0$ at $k^{*}=0$, and as such allows both $a^{(\siii)}=\pm\infty$,
and it is  useful to consider the constraints on $1/a^{(\siii)}$ rather than $a^{(\siii)}$.
The correlated constraints on $1/a^{(\siii)}$ and $r^{(\siii)}$ are shown in Fig.~\ref{fig:NN3s1kcotdeltaER}.
The inverse scattering length and effective range determined from the fit region in Fig.~\ref{fig:NN3s1kcotdeltaER}
are
\begin{eqnarray}
\left( m_\pi a^{(\siii)} \right)^{-1} & = & 
-0.04^{+(0.07)(0.08)}_{-(0.10)(0.17)}
\ \ ,\ \ 
m_\pi r^{(\siii)} \ =\ 
7.8^{+(2.2)(3.5)}_{-(1.5)(1.7)}
\nonumber\\
\left(  a^{(\siii)} \right)^{-1} & = & 
-0.09^{+(0.15)(0.19)}_{-(0.23)(0.39)}~{\rm fm}^{-1}
\ \ ,\ \ 
 r^{(\siii)} \ =\ 
3.4^{+(1.0)(1.5)}_{-(0.7)(0.8)}~{\rm fm}
\ \ \ .
\end{eqnarray}
Further calculations in larger volumes (and hence at smaller $k^{*2}$) will be required to refine these extractions.
There is a potential self-consistency issue raised by the size of the effective range that is within the  uncertainties that are reported.  
L\"uscher's method is valid only for the interaction ranges
$R \ll L/2$, otherwise the exponentially small corrections due to deformation of the inter-hadron forces become large.  
Assuming the range of the interaction is of similar size to the effective range (as expected for "natural" interactions), this requirement is not met and deviations from the assumed linear fitting function should be entertained. 
Higher precision analyses will be required to investigate this further.

%%%%%%%%%%%%%%%%%%%%%%%%%%%%%%%%%%%%%%%%%%%%%%%%
\section{The $\si$ Channel and the Dineutron}
\label{sec:1s0}

The  analysis of LQCD calculations in the $\si$ channel are somewhat simpler than in the 
$\siii$-$\diii$ coupled channels as scattering below the inelastic threshold is described by a single phase shift, 
$\delta^{(\si)}$.  
In FV, the relation between energy eigenvalues of the system at rest in 
the $\mathbb{A}_1$ cubic irrep
and $\delta^{(\si)}$ are  
given by eq.~(\ref{eq:Tone}) with $\delta_{1\alpha}\rightarrow \delta^{(\si)}$
and $k^*_{\mathbb{T}_1}\rightarrow k^*_{\mathbb{A}_1}$.
Unfortunately, 
the  correlation functions  in this channel 
have larger fluctuations and excited state contamination
 than those in the 
$\siii$-$\diii$ coupled channels system.
Consequently, the uncertainties associated with each energy level are larger.

%%%%%%%%%%%%%%%%%%%%%%%%%%%%%%%%%%%%%%%%%%%%%%%%
\subsection{The  dineutron}
\label{sec:dineutron}

\begin{figure}[!ht]
  \centering
  \includegraphics[width=0.32 \columnwidth]{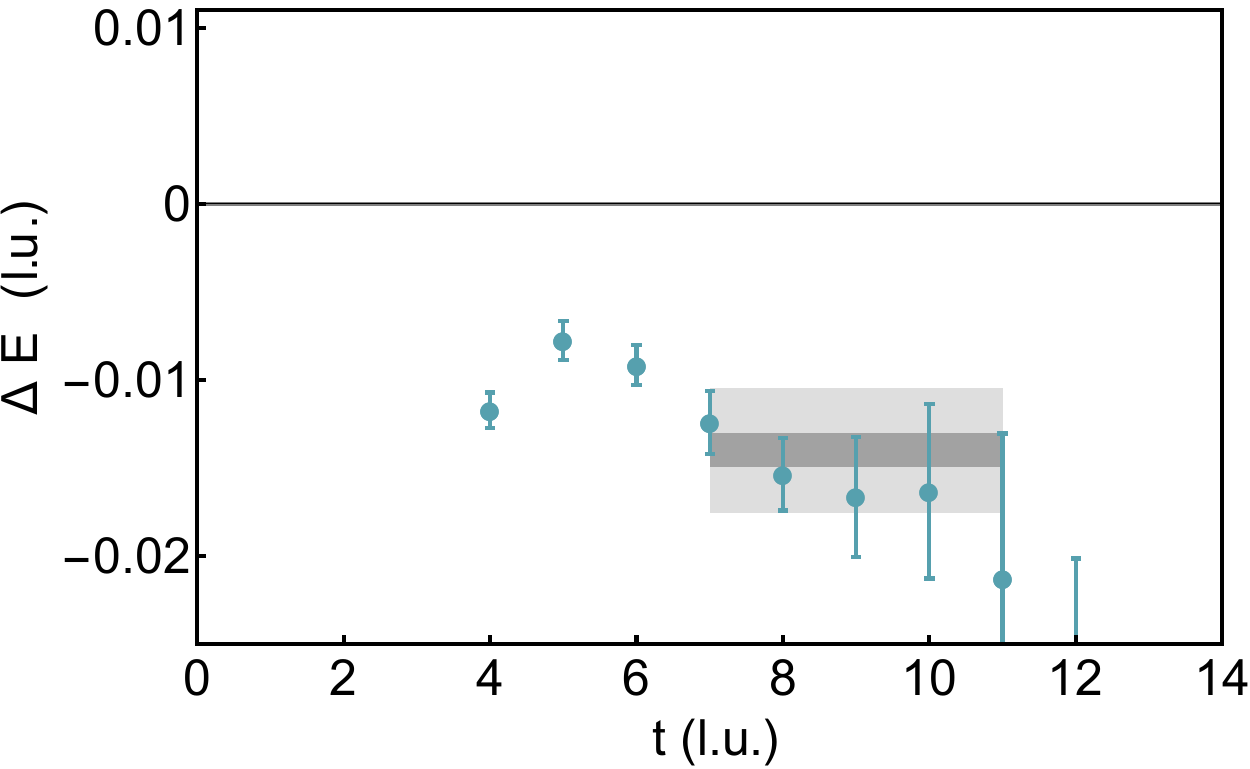}  
   \includegraphics[width=0.32 \columnwidth]{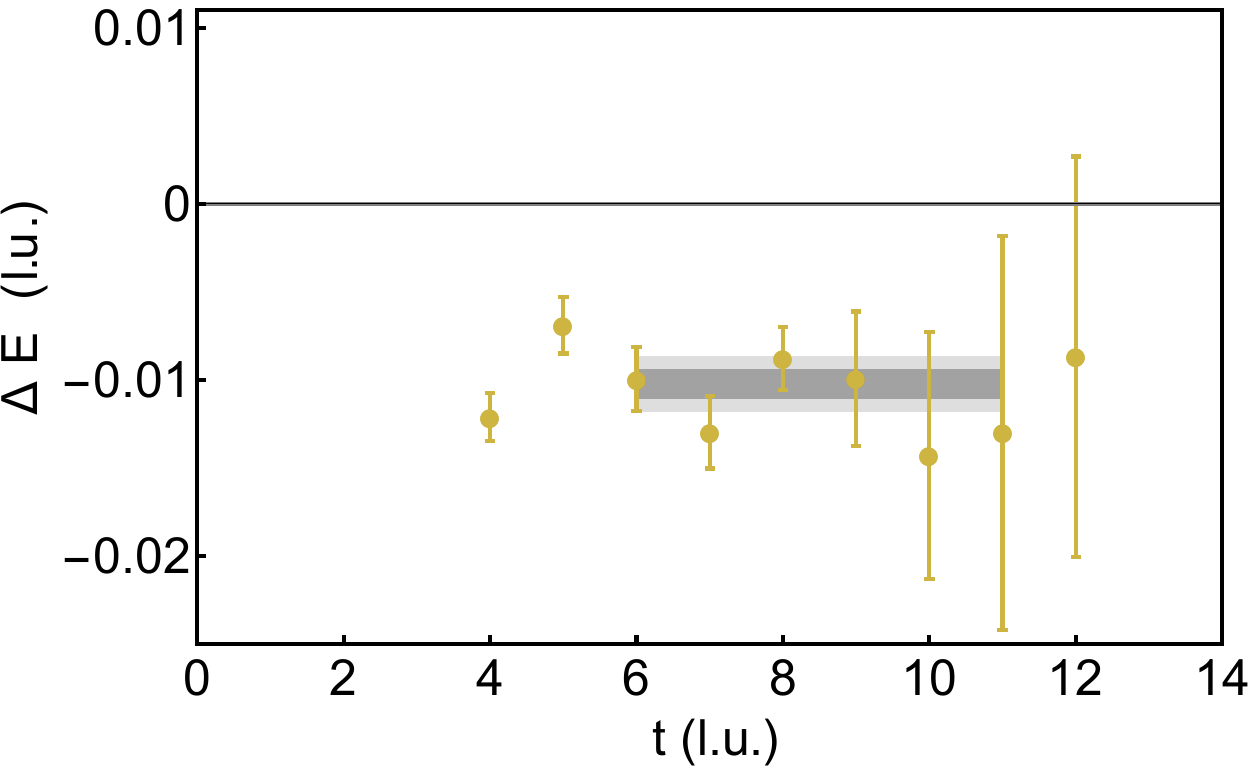}    
   \includegraphics[width=0.32 \columnwidth]{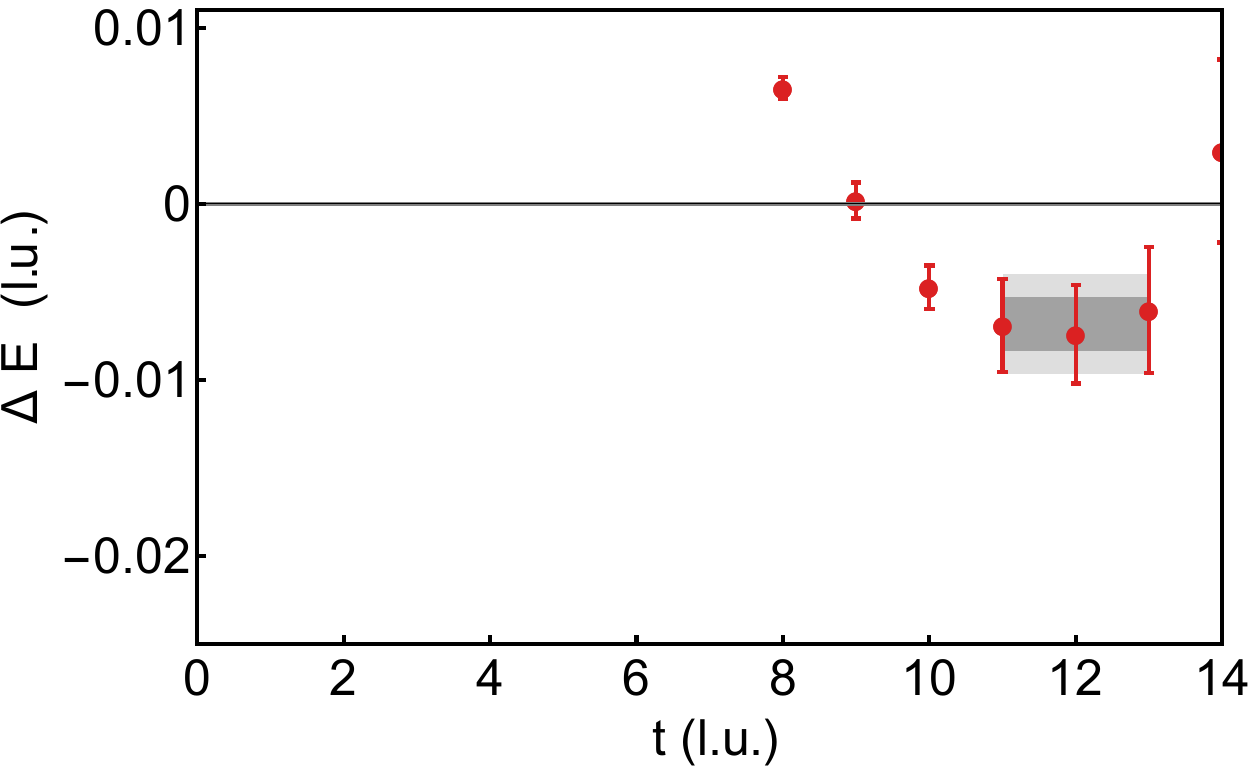}    
  \caption{
EMPs for the dineutron in 
the $L=24$ (left), $L=32$ (center)  and $L=48$ (right) ensembles,
along with fits to the plateau regions.
The extracted binding energies are given in Table~\protect\ref{tab:ppbinding}.
  }
  \label{fig:ppEMPs}
\end{figure}

\textcolor{\revrevrevcolor}{
Unlike in nature, the dineutron is found to be bound at heavier quark 
masses~\cite{Beane:2012vq,Yamazaki:2012hi,Beane:2013br,Yamazaki:2013rna,Yamazaki:2015asa}
by direct calculations of the ground-state energies of two nucleons in  finite lattice volumes.~\footnote{
The HAL QCD  method  appears not to give rise to a bound deuteron or dineutron at these heavier 
pion masses, e.g. Ref.~\cite{Aoki:2012tk}.
}
}
Plateaus identified with a negatively shifted  dineutron were found in all three ensembles,
with the associated EMPs  shown in Fig.~\ref{fig:ppEMPs} and the 
extracted energy shifts
shown in Table~\ref{tab:ppbinding}.
\renewcommand{\arraystretch}{1.5} 
\begin{table}
\begin{center}
\begin{minipage}[!ht]{16.5 cm}
  \caption{
The dineutron binding energies from fitting to the EMPs shown in Fig.~\protect\ref{fig:ppEMPs}.
  }  
\label{tab:ppbinding}
\end{minipage}
\setlength{\tabcolsep}{0.3em}
\begin{tabular}{c|cc|c}
\hline
      Ensemble & 
      $ \Delta E $\ (l.u.) & 
      $B_{nn} $\ (MeV)  &
      $e^{-\kappa L}$
      \\
\hline
\cfga &      -0.0142(09)(27) & 24.1(1.5)(4.5) & 0.088\\
\cfgb &    -0.0109(09)(20) & 18.4(1.5)(3.3) & 0.058\\
\cfgc &      -0.0070(11)(18) & 11.8(1.9)(3.1) & 0.033\\
\hline 
$L=\infty$
& $-0.0074^{+(10)(15)(01)}_{-(11)(27)(01)}$ 
& $\Bnn$ \\
\hline
\end{tabular}
\begin{minipage}[t]{16.5 cm}
\vskip 0.0cm
\noindent
\end{minipage}
\end{center}
\end{table}     
Performing a volume extrapolation  using the form given in eq.~(\ref{eq:Bvol})
leads to a binding energy of~\footnote{
Extrapolating with a form consistent with a scattering state, which would display a volume dependence of $\Delta E\sim 1/L^3$,
results in a poor goodness-of-fit.
}
\begin{eqnarray}
B_{nn}^{(\infty)} & = & \Bnn~{\rm MeV}
\ \ \ .
\label{eq:Bnn}
\end{eqnarray}
Combining the errors in eq.~(\ref{eq:Bnn}) 
in quadrature leads to 
$B_{nn}^{(\infty)} = \BnnSUMMARY~{\rm MeV}$.
\begin{figure}[!ht]
  \centering
  \includegraphics[width=0.48 \columnwidth]{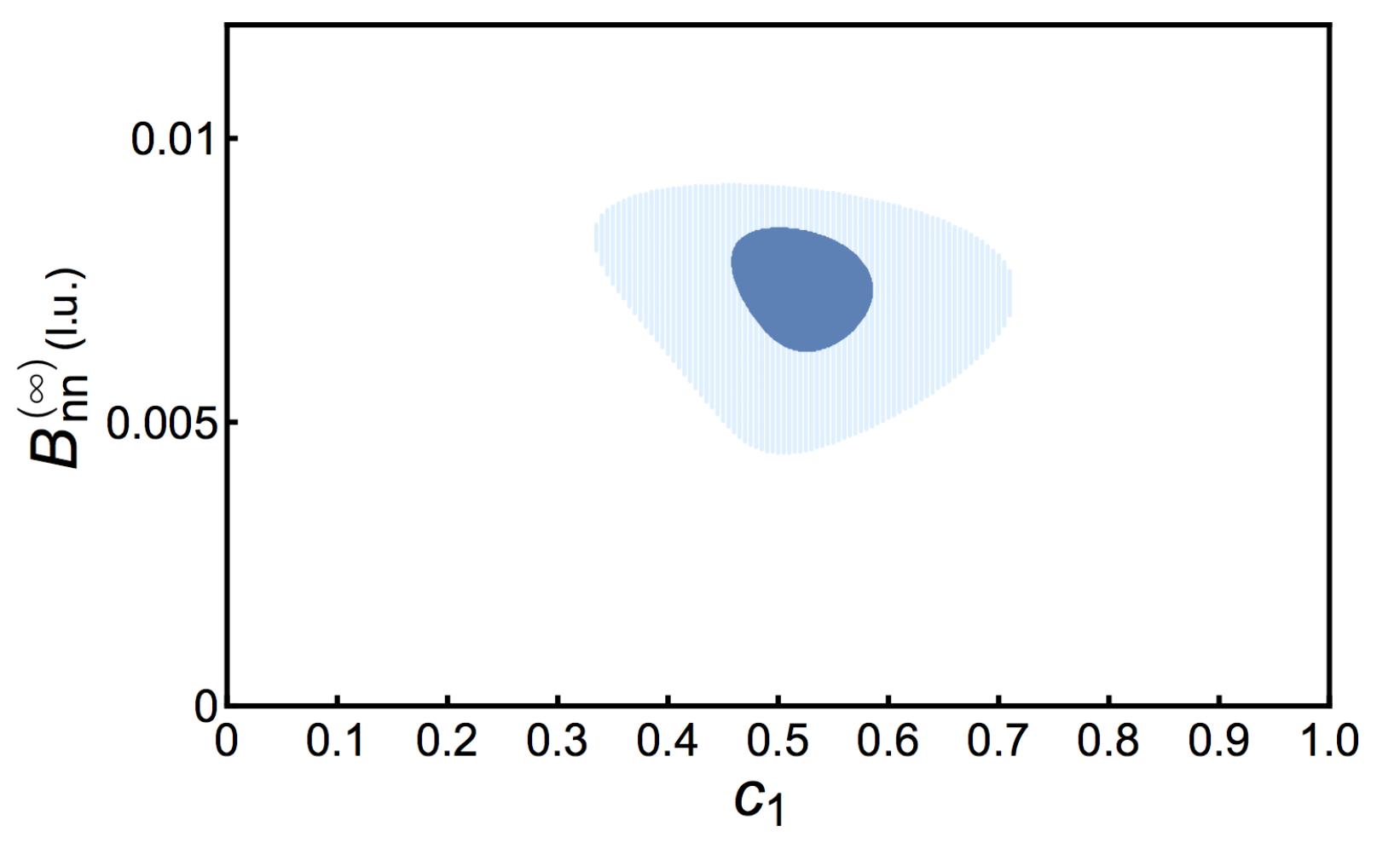}    
  \caption{
 The region in $c_1$-$B_{nn}^{(\infty)}$ parameter space 
 defined by $\chi^2\rightarrow\chi^2_{\rm min}+1$.
 The inner region is defined by the statistical uncertainty, while the outer is defined by the statistical and 
 systematic uncertainties combined in quadrature.
  }
  \label{fig:ppBc}
\end{figure}
The $c_1$-$B_{nn}^{(\infty)}$ parameter space defined by $\chi^2\rightarrow\chi_{\rm min}^2+1$
determined from an uncorrelated fit to the dineutron binding energies in the three volumes
is shown in Fig.~\ref{fig:ppBc}.
This dineutron binding energy
is consistent with the binding energy of the deuteron within uncertainties.
The EMPs associated with the difference between the deuteron and dineutron energies  in each ensemble 
are shown in Fig.~\ref{fig:ppMINUSdeutEMPs},
\begin{figure}[!ht]
  \centering
  \includegraphics[width=0.32 \columnwidth]{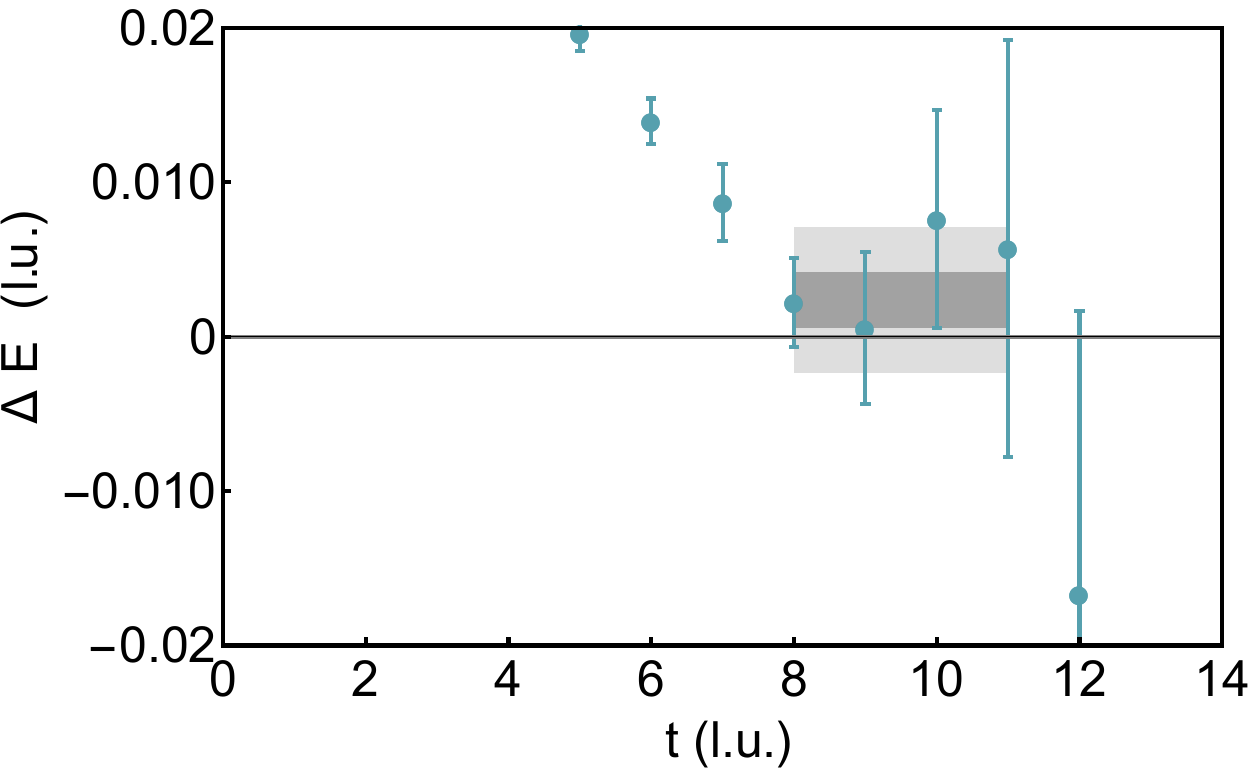}   \includegraphics[width=0.32 \columnwidth]{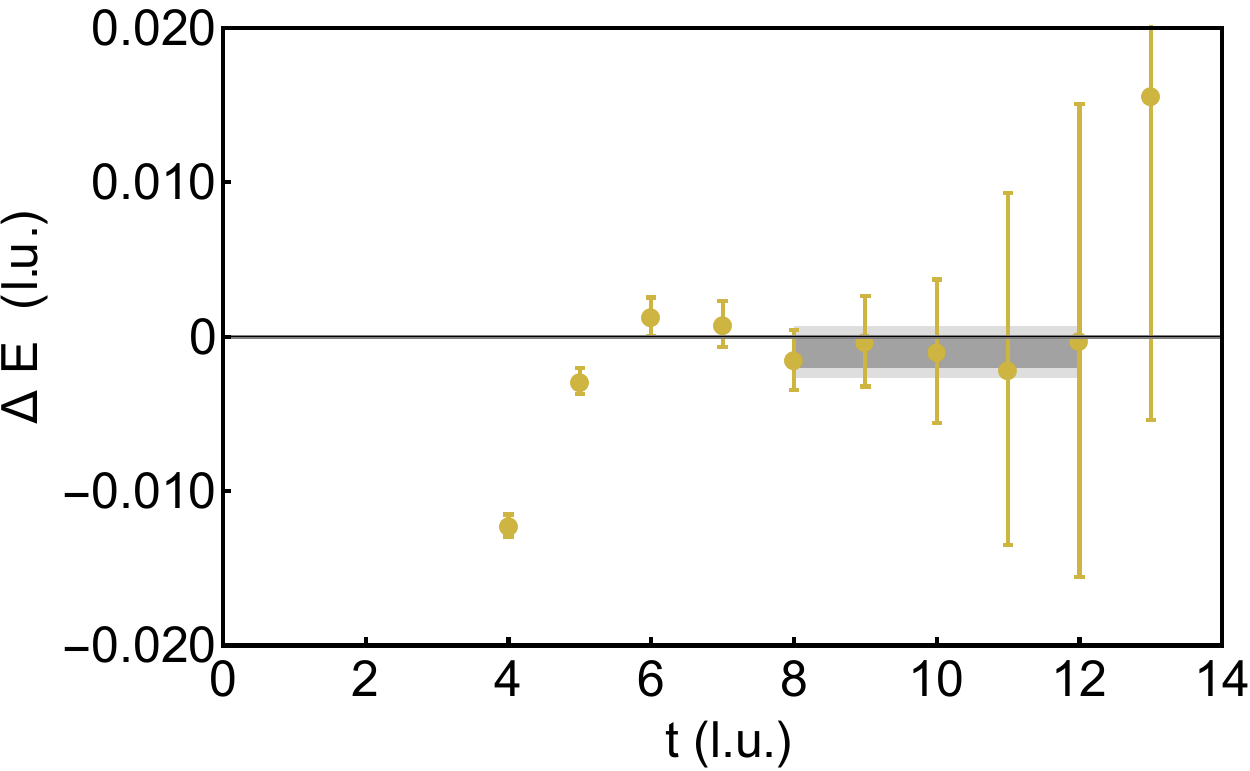}    \includegraphics[width=0.32 \columnwidth]{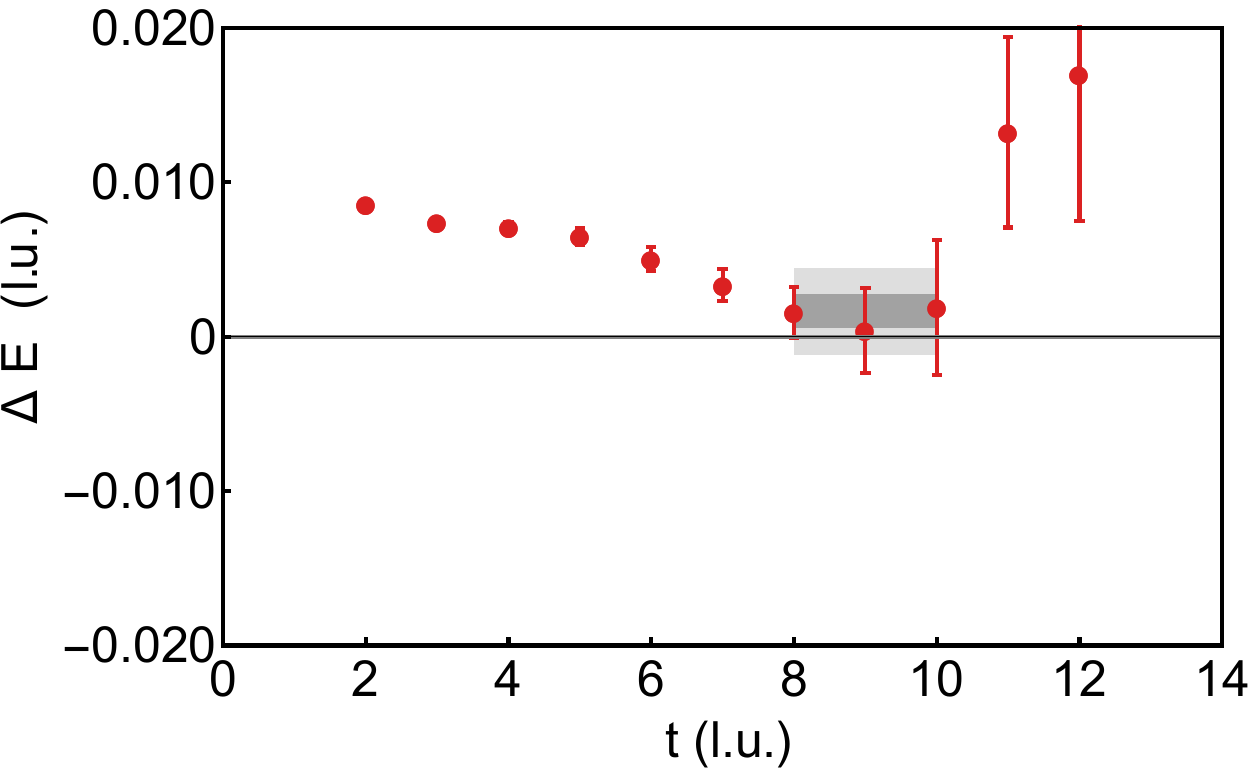}    
  \caption{
EMPs for the energy difference between the dineutron and the deuteron in 
the $L=24$ (left), $L=32$ (center) and   $L=48$ (right) ensembles,
along with fits to the plateau regions.
  }
  \label{fig:ppMINUSdeutEMPs}
\end{figure}
resulting in the energy differences  given in Table~\ref{tab:ppMINUSdeut}.
No significant difference has been extracted.
\renewcommand{\arraystretch}{1.5}% Wider
\begin{table}
\begin{center}
\begin{minipage}[!ht]{16.5 cm}
  \caption{
Energy differences between the dineutron and deuteron 
from fitting to the EMPs shown in Fig.~\protect\ref{fig:ppMINUSdeutEMPs}.
All differences are consistent with zero, as is their infinite-volume extrapolation.
  }  
\label{tab:ppMINUSdeut}
\end{minipage}
\setlength{\tabcolsep}{0.3em}
\begin{tabular}{c|cc}
\hline
      Ensemble & 
      $ E_{\rm nn} - E_{\rm deut} $\ (l.u.) & 
     $ E_{\rm nn} - E_{\rm deut} $\ (MeV) 
      \\
\hline
\cfga &      +0.0022(16)(28) &  +3.7(2.8)(4.7)(0.0)  \\
\cfgb &    -0.0014(09)(15) &  -2.4(1.6)(2.5) \\
\cfgc &       +0.0027(04)(31)& +4.6(0.7)(5.3)\\
\hline 
\hline
\end{tabular}
\begin{minipage}[t]{16.5 cm}
\vskip 0.0cm
\noindent
\end{minipage}
\end{center}
\end{table}     
%

%%%%%%%%%%%%%%%%%%%%%
\subsubsection{A Compilation of Dineutron Binding Energies from LQCD}
\label{subsec:dineuttBindLQCD}

\begin{figure}[!ht]
  \centering
  \includegraphics[width=0.48 \columnwidth]{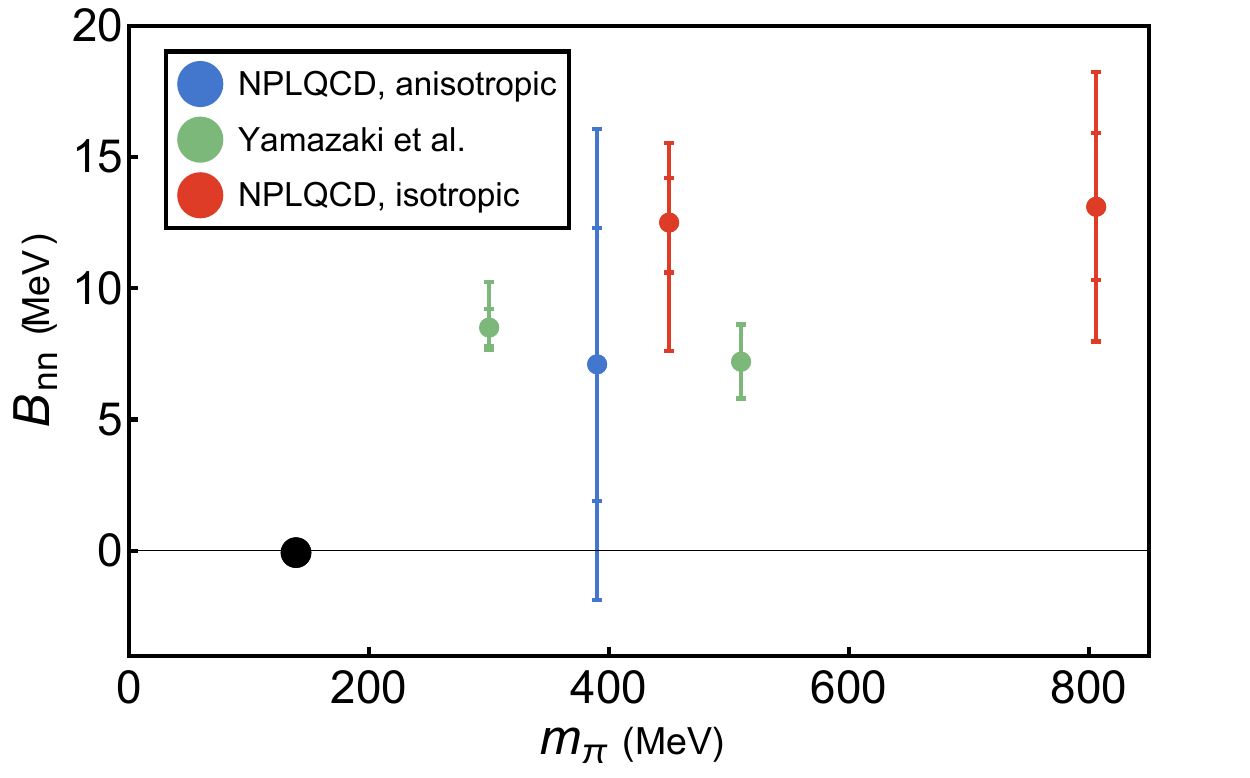}\ \
  \caption{
  The pion-mass dependence of the dineutron binding energy calculated with LQCD.
  The NPLQCD anisotropic-clover  
  result is from Ref.~\protect\cite{Beane:2011iw}, the Yamazaki {\it et al}. results are from Refs.~\protect\cite{Yamazaki:2012hi,Yamazaki:2015asa}
  and the NPLQCD isotropic-clover results are from this work and  Ref.~\protect\cite{Beane:2012vq}.
    The black disk corresponds to the location of the near-bound state at the physical quark masses.
    }
  \label{fig:nnALLlqcd}
\end{figure}

The current calculation of the dineutron binding energy adds to a small number of previous calculations,
a  compilation of which is shown in Fig.~\ref{fig:nnALLlqcd}.
There does not appear to be a clear pattern 
emerging as to how  the dineutron will unbind as the pion mass is reduced.
The results that have been obtained in  Refs.~\protect\cite{Yamazaki:2012hi,Yamazaki:2015asa}
have consistently smaller uncertainties than those found in 
Ref.~\protect\cite{Beane:2011iw,Beane:2012vq} and in the present work.
However, the results are consistent within the uncertainties.

%%%%%%%%%%%%%%%%%%%%%
\subsection{Scattering in the $\si$  Channel}
\label{subsec:singletscattering}

\begin{figure}[!ht]
  \centering
  \includegraphics[width=0.32 \columnwidth]{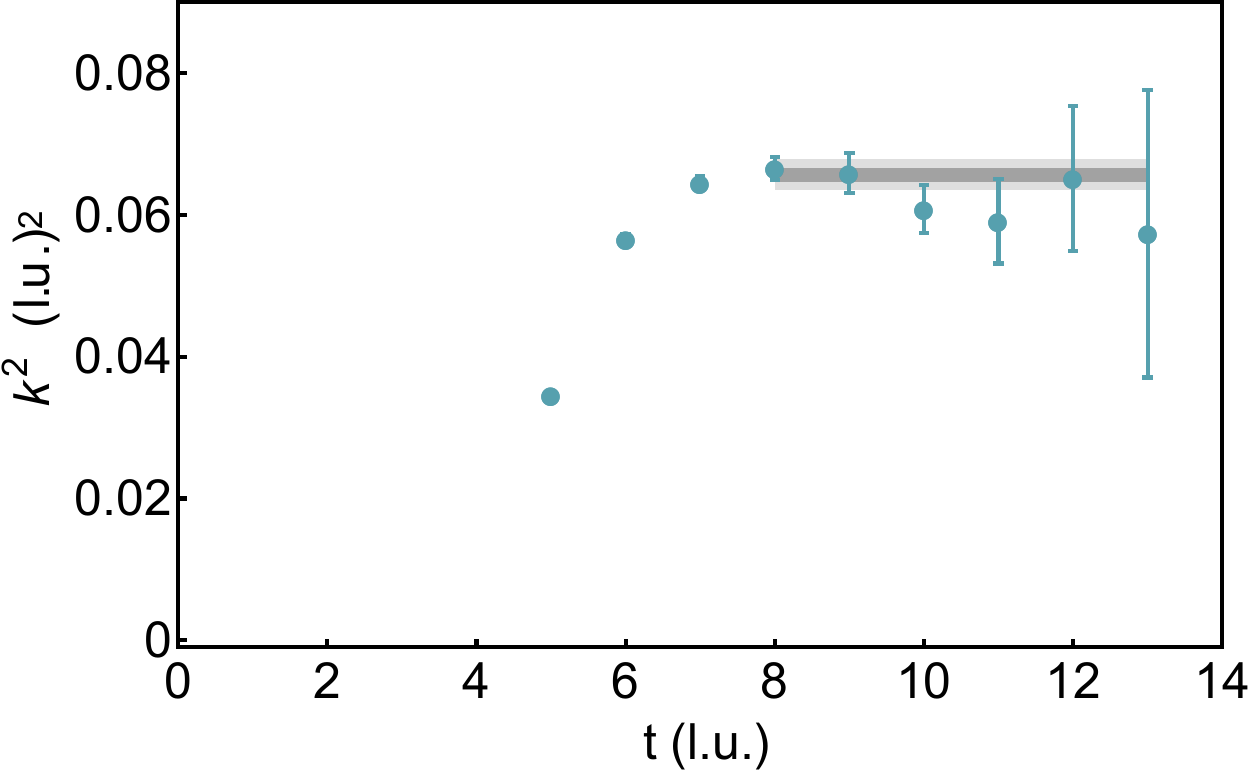}   
  \includegraphics[width=0.32 \columnwidth]{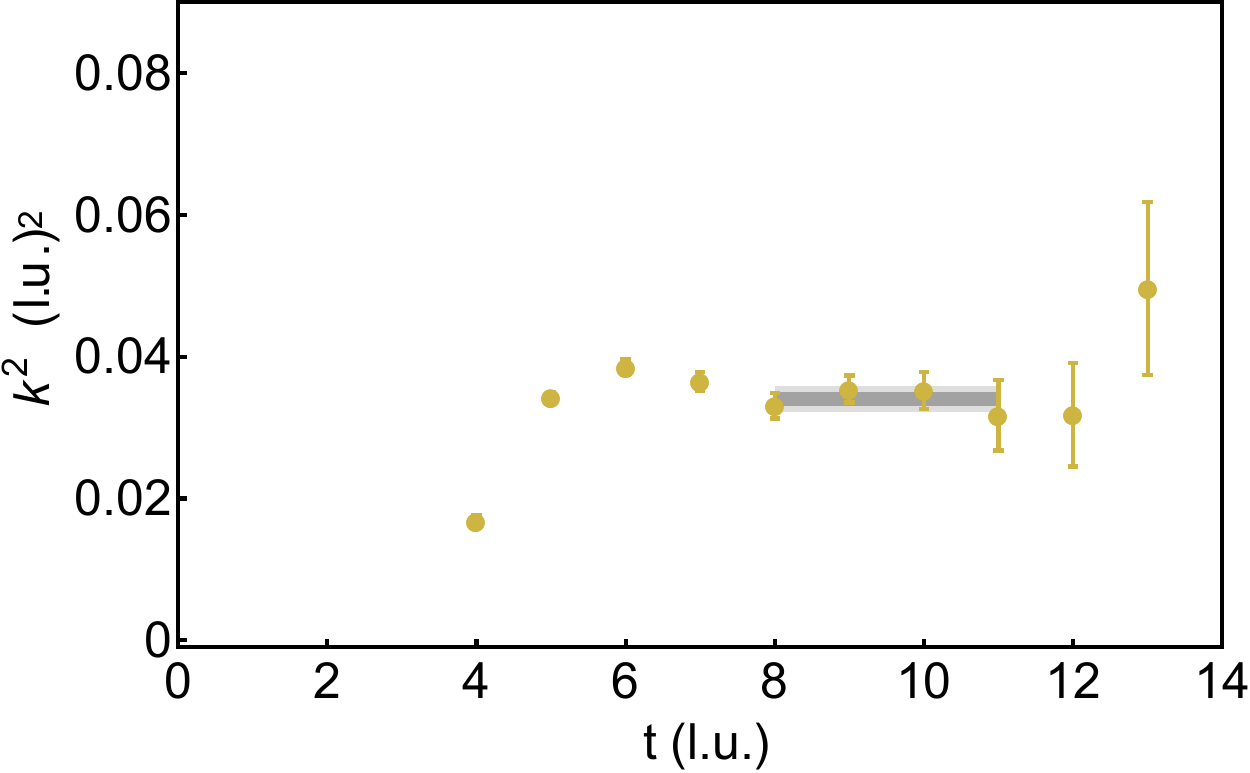}     
  \includegraphics[width=0.32 \columnwidth]{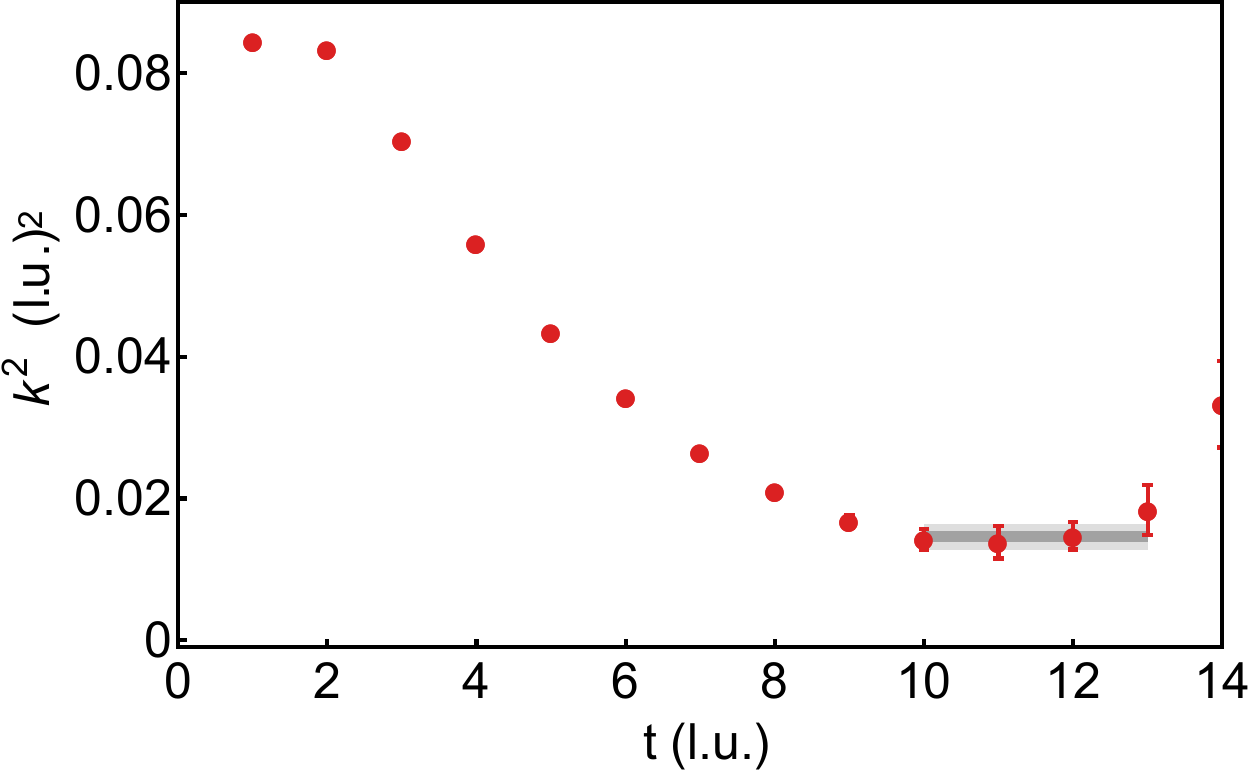}    
  \caption{
Ek2Ps for the lowest lying continuum
$\si$ NN state near 
\textcolor{\revrevcolor}{
$k^*=2\pi/L$
}
 in the $L=24$ (left), $L=32$ (center) and $L=48$ (right) ensembles,
along with fits to the plateau regions.
  }
  \label{fig:1s0neq1EMPs}
\end{figure}

\begin{figure}[!ht]
  \centering
 \includegraphics[width=0.32 \columnwidth]{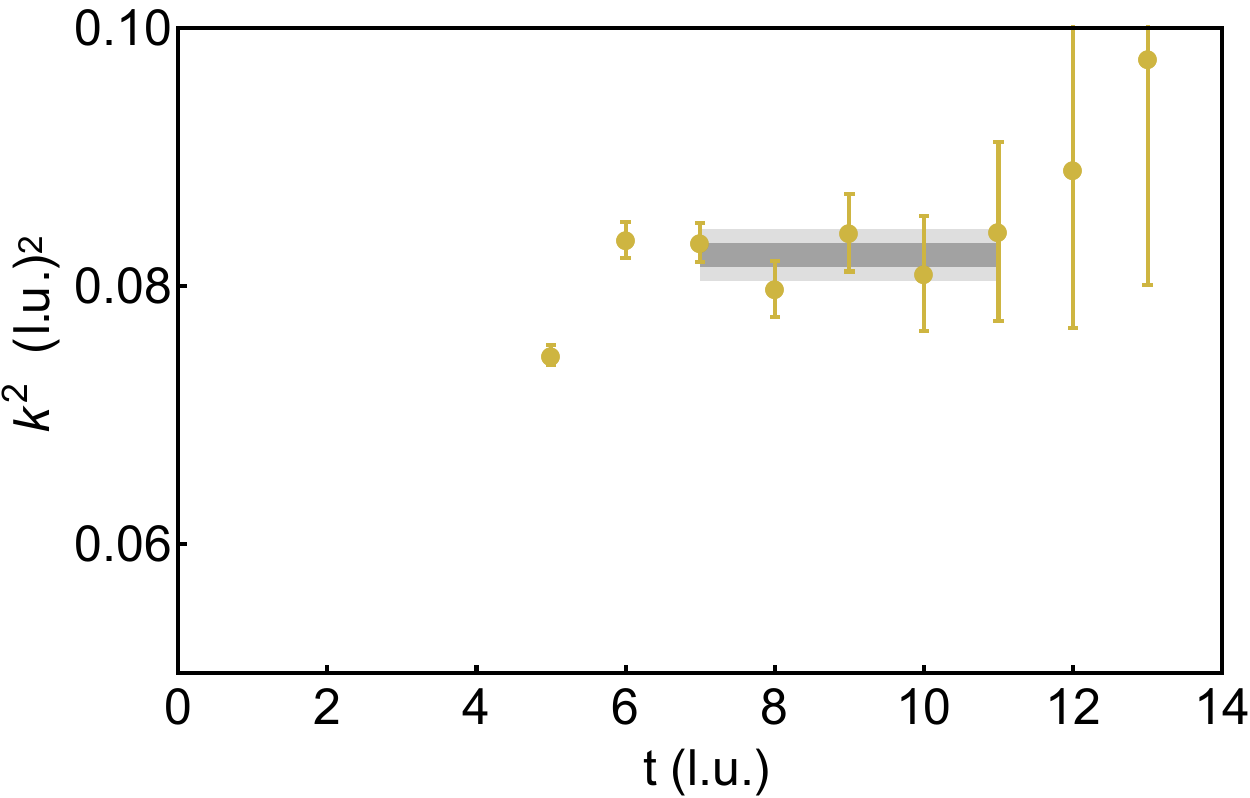}      
 \caption{
Ek2P for the lowest lying continuum
$\si$ NN state near 
\textcolor{\revrevcolor}{
$k^*=2\sqrt{2}\pi/L$
}
 in the $L=32$ ensemble,
along with the fit to the plateau region.
  }
  \label{fig:1s0neq2EMPs}
\end{figure}

Correlation functions for two nucleons in the $\si$ state 
were constructed in the $\mathbb{A}_1$ irrep of the cubic group.
The Ek2Ps associated with the states near the 
\textcolor{\revrevcolor}{
$k^*=2\pi/L$ and  $k^*=2\sqrt{2}\pi/L$ 
}
noninteracting levels are shown in 
Fig.~\ref{fig:1s0neq1EMPs} and Fig.~\ref{fig:1s0neq2EMPs}, respectively.
\begin{figure}[!ht]
  \centering
 \includegraphics[width=0.32 \columnwidth]{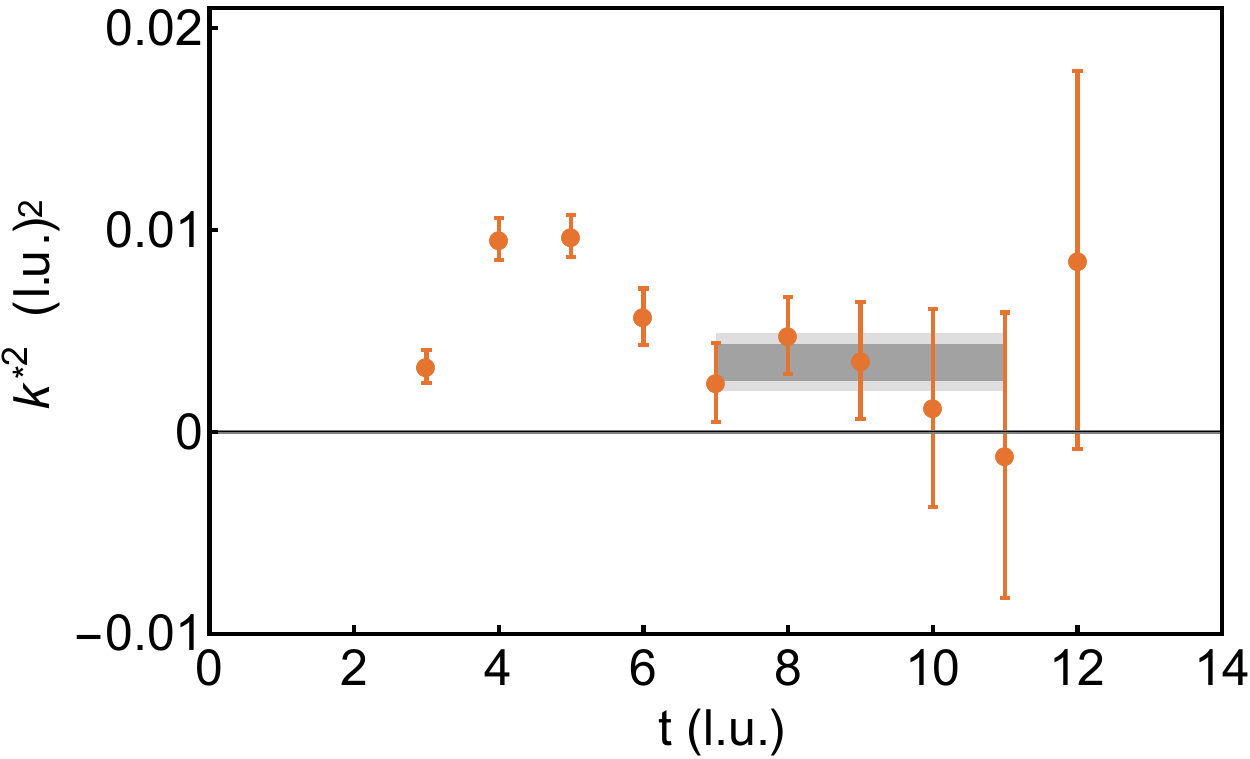}      
  \caption{
Ek2P  for the continuum $\si$ NN states 
with ${\bf d}=(0,0,1)$ near 
\textcolor{\revrevcolor}{
$k^*=0$
}
 in the $L=32$ ensemble,
along with the fit to the plateau region.
  }
  \label{fig:1s0Peq1EMPs}
\end{figure}
For the lowest-lying ``continuum'' state, plateaus were found in all three ensembles, however, 
only the 
\textcolor{\revrevcolor}{
$L=32$
}
ensemble has correlation functions that were sufficiently clean to extract  the next higher level.
A plateau was also identified in the system with one unit of total momentum, as shown in  Fig.~\ref{fig:1s0Peq1EMPs}.
The values of  $k\cot\delta^{(\si)}$ and the phase shift are given in Table~\ref{tab:1s0quantities} and shown in 
Fig.~\ref{fig:NN1s0kcotdelta}.
\renewcommand{\arraystretch}{1.5}% Wider
\begin{table}
\begin{center}
\begin{minipage}[!ht]{16.5 cm}
  \caption{
Scattering information in the $\si$  channel.
The uncertainties are highly correlated, as can be seen from 
Fig.~\protect\ref{fig:NN1s0kcotdelta}.
}  
\label{tab:1s0quantities}
\end{minipage}
\setlength{\tabcolsep}{0.3em}
\begin{tabular}{c|c|ccc}
\hline
      Ensemble &
      $|{\bf P}_{\rm tot}|$ (l.u.) & 
      $ k^*/m_\pi $  & 
      $ k^* \cot\delta^{(\si)} / m_\pi$  &
      $\delta^{(\si)}$ (degrees) 
       \\
\hline
All & 0 & $i 0.274^{+(19)(26)}_{-(20)(44)}  $ & $-0.274^{+(19)(26)}_{-(20)(44)}  $& -  \\
\hline
\cfga &    0 & $0.954^{+(08)(18)}_{-(08)(19)}$ &  $5.0^{+(2.0)(10.0)}_{-(1.1)(1.8)}$ & $10.8^{+(3.0)(6.5)}_{-(3.0)(6.7)}$  \\
\hline
\cfgb &  0 & $0.691^{+(09)(16)}_{-(09)(16)}$ &    $1.7^{+(0.5)(1.1)}_{-(0.3)(0.5)} $ & $22.0^{+(4.2)(7.0)}_{-(4.2)(7.2)}$  \\
\cfgb &  0 & $1.079^{+(05)(10)}_{-(05)(10)}$ &    $-3.3^{+(0.4)(0.7)}_{-(0.6)(1.5)} $ & $-18.3(2.6)(5.2) $  \\
\cfgb &  1 & $0.220^{+(28)(32)}_{-(32)(42)}$ &    $0.13^{+(10)(14)}_{-(08)(08)} $ &   $60^{+(14)(20)}_{-(12)(14)} $ \\
\hline
\cfgc &  0 &  $0.453(11)(29)$ & $0.89^{+(39)(3.7)}_{-(23)(44)}$ & $27^{+(07)(18)}_{-(07)(20)}$    \\
\hline
\end{tabular}
%noalign{\smallskip\hrule}\cr}
\begin{minipage}[t]{16.5 cm}
\vskip 0.0cm
\noindent
\end{minipage}
\end{center}
\end{table}     
\begin{figure}[!ht]
  \centering
  \includegraphics[width=0.48 \columnwidth]{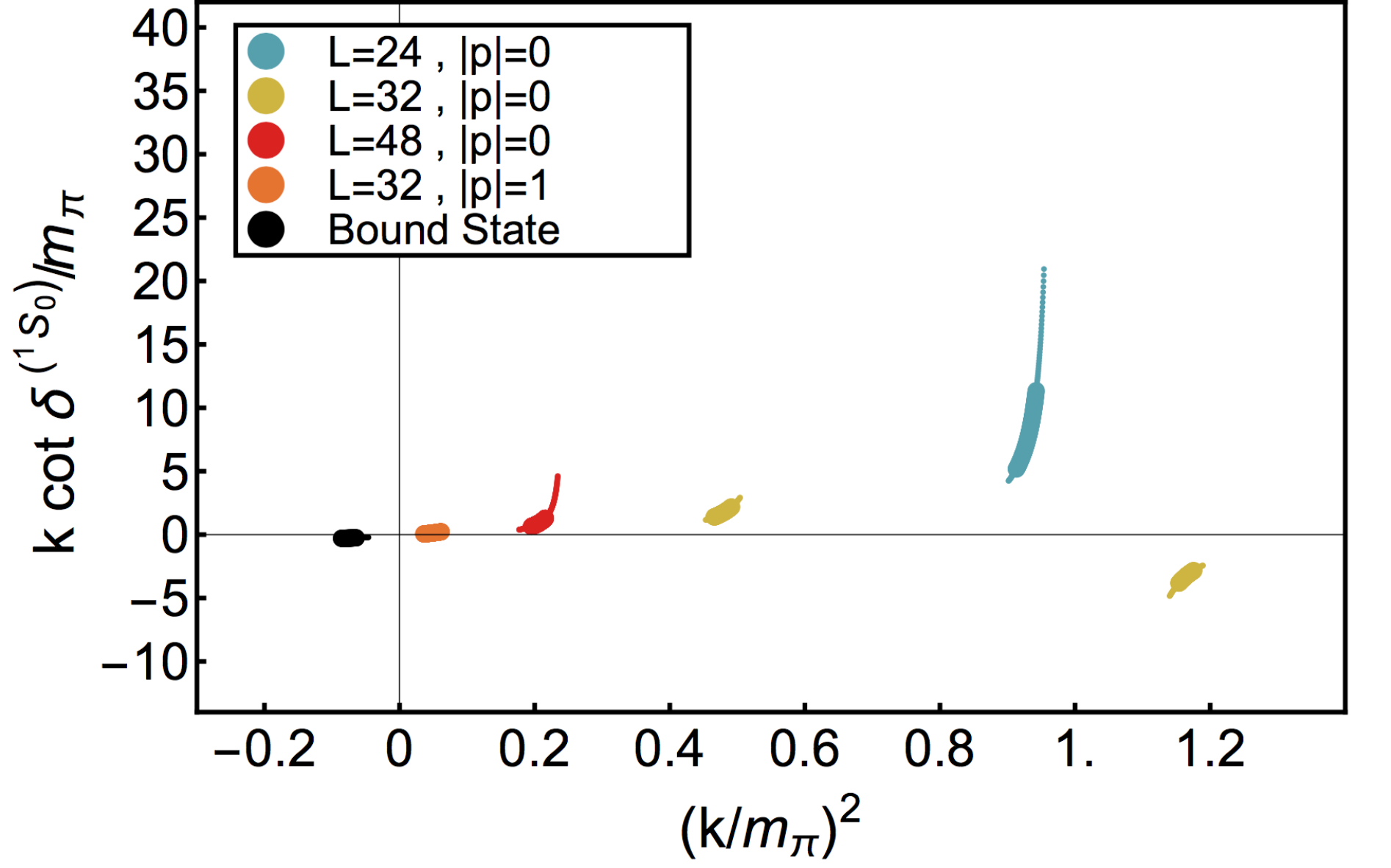}      
  \includegraphics[width=0.48 \columnwidth]{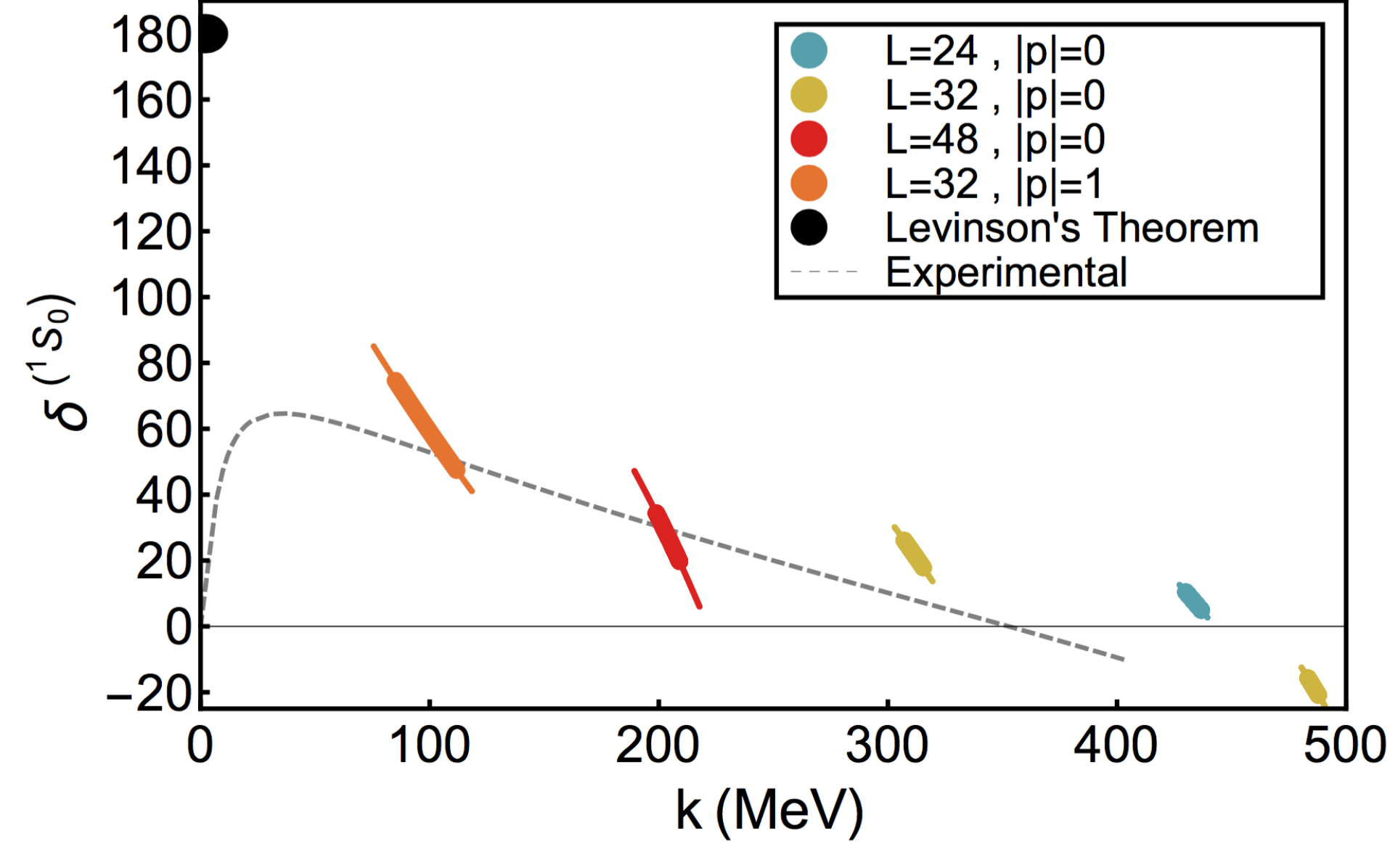}      
  \caption{
Scattering in the $\si$  channel.  
The left panel shows $k^*\cot\delta^{(\si)}/m_\pi$ is a function of $k^{*2}/m_\pi^2$,
while the right panel shows the phase shift as a function of momentum in MeV.  
The thick (thin) region of each result correspond to the statistical uncertainty (statistical and systematic uncertainties combined in quadrature).
The black circle in the right panel corresponds to the known bound-state
result from Levinson's theorem, while
the dashed-gray curve corresponds to the  phase shift extracted from the Nijmegen partial-wave analysis of experimental data~\cite{NijmegenPWA}.
  }
  \label{fig:NN1s0kcotdelta}
\end{figure}
Many of the qualitative features of the results for the scattering amplitude  
in this channel are similar to those in the 
$\siii$-$\diii$ coupled channels.  
A zero of the phase shift near $k\sim m_\pi\sim 450~{\rm MeV}$ is evident and occurs quite close to the zero of the phase shift in nature.
However for $k< 100~{\rm MeV}$, the $\si$ phase shift at $m_\pi\sim 450~{\rm MeV}$ and in nature become significantly different.
In Fig.~\ref{fig:NN1s0kcotdeltaER},  a linear fit is shown to the three results below the start of the t-channel cut, with 
the extracted correlated constraints on the scattering parameters also shown.
\begin{figure}[!ht]
  \centering
  \includegraphics[width=0.48 \columnwidth]{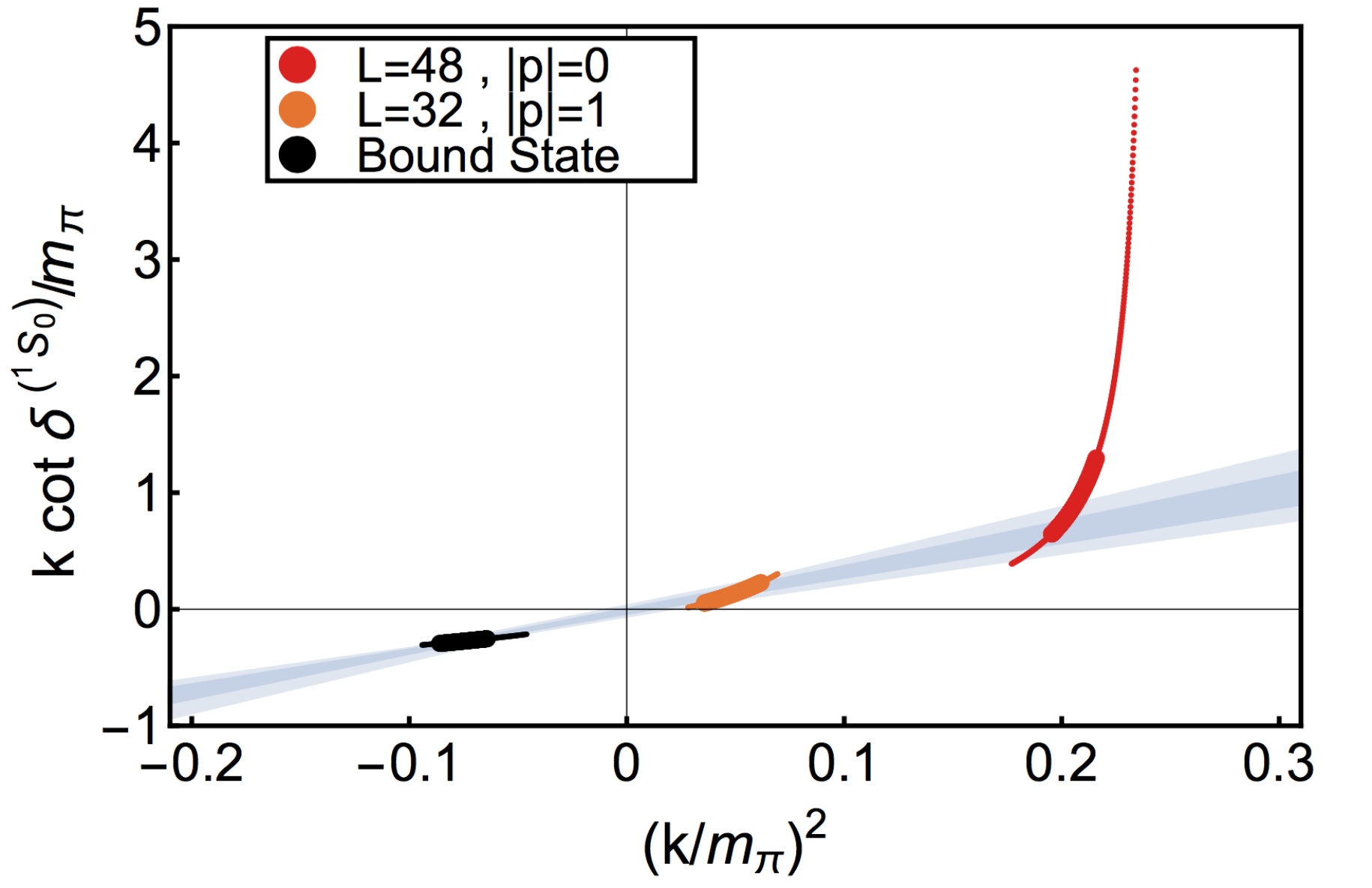}    
  \includegraphics[width=0.44 \columnwidth]{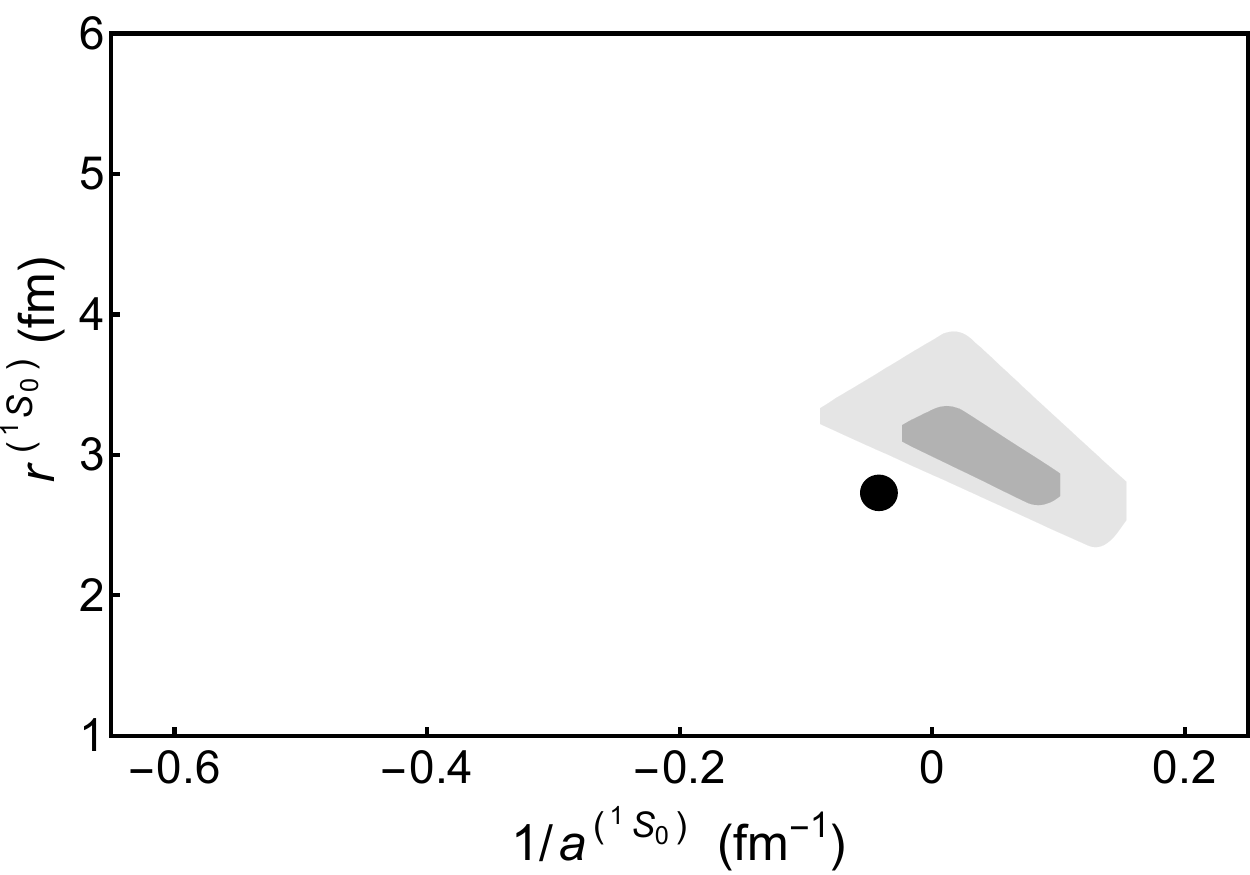}    
  \caption{
Scattering in the $\si$  channel below the start of the t-channel cut,
$k^{*2} < m_\pi^2/4$.  
The left panel 
shows the linear fit  with the darker and lighter shaded regions associated with the  
 statistical uncertainty
and the statistical and systematic uncertainties combined in quadrature.
The right panel shows the scattering parameters, $1/a^{(\si)}$ and $r^{(\si)}$ 
determined from fits to scattering results below the t-channel cut.
The solid circle corresponds to the experimental values.
}
  \label{fig:NN1s0kcotdeltaER}
\end{figure}
The inverse scattering length and effective range determined from the fit region in Fig.~\ref{fig:NN1s0kcotdeltaER}
are
\begin{eqnarray}
\left( m_\pi a^{(\si)} \right)^{-1} & = & 
0.021^{+(28)(32)}_{-(36)(63)}
\ \ ,\ \ 
m_\pi r^{(\si)} \ =\ 
6.7^{+(1.0)(2.0)}_{-(0.8)(1.3)}
\nonumber\\
\left(  a^{(\si)} \right)^{-1} & = & 
0.05^{+(06)(08)}_{-(08)(14)}~{\rm fm}^{-1}
\ \ ,\ \ 
 r^{(\si)} \ =\ 
2.96^{+(43)(87)}_{-(34)(55)}~{\rm fm}
\ \ \ .
\end{eqnarray}
The allowed region of scattering parameters is shown in 
Fig.~\ref{fig:NN1s0kcotdeltaER}
and is close to containing the experimentally determined 
scattering length and effective range.
Since the quark masses are unphysical, the physical values need not be contained in this region and it is interesting how close the current results are to those in nature.
As in the $\siii$-$\diii$ coupled-channels system analyzed in the previous section, 
there is a potential self-consistency issue raised by the region of the extracted effective range.

%%%%%%%%%%%%%%%%%%%%%
\section{  Observations about Nucleon-Nucleon Effective Field Theory  Analyses }
\label{sec:NNEFT}

A modern approach to low-energy nuclear physics rests upon the chiral nuclear forces
arising 
from a non-trivial extension of $\chi$PT into the multi-nucleon sector (see, for instance, Refs.~\cite{Epelbaum:2008ga,Machleidt:2011zz,Binder:2015mbz}). 
Because of the small scales in the two nucleon systems 
($\gamma_d\sim 45~{\rm MeV}$ and $|\gamma_{nn}|\sim8~{\rm MeV}$),  
the NNEFTs are more complicated than a simple expansion in 
quark masses and momenta that defines $\chi$PT, 
and there are additional dynamics that must be considered. 
Following the initial developments by Weinberg~\cite{Weinberg:1990rz,Weinberg:1991um,Ordonez:1992xp}, 
much effort has gone in to understanding the construction and 
behavior of these theories.

NNEFTs provide a powerful means with which to analyze the momentum and quark-mass dependences of the phase shifts and it
is illuminating to consider the LQCD results presented in this work in their context.
As is appropriate, we use
KSW power counting~\cite{Kaplan:1998tg,Kaplan:1998we,Kaplan:1998sz} 
in the $\si$ channel and BBSvK power counting~\cite{Beane:2001bc}, a variant of Weinberg's power 
counting~\cite{Weinberg:1990rz,Weinberg:1991um},  
in the $\siii$-$\diii$ coupled channels.
There are a number of reasons  to undertake this investigation.  
The chiral decomposition of nuclear forces automatically requires the introduction of terms that are only 
distinguishable through variation of the quark masses. 
Comparison of LQCD calculations at unphysical masses allows this previously unavailable ``dial'' to be turned in the 
dual expansion that defines chiral NNEFTs. 
Secondly, the full decomposition of the chiral NN forces, and thereby precise predictions for nuclear observables, 
requires knowledge
of the mass dependence discussed above and it is essential that such calculations be performed to maximize the 
predictive power of NNEFTs. 
Thirdly, the current calculations enable an exploration of the convergence of 
NNEFTs with  pions included as explicit degrees of freedom at relatively large pion masses.

The quality and kinematic coverage of scattering results that have been  presented is not yet sufficient 
to perform a comprehensive analysis of NNEFT matching to LQCD. 
Instead, we present a simplified discussion of the two channels to highlight some of the important 
features and questions that will need to be addressed in order to accomplish a reliable determination of the chiral nuclear forces from LQCD.
Related discussions in the context of pionless EFTs
 for multi-nucleon systems can be found in Ref.~\cite{Kirscher:2015ana,Kirscher:2015yda}
and implicitly in the presentation of the effective range expansion above.

%%%%%%%%%%%%%%%%%%%%%
\subsection{ KSW Analysis of the $\si$ Channel }
\label{subsec:KSW}

The KSW power counting~\cite{Kaplan:1998tg,Kaplan:1998we,Kaplan:1998sz}
 provides a rigorous framework with which to perturbatively expand the two-nucleon scattering amplitude in the $\si$ channel
in the two small-expansion parameters, nominally $p/\Lambda_{NN}$ and $m_\pi/\Lambda_{NN}$.
Here $\Lambda_{NN}=  8\pi f_\pi^2 /g_A^2 M_N$ is the natural scale of validity of the NNEFT.
 At the physical point  $\Lambda_{NN}\sim 289~{\rm MeV}$, while 
at a pion mass of \pionmass\   it is $\Lambda_{NN}\sim 350~{\rm MeV}$.
These scales
 should be compared with the start of the t-channel cut from the next lightest meson, $m_\rho/2\sim 385~{\rm MeV}$ 
 at the physical point, and  
$m_\rho/2\sim 443~{\rm MeV}$  at a pion mass of \pionmass.
This power counting treats the zero-derivative two-nucleon operator nonpertubatively, 
and was developed in order to correctly define a  theory that is finite and renormalization group invariant 
at each order in the expansion. 
An analysis of 
NN interactions at the physical point has been carried out to NNLO in the KSW expansion~\cite{Kaplan:1998tg,Kaplan:1998we,Mehen:1999hz,Fleming:1999ee,Beane:2008bt}, and we have performed the analogous analysis of the present LQCD results.
The LO, NLO and NNLO amplitudes in the $\si$ channel can be found in 
Refs.~\cite{Kaplan:1998tg,Kaplan:1998we,Mehen:1999hz,Fleming:1999ee}, along with the relevant expansion of the phase shift.
\begin{figure}[!ht]
  \centering
  \includegraphics[width=0.48 \columnwidth]{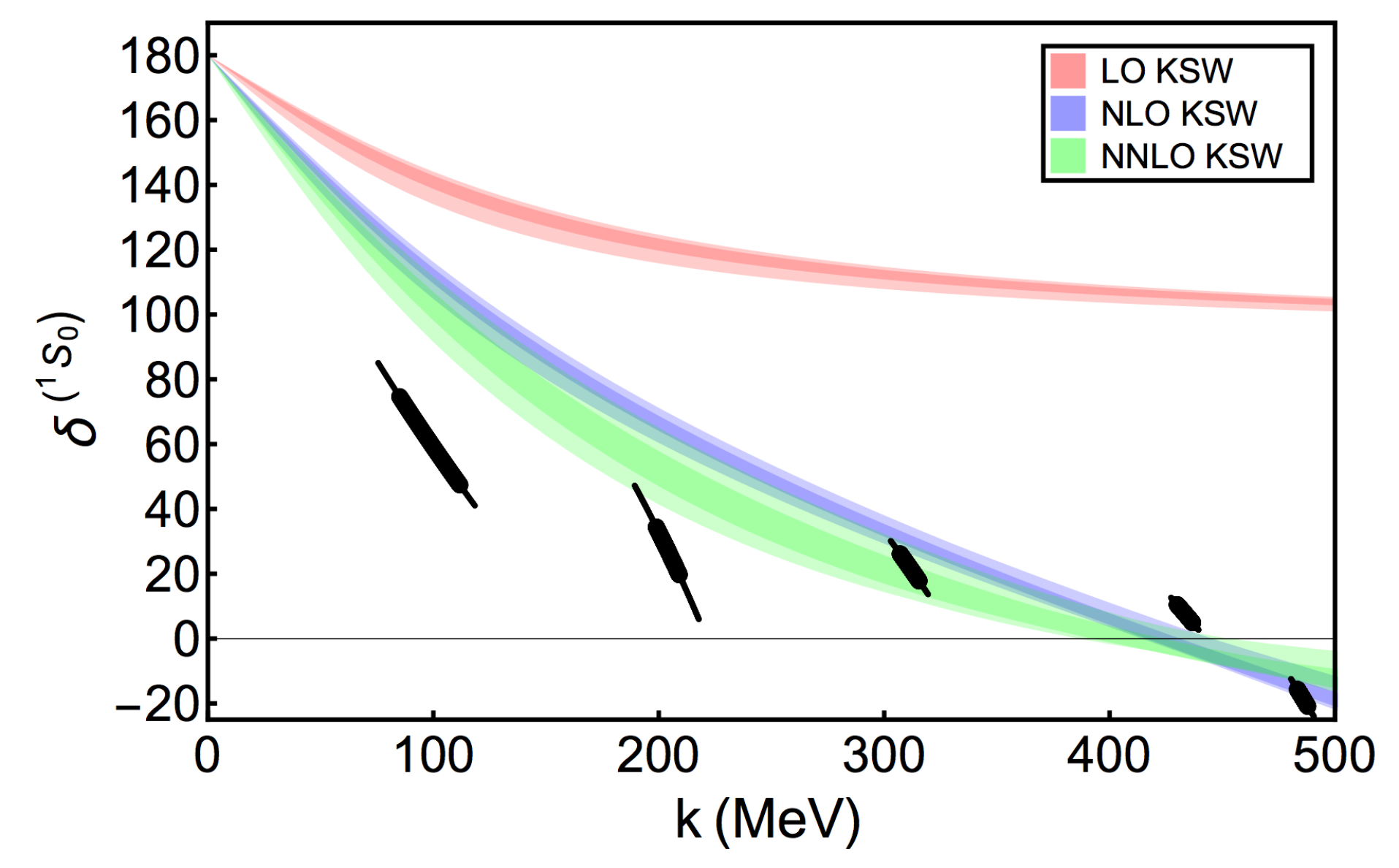}    
  \includegraphics[width=0.48 \columnwidth]{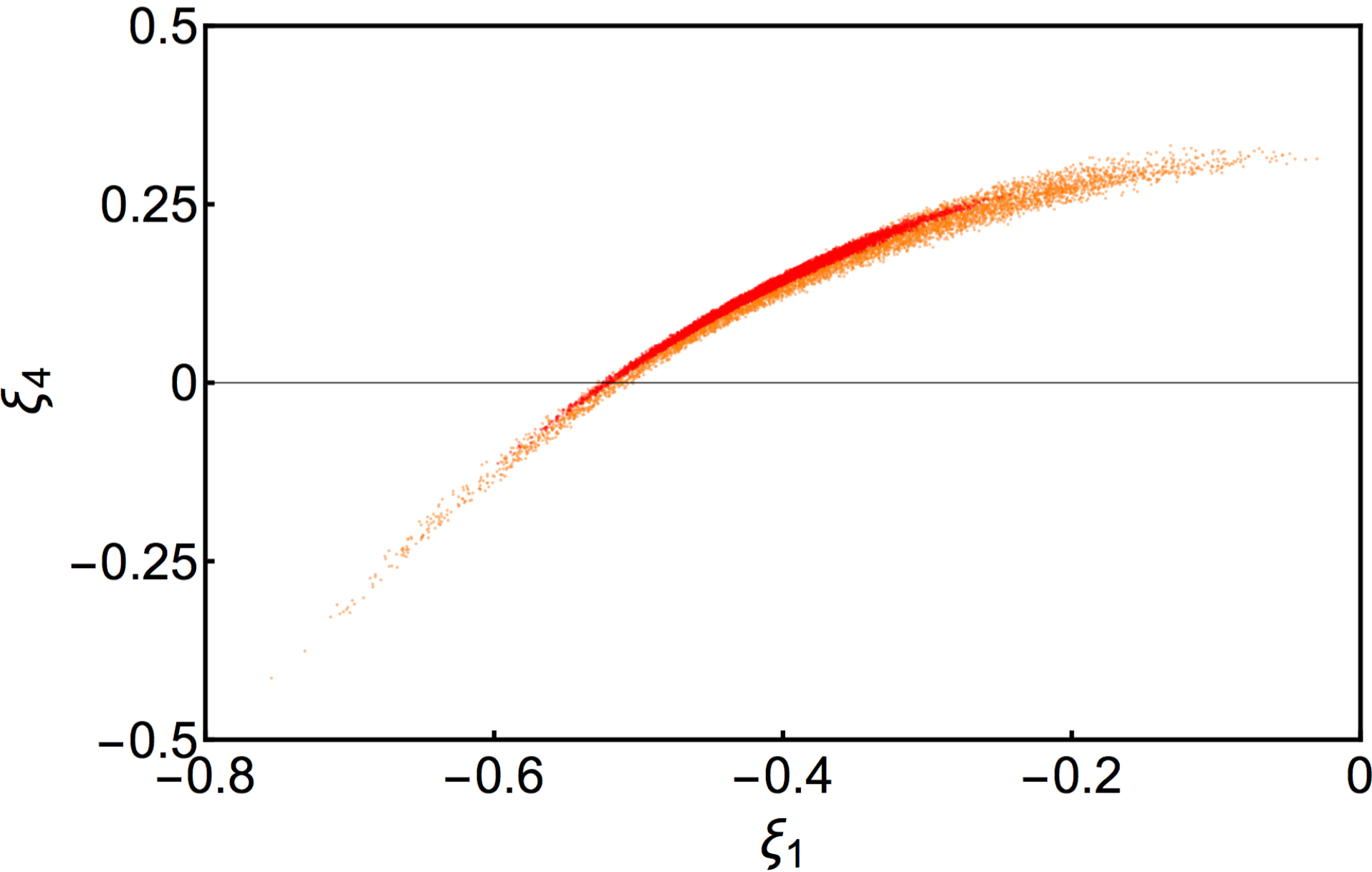}    
  \caption{
The left panel shows the LQCD $\si$ scattering phase shift along with  the KSW NNEFT fits at LO, NLO and NNLO.  
At LO there is one  parameter that is fit to recover the dineutron pole, giving the red-shaded region, 
at NLO there is one additional fit parameter, giving the blue-shaded region, 
and at NNLO there is a further fit parameter, giving the green-shaded region.
The darker (lighter) shaded regions correspond to the statistical (statistical and systematic uncertainties combined in quadrature).
The right panel is a scatter plot of the central values of the extracted NNLO 
fit parameters, $\xi_{1,4}$ over the 1-$\sigma$ range of the dineutron pole.
The red (orange) shaded regions correspond to the statistical (statistical and systematic uncertainties combined in quadrature).
}
  \label{fig:NN1s0KSW}
\end{figure}
At LO, there is only one fit parameter, constrained by the location of the dineutron pole.  
At NLO, there are nominally two additional fit parameters, but requiring the dineutron pole remains 
unchanged reduces the number to  one, $\xi_1$, while the other, $\xi_2$, can be directly related to $\xi_1$.
Finally, at NNLO there are three more parameters, 
but only one parameter, $\xi_4$, is  independent for similar reasons as at NLO.
Therefore, there are only three  fit parameters for a complete analysis at NNLO.
Results of fitting the LO, NLO and NNLO phase shifts are shown in Fig.~\ref{fig:NN1s0KSW}.
The phase shifts at all momenta are utilized in the fits (a more complete analysis would consider the effects of truncations).

Fitting  the location of the dineutron bound state,
the LO fit is clearly inconsistent with the phase shifts at higher energies, as is also seen in fits at the physical point.  
At NLO the fit is quite reasonable at the energies near the zero of the phase shift, but becomes somewhat deficient at  lower energies.  
The NNLO fit is found to move closer to the LQCD results. 
It appears that the KSW expansion is converging to the LQCD results, but fits beyond NNLO are required to reproduce the LQCD results with 
an acceptable  goodness-of-fit.
The values of  $\xi_{1,4}$, are both of natural size, as can be seen in Fig.~\ref{fig:NN1s0KSW}.

The resulting scattering parameters at NLO and NNLO are
\begin{eqnarray}
a^{(\si)}_{\rm NLO} & = & 2.62(07)(16)~{\rm fm}
\qquad
r^{(\si)}_{\rm NLO} \ =\ 1.320(18)(38))~{\rm fm}
\nonumber\\
a^{(\si)}_{\rm NNLO} & = & 2.99(07)(15)~{\rm fm}
\qquad
r^{(\si)}_{\rm NNLO} \ =\ 1.611(42)(83))~{\rm fm}
\ \ \ ,
\label{eq:KSWar}
\end{eqnarray}
From the differences between orders,  it is clear that the systematic uncertainty introduced by the KSW expansion exceeds the uncertainties of the 
LQCD calculations, and orders beyond NNLO are required  to render the ``theory error'' 
(from truncating the KSW expansion)
small compared with the 
uncertainties of the calculation.
As the KSW expansion is a double expansion in both momentum and the pion mass, the 
threshold scattering parameters have chiral expansions order-by-order in the  expansion.
The values of the scattering parameters extracted from fitting the KSW expressions differ from those obtained by fitting a  truncated ERE to 
the phase shifts at the lowest two  momenta and the dineutron pole, i.e. they do not lie in the region presented in Fig.~\ref{fig:NN1s0kcotdeltaER}.
This may indicate that the KSW expansion 
should not be applied to the phase shifts over the full range of momenta; 
indeed the largest two momenta have $ k\gsim \Lambda_{NN}$.
However, removing these points does not change the fit qualitatively due to the relative size of the  uncertainties.
These results could also indicate
that the pion mass is simply too large, as it  exceeds $\Lambda_{NN}$.  
 However, it does appear that the expansion is converging, albeit slowly, to the calculated phase shifts.

As the calculations have been performed at 
only one pion mass 
(previous phase shift  calculations at $m_\pi=806~{\rm MeV}$~\cite{Beane:2012vq,Beane:2013br} 
are expected to be beyond the range of applicability of the NNEFT), 
it is not possible to isolate the explicit short-distance pion-mass dependence 
in $\xi_{1,4}$, which both receive contributions from pion-mass independent  and pion-mass dependent terms.
Hence, a chiral extrapolation to the physical point is not feasible from this work alone.
Calculations that are currently underway will provide 
results at 
a lower pion mass, 
from which predictions at the physical point will become possible.

%%%%%%%%%%%%%%%%%%%%%
\subsection{ BBSvK Analysis of the $\siii$-$\diii$ Coupled Channel }
\label{subsec:BBSvK}

BBSvK power counting~\cite{Beane:2001bc} is similar to Weinberg's power counting~\cite{Weinberg:1990rz,Weinberg:1991um},
and is an appropriate scheme to use in the case of the $\siii$-$\diii$ coupled channels.
An NN interaction (two-particle irreducible) is derived using the familiar rules of $\chi$PT
and,  due to the infrared behavior of the two-nucleon system, is iterated to all orders with the Schr\"odinger equation 
to generate the bound-state pole(s) and scattering amplitude(s). 
See Ref.~\cite{Epelbaum:2014efa} for a review.

At LO in Weinberg's power counting, the NN interactions are determined by momentum-independent and quark-mass-independent two-nucleon 
contact interactions and by one-pion exchange (OPE).  
However, the short-distance nature of the tensor force, resulting from OPE, 
generates renormalization-scale dependence in the 
D-waves that  requires the presence of a counter term at LO, and
BBSvK is the simplest power counting to remedy this situation. 
At NLO in the counting, there are contributions from pion-loop diagrams, from momentum-dependent  two-nucleon 
contact interactions, and from insertions of the light-quark mass matrix into momentum-independent two-nucleon 
contact interactions.
With the parameters in the meson sector, e.g. $g_A$, $f_\pi$, fixed to the results of other LQCD calculations at similar quark masses,
there is one free parameter at LO in BBSvK counting - the coefficient of the momentum-independent two-nucleon contact interaction.
This is common to both the S-waves and D-waves.
\begin{figure}[!ht]
  \centering
  \includegraphics[width=0.48 \columnwidth]{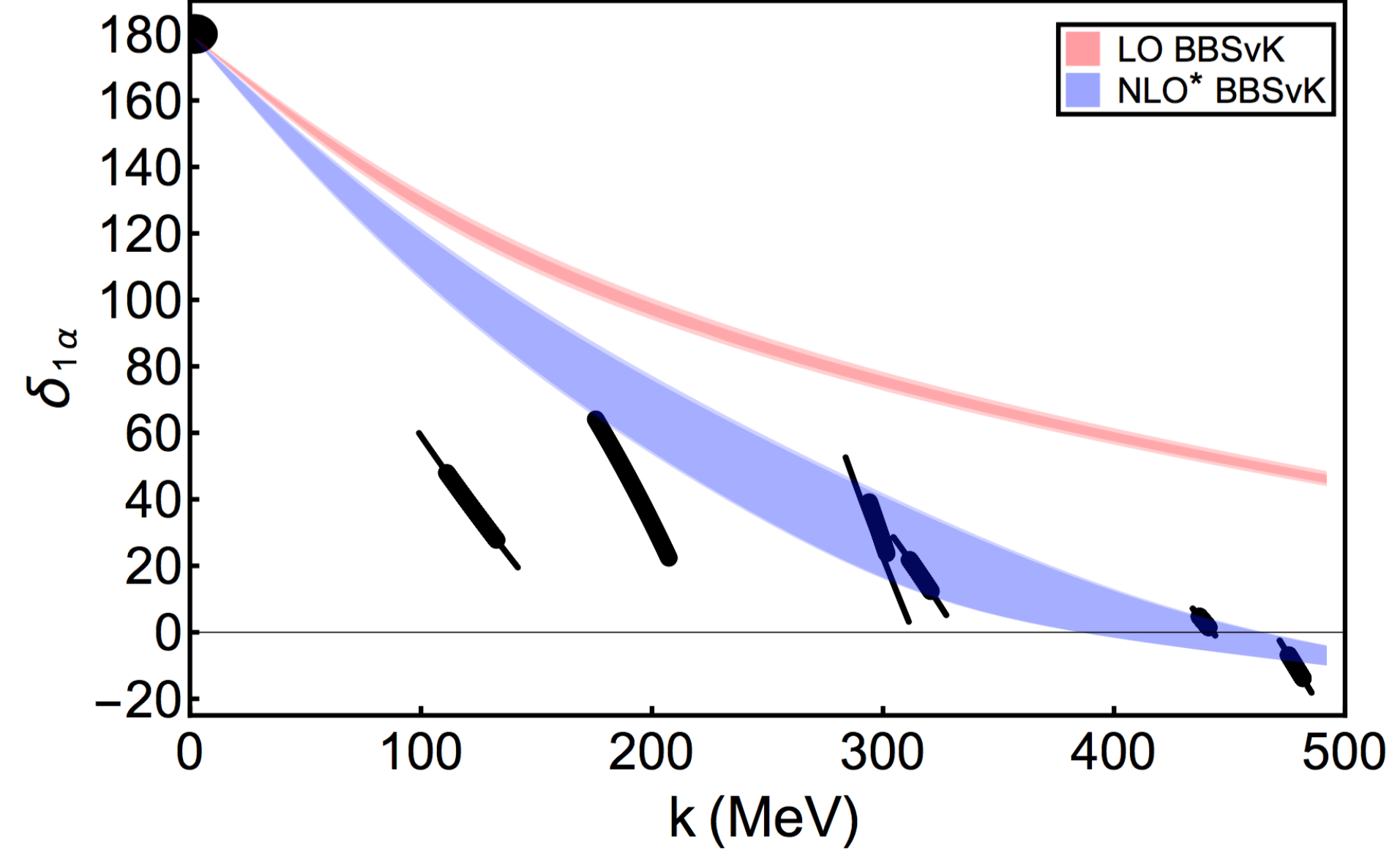}    
  \caption{
The $\siii$-$\diii$ coupled channels  scattering phase shift, $\delta_{1\alpha}$,  
along with BBSvK fits at LO and NLO$^*$.
At LO there is one  parameter that is fit to recover the deuteron pole, giving the red-shaded region, 
while at NLO$^*$ there is one additional fit parameter, giving the blue-shaded region. 
The darker (lighter) shaded regions correspond to the statistical (statistical and systematic uncertainties combined in quadrature).
}
  \label{fig:NN3s1BBSvK}
\end{figure}

At NLO, the expansion becomes more complicated with different interactions in the S-waves and D-waves. 
Without being able to separately resolve the $\delta_{1\alpha}$ and $\delta_{1\beta}$ phase shifts, 
only the common terms can be determined. To this end, we have defined  
 NLO$^*$ to be LO with the inclusion of the leading momentum-dependent two-nucleon contact interaction 
that is also common to both the S-waves and D-waves,  but  omitting other NLO contributions.
NLO$^*$ introduces a single additional parameter beyond LO.

The results of fitting the LO and NLO$^*$ parameters to the results of our LQCD calculations are shown 
in Fig.~\ref{fig:NN3s1BBSvK}.~\footnote{
A square-well 
with a radius of $R=0.30~{\rm fm}$
has been used to regulate the interaction at short distances.
Previous work~\cite{Beane:2001bc}  shows that the observables have corrections that 
depend only on positive powers of $R$ (after refitting coefficients), 
as expected from a Wilsonian renormalization group analysis in the limit $R\rightarrow 0$.
}
The LO fit to the deuteron binding energy leads to phase shifts that significantly over estimate the LQCD results
(this is slso seen in analyses at the physical point).
However, by including the contact-$p^2$ interaction, relatively good agreement is 
found in the NLO$^*$ fit to all the LQCD phase-shift extractions, 
with the exception of the lowest energy point 
\textcolor{\revrevrevcolor}{
(which we attribute to a downward statistical fluctuation whose significance is likely to be reduced at higher orders in the expansion).
}

The values of the scattering parameters resulting from the fits are
\begin{eqnarray}
a^{(\siii)}_{\rm LO} & = & 1.94(09)(17)~{\rm fm}
\qquad
r^{(\siii)}_{\rm LO} \ =\ 0.674(17)(29))~{\rm fm}
\nonumber\\
a^{(\siii)}_{\rm NLO^*} & = & 2.72(22)(27)~{\rm fm}
\qquad
r^{(\siii)}_{\rm NLO^*} \ =\ 1.43(12)(13))~{\rm fm}
\ \ \ ,
\label{eq:BBSvKar}
\end{eqnarray}
which are consistent, within uncertainties, with those obtained in the $\si$ channel with KSW counting.
It is interesting to note that the ratio of scattering length to effective range is $a/r\sim 2$, as was found to be the case at the 
SU(3) symmetric point~\cite{Beane:2012vq,Beane:2013br} .

\begin{figure}[!ht]
  \centering
  \includegraphics[width=0.48 \columnwidth]{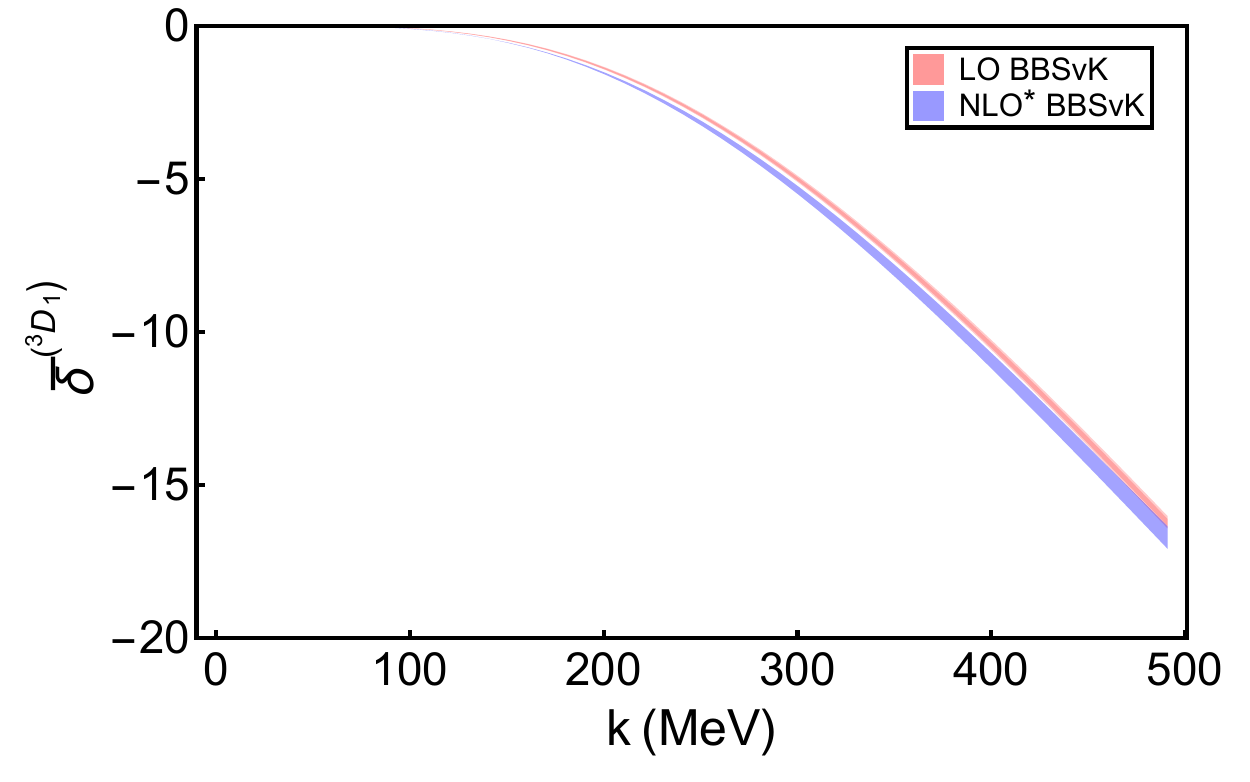}    
  \includegraphics[width=0.48 \columnwidth]{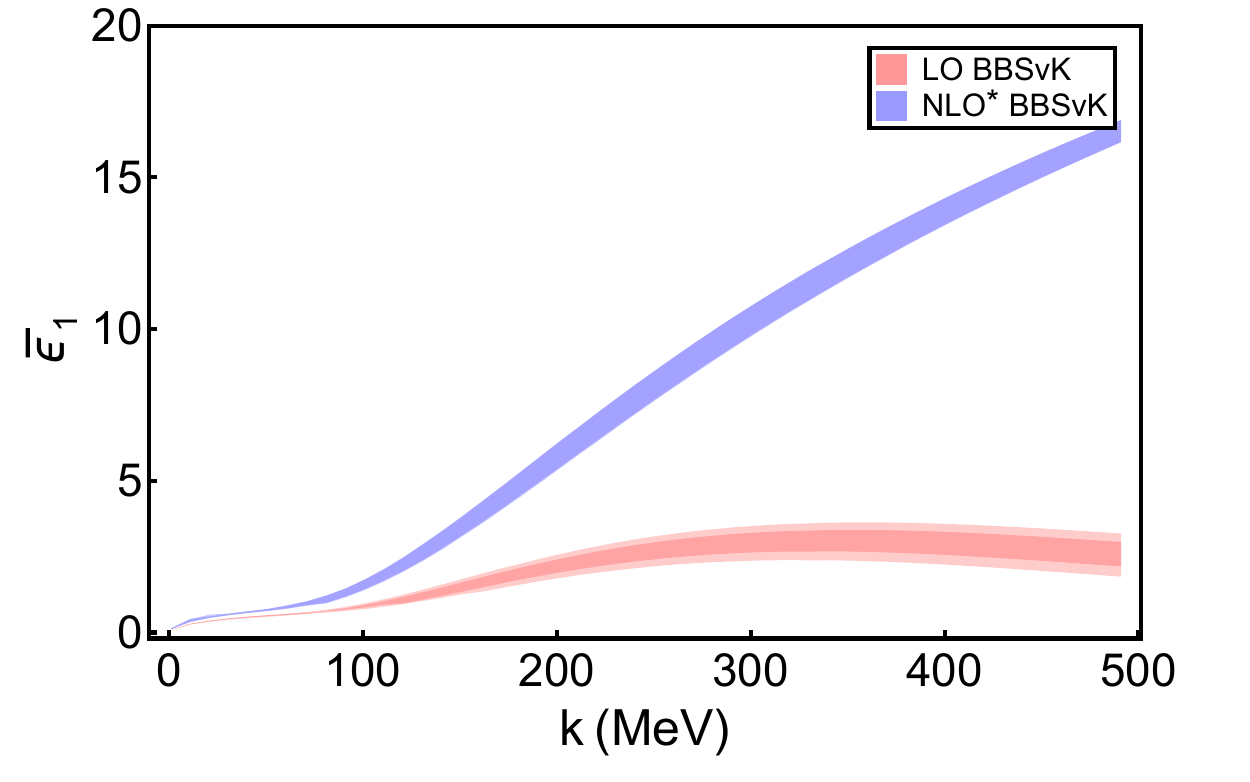}    
  \caption{
The left panel shows
the $\diii$ scattering phase shift,
$\overline{\delta}^{(\diii)}$, in the Stapp Parameterization~\protect\cite{Stapp:1956mz}
along with BBSvK fits at LO and NLO$^*$,
while the right panel shows the mixing parameter, $\overline{\epsilon}_1$.
The darker (lighter) shaded regions correspond to the statistical (statistical and systematic uncertainties combined in quadrature).
}
  \label{fig:NN3s1BBSvKdsd}
\end{figure}

A feature of BBSvK counting  is that predictions can be made for the 
mixing parameter, $\epsilon_1$ and $\delta_{1\beta}$, 
or 
$\overline{\epsilon}_1$ and $\overline{\delta}^{(\diii)}$ in the more familiar Stapp~\cite{Stapp:1956mz} parameterization of the S-matrix.
These are  shown in Fig.~\ref{fig:NN3s1BBSvKdsd}, 
and it is important to keep in mind that   
the coefficients determined from the deuteron pole and S-wave phase shift, contribute to both these quantities.
While the D-wave phase shift is only slightly modified by the NLO$^*$ interaction, $\overline{\epsilon}_1$ is changed dramatically.
In 
this initial investigation, the range  
of the square well interaction has not been varied and estimates of contributions from higher orders have not been included.
It is clear that the ``theory error'' due to truncation of the BBSvK expansion is large for $\overline{\epsilon}_1$, but not for the D-wave phase shift.
In fact, this expansion of $\overline{\epsilon}_1$ is found to be less convergent 
at this pion mass than at the physical point~\cite{Beane:2001bc}.

%%%%%%%%%%%%%%%%%%%%%%%%%%%%%
\section{Conclusions}
\label{sec:concs}

Recovering the experimentally known properties of the two-nucleon systems,
such as
the deuteron bound state, the dineutron virtual-bound state and scattering observables,
from QCD represents a major challenge for Lattice QCD calculations.  
Once verified by comparison to known experimental extractions, 
LQCD calculations hold the promise of refining our knowledge of these systems beyond what 
is possible experimentally, particularly in the neutron-neutron system and more exotic processes involving hyperons.
LQCD calculations have steadily developed in recent years and in the near future calculations of multi-nucleon systems 
with physical quark masses will be available~\cite{Yamazaki:2015LQCD}. 
Eventually these calculations will also include the effects of isospin-breaking and QED.
 In this work, we report the results of calculations of nucleon-nucleon interactions in the $\siii$-$\diii$ 
 coupled channels and the $\si$ channel at a pion mass of 
$m_\pi\sim\pionmass$ in three lattice volumes and at a single 
lattice spacing.
\textcolor{\revrevrevcolor}{
The lattice-spacing artifacts are estimated to be small, entering at ${\cal O}(\Lambda_{\rm QCD}^2 b^2)$,  
and are expected to modify the binding energies and 
phase shifts by amounts that are small compared with the quoted statistical and systematic uncertainties.
}
Both the deuteron and dineutron are found to be bound at this pion mass, consistent with expectations based upon previous calculations.  
The phase shifts in both channels are determined at a few discrete momenta and, in both channels, 
a zero in the phase shift is found to occur near the momentum at which a zero is observed in nature.  
Calculations of increased precision and kinematic coverage will further our understanding of the two-nucleon systems 
at this set of quark masses. 
Further calculations at other quark masses will enable direct  comparison with experimental  
extractions and will elucidate important features of the chiral nuclear forces that are not accessible in experiment alone.

%%%%%%%%%%%%%%%%%%%%%%%%%%%%%%%%%%%%%%%%%%%%%%%

%%%%%%
\begin{acknowledgments}
We thank  Andre Walker-Loud and  Thomas Luu for collaboration during initial stages of this work,
and Zohreh Davoudi  
\textcolor{\revrevcolor}{and Raul Briceno}
 for comments on the manuscript.
  Calculations were performed using computational resources provided
  by the Extreme Science and Engineering Discovery Environment
  (XSEDE), which is supported by National Science Foundation grant
  number OCI-1053575, NERSC (supported by U.S. Department of
  Energy Grant Number DE-AC02-05CH11231),
  and by the USQCD collaboration.  
  This research used resources of the Oak Ridge Leadership 
  Computing Facility at the Oak Ridge National Laboratory, which is supported 
  by the Office of Science of the U.S. Department of Energy under Contract 
  No. DE-AC05-00OR22725.
  Parts of the calculations used the Chroma software
  suite~\cite{Edwards:2004sx}.  
  SRB was partially supported by NSF continuing grant PHY1206498 
  and by U.S. Department of Energy through Grant Number DE-SC001347.  
  WD was supported in part by the U.S. Department of Energy Early Career Research Award DE-SC0010495.
  KO was supported by the U.S. Department of Energy through Grant
  Number DE-FG02-04ER41302 and through contract Number DE-AC05-06OR23177
  under which JSA operates the Thomas Jefferson National Accelerator
  Facility.  
  The work of AP was supported by the contract
  FIS2011-24154 from MEC (Spain) and FEDER. 
  MJS was supported in part
  by DOE grant No.~DE-FG02-00ER41132.  
  This research was supported in part by the National Science Foundation under Grant No. NSF PHY11- 25915 
  and WD and MJS acknowledge the Kavli Institute for Theoretical Physics for hospitality during completion of this work. 
 \end{acknowledgments}
%

%%%%%%%%%%%%%%%%%%%%%%%%%%%%%%%%%%%%%%%%%%%%%%%%%%

\pagebreak

\appendix
\section{Erratum - 2020}

\subsection*{Erratum - 2020: Abstract}
\noindent
An error has been found in the analysis code that processed the deuteron binding energy from the $L=24$ 
ensemble at a pion mass of $m_\pi\sim 450~{\rm MeV}$
presented in Ref.~\cite{Orginos:2015aya} from 2015. We correct the impacted aspects of the paper.
In addition, 
several typographical errors have been corrected in this erratum, none of which influence the discussions and conclusions appearing in 
Ref.~\cite{Orginos:2015aya}.

\subsection*{Erratum - 2020: Introduction}

\noindent
In the process of NPLQCD's ongoing re-analysis of the baryon-baryon  correlation functions at $m_\pi \sim 450$ MeV, 
a subset of which were the subject of  Ref.~\cite{Orginos:2015aya}, an error was identified.~\footnote{
We thank Marc Illa, a recent member of NPLQCD collaboration, for his dedicated effort to compare present and past results, and for bringing to our attention discrepancies which led to the corrections presented in this erratum.}
In particular, it was found that while the  effective mass plots and fits shown in Fig.~11 of the manuscript are correct, an incorrect value of the binding energy and associated 
uncertainties was propagated through parts of the downstream analysis.  
The deuteron binding energy obtained from the $L=24$ ensemble at a pion mass of $m_\pi\sim 450~{\rm MeV}$ that was published and analyzed 
was  (incorrectly) $B_d=19.6(1.2)(1.6)(0.2)~{\rm MeV}$, 
which when combined with results obtained from the $L=32, 48$ ensembles, 
incorrectly produced an infinite-volume extrapolated binding energy of  $B_d^{(\infty)} = \Bdo$ .
The correct value of the deuteron binding energy in the $L=24$ volume is   ${\bf B_d=20.9(1.2)(3.3)(0.2)~{\rm MeV} }$, 
which leads to an infinite volume extrapolated binding energy of
${\bf B_d^{(\infty)} = \Bdn }$.   
The effect of this correction is to increase the systematic uncertainty of the deuteron binding 
energy extracted from the $L=24$ volume used  in the downstream analysis by approximately a factor of two.

The corrected systematic uncertainty was propagated through the downstream analysis, 
and fortunately, it does not invalidate  conclusions or discussions related to the 
results of these lattice QCD calculations.   A complete and detailed summary of the impact of this error on the published
Tables and Figures are detailed  in the  lists that follow.

The re-analysis also uncovered a small number of typographical errors (transcription errors) that also appear in the published paper~\cite{Orginos:2015aya}.  These 
errors are corrected in the last section.

%%%%%%%%%%%%%%%%%%%%%%%%%%%%%%%%%%%%%%%%%%%%%%%%
\subsection*{Erratum - 2020: IV. The $\siii$-$\diii$ Coupled Channels and the Deuteron}

%%%%%%%%%%%%%%%%%%%%%
\subsubsection*{Erratum - 2020: A. The Deuteron}

\noindent
The following lists the correct values and figures in this subsection:
\begin{enumerate}

\item
Published numbers presented in the first and fourth rows of
Table~VI 
are incorrect.
The caption is unchanged.
The revised Table~VI follows.
\renewcommand{\arraystretch}{1.5}
\begin{table}[!h]
   \renewcommand\thetable{VI}
   \begin{center}
\begin{minipage}[h]{16.5 cm}
  \caption{
 The deuteron binding energies extracted from plateaus in the EMPs 
shown in Fig.~11,
along with the infinite-volume extrapolated value.
The size of the FV effects is characterized by $e^{-\kappa L}$, shown in the last column.
The first uncertainty corresponds to the statistical uncertainty associated with the fit, the second corresponds 
to the systematic uncertainty associated with the selection of the fitting interval (determined by varying this range).  
In the case of dimensionful quantities, the third uncertainty is associated with scale setting.
For the infinite-volume values of the binding energy, the last uncertainty 
is introduced by the finite-volume extrapolation in Eq.~(11), 
and is estimated by considering the effect of omitted terms scaling as 
$e^{-2 \kappa_0 L}/ L$.
 }  
\label{tab:deutbindingE}
\end{minipage}
\setlength{\tabcolsep}{0.3em}
\begin{tabular}{c|cc|c}
\hline
      Ensemble & 
      $ \Delta E $\ (l.u.) & 
      $B_d $\ (MeV) &
      $e^{-\kappa L}$
      \\
\hline
\cfga &     -0.0123(07)(19) & 20.9(1.2)(3.3)(0.2) & 0.103\\
\cfgb &     -0.01037(89)(96) & 17.5(1.5)(1.6)(0.2) & 0.063\\
\cfgc &     -0.0078(12)(19) & 13.3(2.0)(3.2)(0.2) & 0.027\\
\hline 
$L=\infty$
& $-0.0084^{+(10)(22)(01)}_{-(12)(22)(01)}$ 
& $\Bdn$  
& \\
\hline
\end{tabular}
\begin{minipage}[h]{16.5 cm}
\vskip 0.0cm
\noindent
\end{minipage}
\end{center}
\end{table}     

\item
The region in the $c_1-B_d^{(\infty)}$ parameter space bounded by $\chi^2\rightarrow\chi^2_{\rm min}+1$
is slightly modified. 
The figure caption is unchanged.
The revised  Fig.~12 follows.
\renewcommand\thefigure{\arabic{figure}}
\setcounter{figure}{11}
\begin{figure}[h]
  \centering
  \includegraphics[width=0.48 \columnwidth]{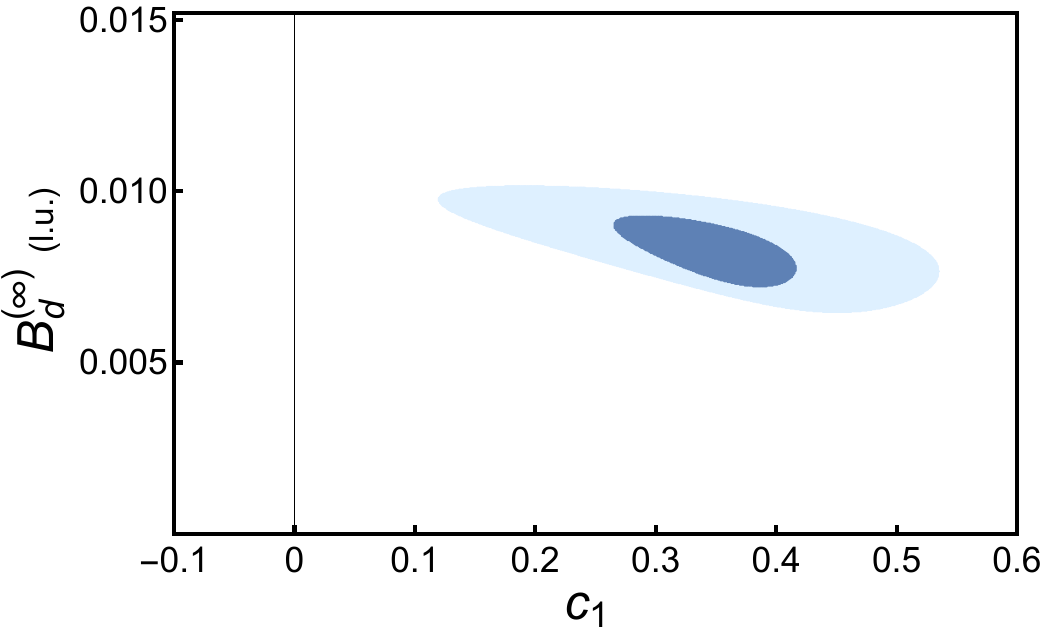}    
  \caption{
 The region in $c_1$-$B_d^{(\infty)}$ parameter space 
 defined by $\chi^2\rightarrow\chi^2_{\rm min}+1$.
 The inner region is defined by the statistical uncertainty, while the outer region is defined by the statistical and 
 systematic uncertainties combined in quadrature.
  }
  \label{fig:DeutBcE}
\end{figure}

\item
Equation~(12) should be:
\begin{eqnarray}
B_d^{(\infty)} & = & \Bdn~{\rm MeV}
\ \ \ .
\label{eq:BdE}
\end{eqnarray}

\item
The deuteron binding energy given in the text  below Eq.~(12) should be:
$B_d^{(\infty)} = \BdSUMMARYn~{\rm MeV}$.

\end{enumerate}

\vskip 0.1in
\noindent
In the subsubsection entitled {\it A Compilation of Deuteron Binding Energies from LQCD}, 
the only correction, resulting from the previously discussed error,
is to the deuteron binding energy at this pion mass that is shown in Fig.~14.
The figure caption is unchanged.
The revised Fig.~14 follows.
\renewcommand\thefigure{\arabic{figure}}
\setcounter{figure}{13}
\begin{figure}[!ht]
  \centering
  \includegraphics[width=0.48 \columnwidth]{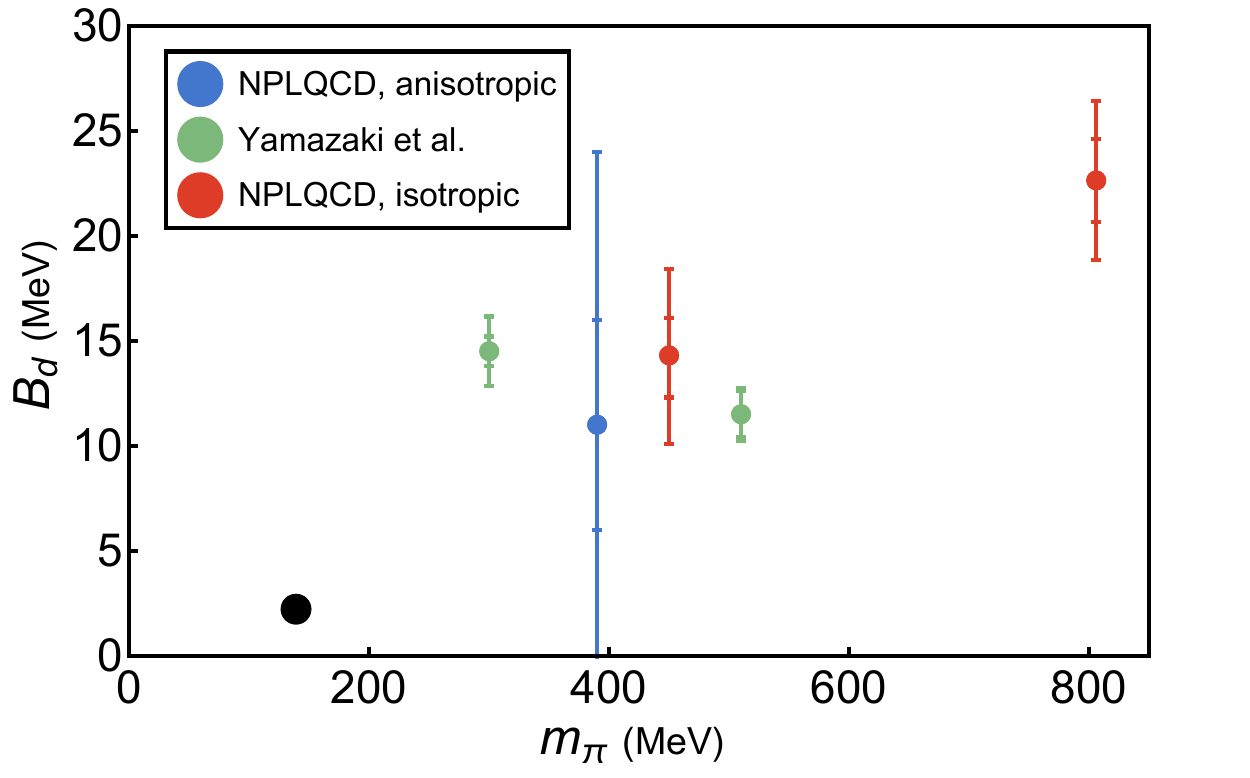}\ \
  \caption{
  The pion mass dependence of the deuteron binding energy calculated with LQCD.
  The NPLQCD anisotropic-clover  
  result is from Ref.~[10], the Yamazaki {\it et al}. results are from Refs.~[14,20]
  and the NPLQCD isotropic-clover results are from this work and  Ref.~[13].
  The black disk corresponds to the experimental binding energy.
    }
  \label{fig:DeutALLlqcdE}
\end{figure}
%

%%%%%%%%%%%%%%%%%%%%%
\subsubsection*{Erratum - 2020: B. Scattering in the $\siii$-$\diii$ Coupled Channels}
\noindent
As a result of the previously described error, 
the value of the deuteron binding energy and binding momentum  shown in Tables and Figures in this subsection are incorrect.

\begin{enumerate}

\item
The deuteron binding momentum given in Table~VII is modified (in the row labelled as ensemble "All").
The Table caption is unchanged.
The corrected value is $i 0.284^{+(28)(24)}_{-(30)(34)}$, as shown in Table~\ref{tab:errTabs7}.

\item
The binding momentum and $k\cot\delta$ values shown in the left panel of Fig.~18 are modified.
There are no other changes to the figure or caption. 
The revised  Fig.~18 is shown below.
\renewcommand\thefigure{\arabic{figure}}
\setcounter{figure}{17}
\begin{figure}[!ht]
  \centering
  \includegraphics[width=0.48 \columnwidth]{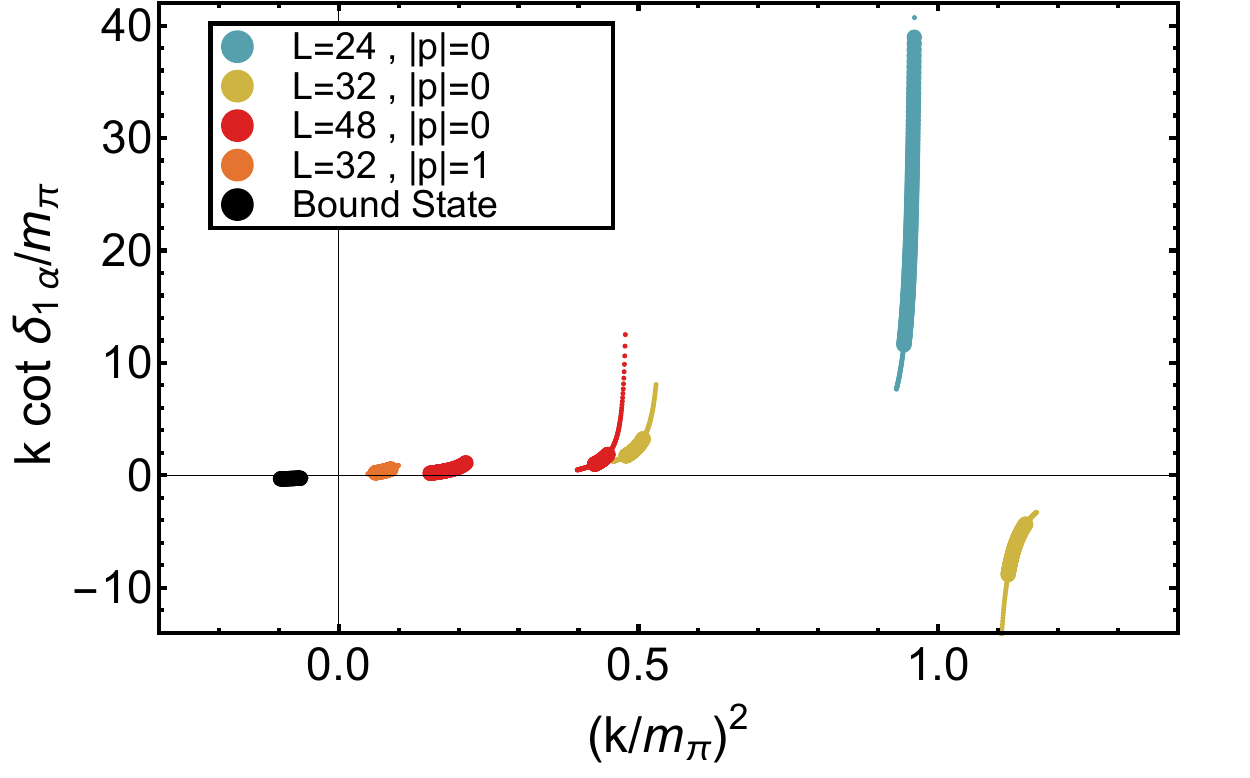}      
  \includegraphics[width=0.48 \columnwidth]{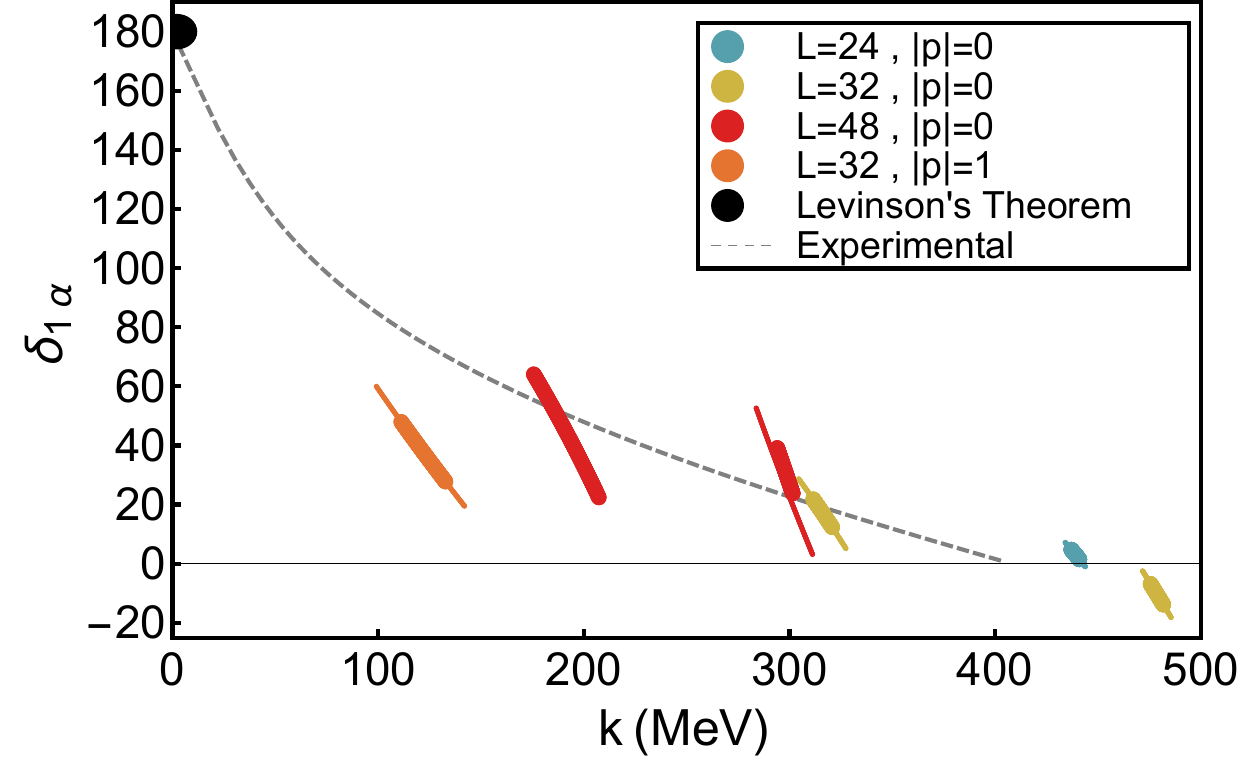}    
  \caption{
Scattering in the $\siii$-$\diii$ coupled channels.  
The left panel shows $k^*\cot\delta_{1\alpha}/m_\pi$ as a function of $k^{*2}/m_\pi^2$,
while the right panel shows the phase shift as a function of momentum in MeV,
assuming that $\delta_{1\beta}$ and the D-wave and higher partial-wave phase shifts vanish.  
The thick (thin) region of each result correspond to the statistical uncertainty (statistical and systematic uncertainties combined in quadrature).
The black circle in the right panel corresponds to the known result from Levinson's theorem, while
the dashed-gray curve corresponds to the  phase shift extracted from the Nijmegen partial-wave analysis of experimental data~[61]
  }
  \label{fig:NN3s1kcotdeltaE}
\end{figure}

\item
The correct value of the deuteron binding momentum leads to changes to the corresponding point in the left panel shown in Fig.~19, 
and the associated fit regions,
and to the 
region of scattering parameters shown in the right panel of the same figure.
The caption is unchanged.
The revised Figure 19 is shown below.
\renewcommand\thefigure{\arabic{figure}}
\setcounter{figure}{18}
\begin{figure}[!h]
  \centering
  \includegraphics[width=0.48 \columnwidth]{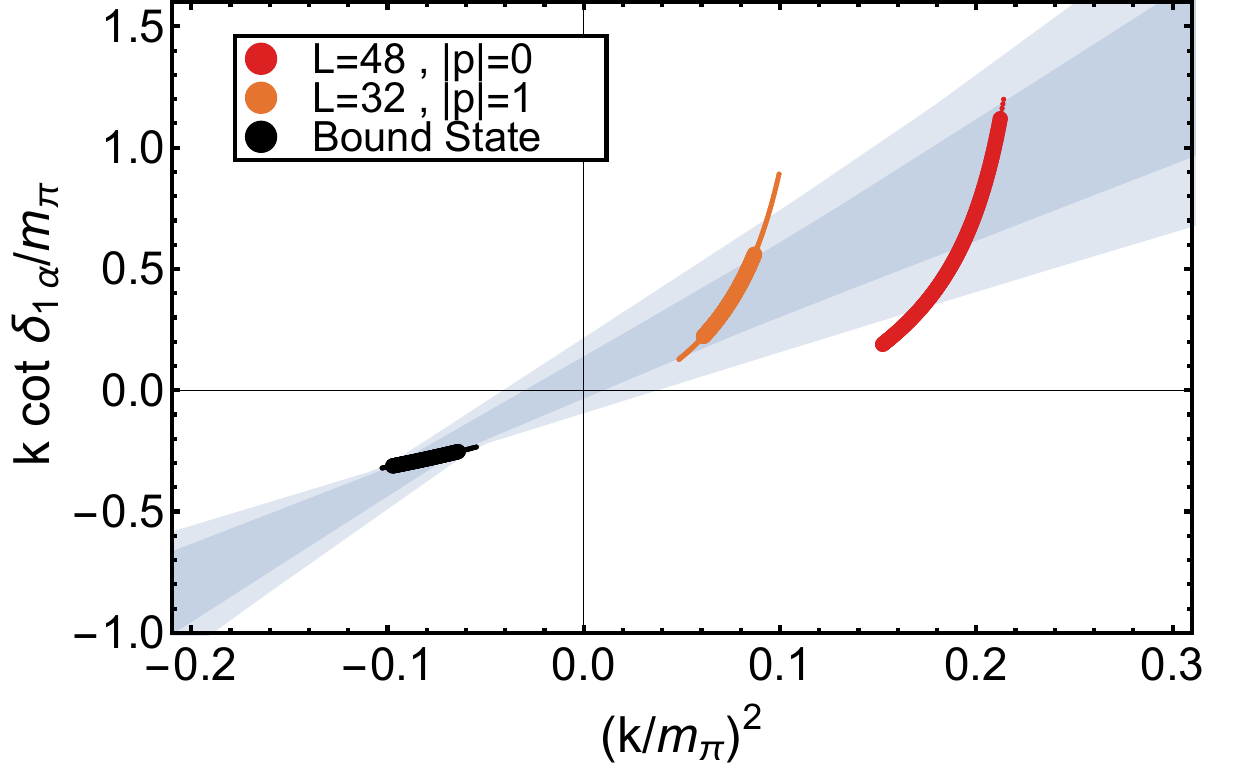}    
  \includegraphics[width=0.44 \columnwidth]{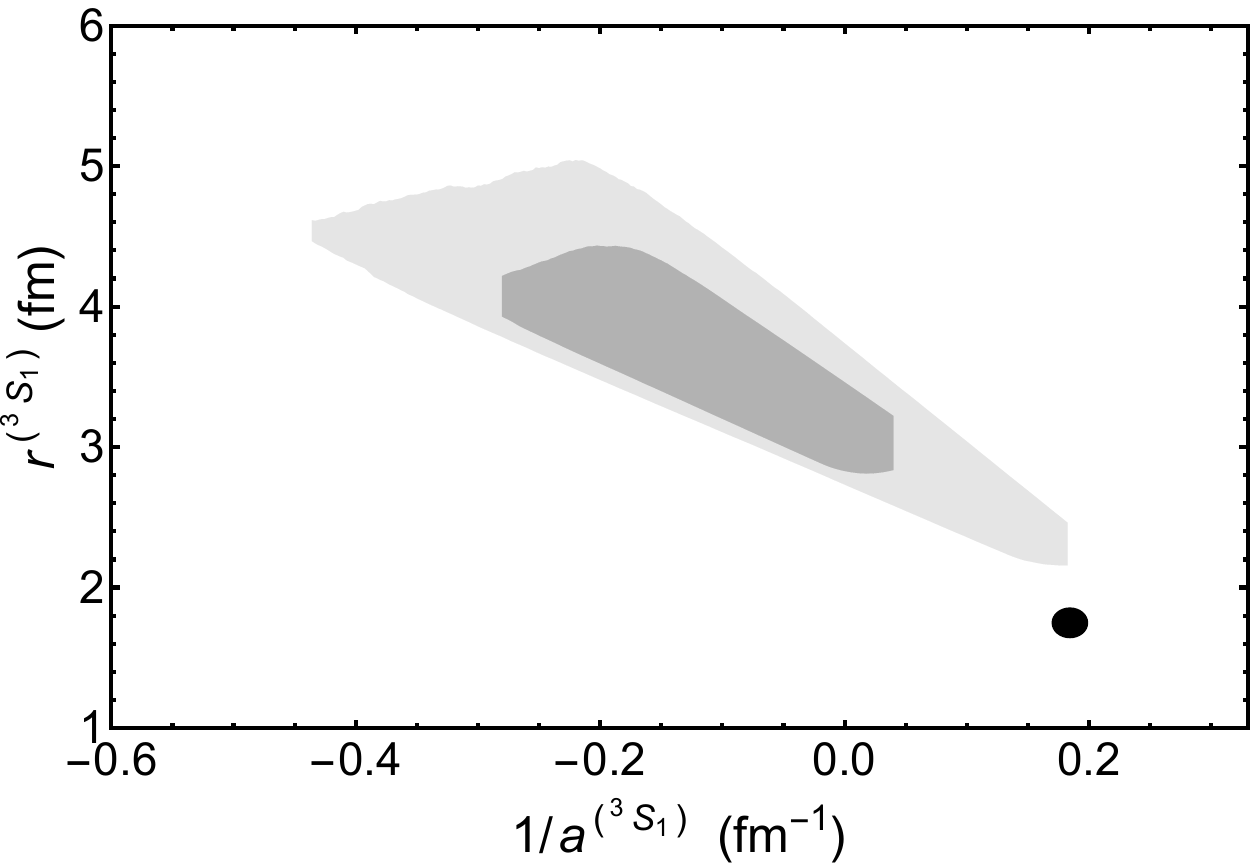}    
  \caption{
Scattering in the $\siii$-$\diii$ coupled channels below the start of the t-channel cut,
$k^{*2} < m_\pi^2/4$,
assuming that $\delta_{1\beta}$ and the D-wave and higher partial-wave phase shifts vanish.  
The left panel shows solid region corresponding to linear fits associated with the statistical uncertainty
and the statistical and systematic uncertainties combined in quadrature.
The right panel shows the scattering parameters, $1/a^{(\siii)}$ and $r^{(\siii)}$ 
determined from fits to scattering results below the t-channel cut.
The solid circle corresponds to the experimental values.
}
  \label{fig:NN3s1kcotdeltaERE}
\end{figure}

\item
Equation~(17) should be replaced with:
\setcounter{equation}{16}
\begin{eqnarray}
\left( m_\pi a^{(\siii)} \right)^{-1} & = & 
-0.03^{+(0.07)(0.07)}_{-(0.10)(0.18)}
\ \ ,\ \ 
m_\pi r^{(\siii)} \ =\ 
7.9^{+(2.5)(3.8)}_{-(1.6)(1.8)}
\nonumber\\
\left(  a^{(\siii)} \right)^{-1} & = & 
-0.08^{+(0.16)(0.17)}_{-(0.24)(0.42)}~{\rm fm}^{-1}
\ \ ,\ \ 
 r^{(\siii)} \ =\ 
3.5^{+(1.1)(1.7)}_{-(0.7)(0.8)}~{\rm fm}
\ \ \ .
\end{eqnarray}

\end{enumerate}

%%%%%%%%
\subsection*{Erratum - 2020: Typographical Errors}
\noindent
A number of typographical errors in the published paper  have been identified.  
Typographical errors were found in Tables II, IV, VII and X.  
These  errors were made in preparing for publication, and do not appear in the analysis pipeline from which physics conclusions were deduced.
The corrections to Tables II and IV of the 2015 paper, Ref.~\cite{Orginos:2015aya},
are given in Table~\ref{tab:errTabs24}, 
corrections to Table VII are given in Table~\ref{tab:errTabs7}, and 
corrections to Table X are given in Table~\ref{tab:errTabs10}.
\begin{table}[ht!]
\renewcommand{\thetable}{A}
\renewcommand*{\arraystretch}{1.5}
\centering
\caption{Correcting typographical errors appearing in Tables II and IV of our 2015 paper~\cite{Orginos:2015aya}. }
\begin{tabular}{c|c|cc|cc}
\hline
 &  & 
 \multicolumn{2}{c|}{Entries in 2015 paper~\cite{Orginos:2015aya}} & \multicolumn{2}{c}{Correct Entries}  \\ \hline
Meson & Ensemble & $|\mathbf{n}|=0$ (l.u.) & MeV & $|\mathbf{n}|=0$ (l.u.) & MeV \\ \hline
$\pi^{\pm}$ & $L=\infty$ (Table II) & $0.26606(14)(08)$ & - & $0.26607(13)(11)$ & -   \\ 
 & $L=\infty$ (Table IV)  & $0.26614(15)(15)$ & $449.9(0.3)(0.3)(4.6)$ &    $0.26607(13)(11)$ & $449.9(0.2)(0.2)(4.6)$
 \\ \hline
$K^{\pm}$ & $L=48$ (Table II) & $0.35236(16)(25)$ & - & $0.35236(16)(05)$ & - \\  \cline{2-6}
  & $L=\infty$ (Table IV)  & $0.35241(12)(11)$ & $595.9(0.2)(0.2)(6.1)$ & $0.35240(11)(03)$ &  $595.4(0.2)(0.1)(6.1)$ 
   \\  \hline 
\end{tabular} 
\label{tab:errTabs24}
\end{table}
\begin{table}[ht!]
\renewcommand{\thetable}{B}
\renewcommand*{\arraystretch}{2}\centering
\caption{
Correcting typographical errors appearing in Table VII of our 2015 paper~\cite{Orginos:2015aya}.
Corrected values are highlighted in bold-red for convenience.
The entries highlighted in bold-blue result from the error described in the previous section and are not typographical in origin.
}
\resizebox{\textwidth}{!}{
\begin{tabular}{c|c|ccc|ccc}
\hline
 &  & \multicolumn{3}{c|}{Entries in 2015 paper~\cite{Orginos:2015aya}} &\multicolumn{3}{c}{Correct Entries} \\ \hline
 Ensemble & $|\mathbf{P}_{\text{tot}}|$ (l.u.)  & $k^*/m_{\pi}$ & $k^*\cot\delta_{1\alpha}/m_{\pi}$ & $\delta_{1\alpha}$ (degrees) & $k^*/m_{\pi}$ & $k^*\cot\delta_{1\alpha}/m_{\pi}$ & $\delta_{1\alpha}$ (degrees) \\ \hline
All & 0 & $i 0.294^{+(17)(27)}_{-(18)(24)}  $ & $- 0.294^{+(17)(27)}_{-(18)(24)}  $ & - & \textcolor{blue}{$ \bm {i 0.284^{+(28)(24)}_{-(30)(34)}  }$}  & \textcolor{blue}{$ \bm { - 0.284^{+(28)(24)}_{-(30)(34)}   }$} & - \\
 \hline
$24^3 \times 64$ & 0 & $0.9754^{+(44)(98)}_{-(45)(99)}$  & - & $3.1(1.7)(3.7)$ & $0.9754^{+(44)(98)}_{-(45)(99)}$  & - & $3.1(1.7)(3.7)$ \\ \hline
$32^3\times 96$ & 0 & $0.702^{+(10)(23)}_{-(10)(24)}$ & $2.3^{+(1.0)(5.7)}_{-(0.55)(0.89)}$ & $17(5)(11)$ & $0.702^{+(10)(23)}_{-(10)(24)}$ & $2.3^{+(1.0)(5.7)}_{-(0.55)(0.89)}$ & $17(5)(11)$  \\ 
$32^3\times 96$ & 0 & $1.065^{+(07)(16)}_{-(08)(17)}$ & $-5.4^{+(1.4)(2.1)}_{-(2.9)(29.5)}$ & $-11.1(3.8)(8.5)$ & $1.065^{+(07)(16)}_{-(08)(17)}$ & $\color{red} {  \bm { -5.4^{+(1.1)(1.8)}_{-(3.4)(18.7)} }  }$ & $\color{red} {  \bm { -11.1^{+(4.3)(7.5)}_{-(2.8)(6.5)}}  }$ \\ 
$32^3\times 96$ & 1 & $0.270^{+(26)(29)}_{-(40)(51)}$ & $0.35^{+(24)(15)}_{-(59)(20)}$ & $38^{+(13)(23)}_{-(11)(16)}$ & \textcolor{red}{  \bm {  $0.270^{+(25)(38)}_{-(23)(44)}  }$ }& $\color{red} {  \bm { 0.35^{+(21)(50)}_{-(13)(18)} }  }$ & $38^{+(13)(23)}_{-(11)(16)}$
\\ \hline
$48^3\times 96$ & 0 & $0.426(03)(12)$ & $0.45^{+(67)(34)}_{-(26)(08)}$ & $44^{+(21)(07)}_{-(21)(08)}$ & 
$\color{red} {\bm{0.426 (35)(12)}}$ & $0.45^{+(67)(34)}_{-(26)(08)}$ & $44^{+(21)(07)}_{-(21)(08)}$ \\ 
$48^3\times 96$ & 0 & $0.662(08)(29)$ & $0.35^{+(0.14)(3.0)}_{-(0.09)(0.21)}$ & $26^{+(07)(25)}_{-(07)(22)}$ & 
$0.662(08)(29)$ & $\color{red} {  \bm {  1.35^{+(0.51)(11.2)}_{-(0.36)(0.79)}  }  }$ & $26^{+(07)(25)}_{-(07)(22)}$ \\
\hline 
\end{tabular} 
}
\label{tab:errTabs7}
\end{table}
\begin{table}[ht!]
\renewcommand{\thetable}{C}
\renewcommand*{\arraystretch}{2}
\centering
\caption{
Correcting typographical errors appearing in Table X of our 2015 paper~\cite{Orginos:2015aya}.
Corrected values are highlighted in bold-red for convenience.}
\resizebox{\textwidth}{!}{
\begin{tabular}{c|c|ccc|ccc}
\hline
 &  & \multicolumn{3}{c|}{Entries in 2015 paper~\cite{Orginos:2015aya}} &\multicolumn{3}{c}{Correct Entries} \\ \hline
 Ensemble & $|\mathbf{P}_{\text{tot}}|$ (l.u.)
  & $k^*/m_{\pi}$ & $k^*\cot\delta^{(\si)}/m_{\pi}$ & $\delta^{(\si)}$ (degrees) & $k^*/m_{\pi}$ & $k^*\cot\delta^{(\si)}/m_{\pi}$ & $\delta^{(\si)}$ (degrees) 
 \\ \hline
All & 0 & $i 0.274^{+(19)(26)}_{-(20)(44)}  $ & $-0.274^{+(19)(26)}_{-(20)(44)}  $& - & $i 0.274^{+(19)(26)}_{-(20)(44)}  $ & $-0.274^{+(19)(26)}_{-(20)(44)}  $& -  \\
\hline
$24^3 \times 64$ &    0 & $0.954^{+(08)(18)}_{-(08)(19)}$ &  $5.0^{+(2.0)(10.0)}_{-(1.1)(1.8)}$ & $10.8^{+(3.0)(6.5)}_{-(3.0)(6.7)}$ 
& \textcolor{red}{ $ \bm { 0.963^{+(08)(11)}_{-(08)(11)}}$}
&  \textcolor{red}{ $  \bm { 7.1^{+(4.2)(13.2)}_{-(2.0)(2.1)} }$} 
& \textcolor{red}{ $ \bm { 7.7^{+(2.8)(4.1)}_{-(2.8)(4.2)}  }  $}  
\\ \hline
$32^3\times 96$ &  0 & $0.691^{+(09)(16)}_{-(09)(16)}$ &    $1.7^{+(0.5)(1.1)}_{-(0.3)(0.5)} $ & $22.0^{+(4.2)(7.0)}_{-(4.2)(7.2)}$ & $0.691^{+(09)(16)}_{-(09)(16)}$ &    $1.7^{+(0.5)(1.1)}_{-(0.3)(0.5)} $ & $22.0^{+(4.2)(7.0)}_{-(4.2)(7.2)}$ \\
$32^3\times 96$  &  0 & $1.079^{+(05)(10)}_{-(05)(10)}$ &    $-3.3^{+(0.4)(0.7)}_{-(0.6)(1.5)} $ & $-18.3(2.6)(5.2) $ & $1.079^{+(05)(10)}_{-(05)(10)}$ &    $-3.3^{+(0.4)(0.7)}_{-(0.6)(1.5)} $ & $-18.3(2.6)(5.2) $  \\
$32^3\times 96$  &  1 & $0.220^{+(28)(32)}_{-(32)(42)}$ &    $0.13^{+(10)(14)}_{-(08)(08)} $ &   $60^{+(14)(20)}_{-(12)(14)} $ & $0.220^{+(28)(32)}_{-(32)(42)}$ &    $0.13^{+(10)(14)}_{-(08)(08)} $ &   $60^{+(14)(20)}_{-(12)(14)} $
 \\ \hline
$48^3\times 96$  &  0 &  $0.453(11)(29)$ & $0.89^{+(39)(3.7)}_{-(23)(44)}$ & $27^{+(07)(18)}_{-(07)(20)}$ &  $0.453(11)(29)$ & $0.89^{+(39)(3.7)}_{-(23)(44)}$ & $27^{+(07)(18)}_{-(07)(20)}$   \\ \hline
\end{tabular}
}
\label{tab:errTabs10}
\end{table}

%%%%%%%%
\subsection*{Erratum - 2020: Acknowledgements}
\noindent
We are indebted to Marc Illa for his exceptional work in thoroughly re-analyzing and better understanding the NPLQCD
$m_\pi\sim 450~{\rm MeV}$ lattice QCD data set, and bringing defects in our published work to our attention, which led to this erratum.

%%%%%%%%%%%%%%%%%%%%%%%%%%%%%%%%%%%%%%%%%%%%%%%%%%
\bibliography{bib_AoneAndTwo,bib_AoneAndTwo_2020}
\end{document}